\documentclass[aps,prd,twocolumn,superscriptaddress]{revtex4-1}
\usepackage{graphicx}% Include figure files
\usepackage{amsmath,amssymb}
\usepackage{aas_macros} % Abbreviations for journals
\begin{document}

\title{Parametrization of gamma-ray production cross-sections for $pp$ interactions in a broad proton energy range from the kinematic threshold to PeV energies}

\author{Ervin~Kafexhiu}
\affiliation{Max-Planck-Institut f\"ur Kernphysik, 
             Saupfercheckweg 1, D-69117 Heidelberg, GERMANY}
\email[E-mail us at: ]{ervin.kafexhiu@mpi-hd.mpg.de}

\author{Felix~Aharonian}
\affiliation{Dublin Institute for Advanced Studies,
31 Fitzwilliam Place, Dublin 2, IRELAND}
\affiliation{Max-Planck-Institut f\"ur Kernphysik, 
             Saupfercheckweg 1, D-69117 Heidelberg, GERMANY}

\author{Andrew~M.~Taylor}
\affiliation{Dublin Institute for Advanced Studies,
31 Fitzwilliam Place, Dublin 2, IRELAND}

\author{Gabriela~S.~Vila}
\affiliation{Instituto Argentino de Radioastronom\'ia (IAR - CONICET),
         C.C. N$^\circ$ 5 (1894), Villa Elisa, Buenos Aires, ARGENTINA}

%\date{}
\begin{abstract}
%\begin{center}
Using publicly available Monte Carlo codes as well as compilation of published data on p--p interactions for proton kinetic energy below 2 GeV, we parametrize the energy spectra and production rates of $\gamma$-rays by simple but quite accurate ($\leq 20 \%$) analytical expressions in a broad range from the kinematic threshold to PeV energies. 
%\end{center}
\end{abstract}

%\keywords{some keywords here ...}

\maketitle

\section{Introduction \label{intro}}
Gamma-ray production through p--p inelastic collisions plays an important role in our understanding of both fundamental physics and high energy astrophysical phenomena. The main source of these $\gamma$-rays is the decay of light mesons, such as, $\pi^0\to2\gamma$, $\eta\to2\gamma$, etc. Heavy particles that are produced through p--p inelastic collisions will decay quickly and produce lots of pions. Therefore, the main source of $\gamma$-rays is the $\pi^0\to2\gamma$ decay. In practice, we are interested in the $\gamma$-ray spectrum for a given proton collision energy. For this we have to know the $\pi^0$ spectrum for that specific proton energy and then compute the $\gamma$-ray spectrum using the kinematics of the two-body decay.

The $\pi^0$ production spectrum is a result of different hadronic mechanisms that underlie p--p inelastic collisions. Experiments show that for proton kinetic energies $T_p<3$~GeV the $\pi^0$ production is dominated by the baryon resonance production. At higher energies observations show that Feynman scaling is violated due to multiple partonic interactions which lead to an increase of the particle rapidity distribution plateau. Also at such energies the diffraction dissociation mechanism becomes important. This mechanism is responsible for producing large rapidity gaps in the particle rapidity distribution, it affects the particle production in the very forward region (i.e. very high energy secondary particle production) and accounts for the logarithmic increase of the total cross section. 

Different models of pion production were first developed to explain cosmic ray and low energy accelerator data. The isobar model \citep[see e.g.][]{Yuan1956isobar, Lindenbaum1957isobar}, was introduced to describe pion production through a group of baryon resonances $N^*$ and $\Delta$, the most important of which is the $\Delta(1232)$ baryon. Very close to the kinematic threshold, pions are produced with very low speeds. As a result, their spectra are modified due to strong interaction with the final state particles. This effect of final state interaction (FSI) has been studied by \citep{Watson1952,Gell-Mann1954}.

Another model that was developed to explain the cosmic ray data with proton kinetic energies $T_p>5$~GeV is the statistical model or the so called fireball model \citep{Lewis1948fireball, Fermi1950fireball, Umezawa1951fireball,Heisenberg1952fireball, Landau1956fireball, Cocconi1958_2fireball}. This model considers the formation of an intermediate state, a ``hot gas of pions'', which then decays isotropically by producing many pions with a quasi-thermal energy distribution.

For calculations with broader proton energy spectra, ref.~\cite{Stecker1968, Stecker1971} married the isobar and the fireball models in a single isobar-fireball model. Pion production at low energies in this model is calculated using the $\Delta(1232)$ baryon resonance. At higher energies it uses the fireball model calculations with a smooth connection to the low energy region.

A different approach was introduced by e.g. \cite{Badhwar1977,Stephens1981,TanNg1983}. They did not model pion production, instead, they fitted the invariant $\pi^0$ production differential cross section ($E\,d\sigma/dp^3$) using accelerator data. The formula that was introduced obeyed the scaling hypothesis \cite{Feynman1969scaling, KNOscaling}. By integrating this fit formula one obtains the $\pi^0$ spectrum and the $\gamma$-ray spectrum. The same approach but in a more comprehensive way, was later considered by \cite{Blattnig2000}. In the same philosophy \cite{Sato2012} introduced a formula for the $\gamma$-ray invariant differential cross section based on the recent Large Hadron Collider (LHC) data.

Ref.~\cite{Dermer1986a} compared the isobar-fireball model and the scaling formula of \cite{Stephens1981}, against accelerator data of $\pi^0$ production spectra that were available at that time. It was concluded that the isobar model \cite{Stecker1971} works better at low energies ($T_p<3$~GeV), whereas the scaling model \cite{Stephens1981} works better at higher energies. Thus, \cite{Dermer1986a} married these two models smoothly in an isobar-scaling model. 

Nowadays, many phenomenological approaches have been developed to calculate hadronic interactions in both perturbative and non-perturbative regimes. Sophisticated Monte Carlo (MC) codes are developed which handle complex calculations and predict many quantities that agree well with experimental data. There are different classes of MC codes. The general high energy interaction event generator class, which include both Standard Model and beyond Standard Model physics, contain codes such as HERWIG \cite{Herwig}, SHERPA \cite{Sherpa1} and PYTHIA \citep{Pythia1, Pythia2}. Other universal codes that include sophisticated high energy hadronic physics models are e.g. Phojet \cite{phojet1,phojet2,phojet3} DPMJET \cite{dpmjet} and EPOS \cite{epos}.

Another class of MC codes are specialized in air shower simulations. They are concerned about high energy secondary particle production, reproducing well the existing accelerator data and extrapolating the results to the very high energy region. SIBYLL \cite{SIBYLL1992,SIBYLL1994,SIBYLL2009} and QGSJET \cite{QSJET1993,GGSJET11997,QGSJET-IIc,QGSJET-IIa,QGSJET-IId} are part of this set of codes, which are used in the CORSIKA extensive air shower simulations (EAS) code \cite{Thakuria2012}.

Calculation of the galactic diffuse $\gamma$-ray spectrum using results of MC codes such as HADRIN \cite{hadrin}, FRITIOF \cite{fritiof} and PYTHIA were done by \cite{Mori1997}.
Ref.~\cite{Kamae2005} noticed that previous calculations of the $pp\to\gamma$-ray spectrum for an astrophysical environment did not consider diffractive interaction, Feynman scaling violation and the logarithmic increase of the inelastic cross section. They stressed the importance of diffractive dissociation, which is the source of very high energy $\gamma$-rays, and the importance of the Feynman scaling violation, which increases the $\gamma$-ray yield in the GeV to multi-GeV energy range. Their first model ``Model A'' and later the ``Readjusted Model A'' \cite{Kamae2006} included these processes by using PYTHIA~6 event generator for $52.6~{\rm GeV}\leq T_p\leq 512~{\rm TeV}$. For $0.488~{\rm GeV}\leq T_p< 52.65$~GeV they included calculations from \cite{Blattnig2000} and extended the model to the resonance region by including two resonances, the $\Delta(1232)$ and a group of resonances called $res(1600)$, readjusting their calculations to be accurate below $T_p<3$~GeV. The results of the ``Readjusted Model A'' are given in a parametrized form in \cite{Kamae2006}.

Ref.~\cite{Kelner2006} used the results of publicly available codes to find an accurate parametrization of different secondary particle spectra at high proton energies, with $0.1~{\rm TeV}\leq T_p\leq 10^5$~TeV. With the $<10$~GeV component of astrophysical $\gamma$-ray sources becoming probeable by present and next-generation instruments, the limitations in energy range of this parametrization are becoming apparent. Therefore, a parametrization that is both accurate and that spans from threshold up to very high energies is needed. Moreover, very hot astrophysical plasma calculations as well as the increasing sensitivity of the $\gamma$-ray instruments such as Fermi--LAT satellite, which has recently observed $\gamma$-ray spectra that reveal a sub-GeV bump, require accurate $\gamma$-ray production cross sections at low energies near the p--p kinematic threshold.

In the present work, we adopt an approach similar to that in \cite{Kelner2006}. We focus on producing a simple but accurate parametrization of the $\gamma$-ray differential cross section for a wide range of proton energies. For this, we combine experimental data of $\pi^0$ production below $T_p\leq 2$~GeV and publicly available sophisticated MC codes which combine theoretical and parametrization driven models, at higher energies. We thus revise the low energy $\pi^0$ production cross section data, and use Geant~4.10.0 \citep{Geant42003, Geant42006} parametrization driven models to connect those experimental data with very high energy models predictions such as Pythia, SIBYLL, QGSJET, etc. Since at very high energies the different models available disagree with each other, we have picked GEANT~4, PYTHIA~8, SIBYLL and QGSJET as representatives of their respective MC model classes. %These MC codes use sophisticated theoretical and phenomenological models and are tuned to the experimental data.

In this way, we introduce here a simple and accurate parametrization that spans from the kinematic threshold to PeV energies and that has the flexibility to switch between different high energy models. We have applied this parametrization to different proton spectra with a wide energy range and show that the parametrization has a smooth transition between different energy regions. We have related the proton and $\gamma$-ray spectra parameters and have fitted these parameters with quite simple functions. We also discuss the effect of the nuclei on the $\gamma$-ray spectrum. We introduce a practical method to calculate the nuclear enhancement factor to very high energies and notice that at low energies this factor loses its meaning. We comment on the nucleus--nucleus subthreshold $\pi^0$ production effect which can produce very efficiently high energy $\gamma$-rays for nuclei with kinetic energy $T_p<0.28$~GeV/nucleon.

Although we do not parametrize the spectra of secondary particles such are muons, electrons, neutrinos and their antiparticles, their spectra can be derived above a certain proton energy using the symmetries and ratios that exist between charged and neutral pion production that are found experimentally. Hence, one can potentially derive useful information on other secondary particles using their ratios with respect to the $\gamma$-rays that exist at high energies, as was shown in \cite{Kelner2006}.

%oooooooooooOOOOOOOOOOOOOOOOoooooooooooooooOOOOOOOOOOOOOOOoooooooooo
% end of introduction
%OOOOOOOOOOOOOOOOoooooooooooOOOOOOOOOOOOOOOooooooooooOOOOOOOOOOOOOOO

\section{Inclusive $\pi^0$ production cross section and multiplicity \label{sec:xs}}

%oooooooooooOOOOOOOOOOOOOOOOoooooooooooooooOOOOOOOOOOOOOOOoooooooooo
% INELASTIC XS
%OOOOOOOOOOOOOOOOoooooooooooOOOOOOOOOOOOOOOooooooooooOOOOOOOOOOOOOOO
\subsection{Total p--p inelastic cross section}
The theoretically motivated parametrization of the high energy p--p inelastic cross section has a quadratic functional form of the logarithm of the proton energy. We take here the total and elastic p--p cross section data compiled by the Particle Data Group (PDG) \cite{PDG}. These data include the recent measurements at center-of-mass energies $\sqrt{s}=7$ and 8 TeV published by the TOTEM collaboration at the LHC \citep[see e.g.][]{TOTEMxs7TeV, TOTEMxs8TeV}. We suggest the following parametrization for the p--p total inelastic cross section
\begin{equation}
\label{eq:XSinel}
\begin{split}
\sigma_{\rm inel}=&\left[ 30.7 - 0.96\,\log\left(\frac{T_p}{T_p^{\rm th}}\right) + 0.18\,\log^2\left(\frac{T_p}{T_p^{\rm th}}\right) \right] \\ & ~~~~~~~~~~~~~~~~~~~~~~~\times\left[1-\left(\frac{T_p^{\rm th}}{T_p} \right)^{1.9}\right]^3~{\rm mb}.
\end{split}
\end{equation}

\noindent Here, $T_p$ is the proton kinetic energy in the laboratory frame, $T_p^{\rm th}= 2m_\pi+m_\pi^2/2m_p\approx0.2797$~GeV is the threshold kinetic energy. $m_p$ and $m_\pi$ are the proton and $\pi^0$ masses, respectively. The units that we use throughout are the natural units (i.e. $\hbar=c=k_B=1$).

\begin{figure}[h]
\includegraphics[scale=0.45]{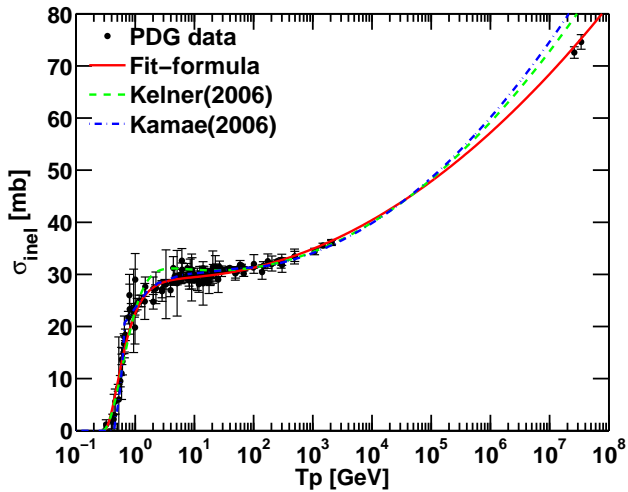}
\caption{The total p--p inelastic cross section as a function of the proton kinetic energy in the laboratory system. The data points are taken from the PDG \cite{PDG} cross section data. The red line is our $\chi^2$-fit formula shown in eq.~(\ref{eq:XSinel}). The dashed green line is the formula given in \cite{Kelner2006} and the dot-dashed blue line is the formula given in \cite{Kamae2006}. \label{fig:XSinel}}
\end{figure}

Figure~\ref{fig:XSinel} compares the parametrization of the \ref{eq:XSinel} against two other parametrizations \cite{Kelner2006} and \cite{Kamae2006} often used in astrophysical contexts, as well as the observational data from \cite{PDG}.

%oooooooooooOOOOOOOOOOOOOOOOoooooooooooooooOOOOOOOOOOOOOOOoooooooooo
% pi0 XS
%OOOOOOOOOOOOOOOOoooooooooooOOOOOOOOOOOOOOOooooooooooOOOOOOOOOOOOOOO
\subsection{Inclusive $\pi^0$ production cross section and average multiplicity}
To obtain the inclusive $\pi^0$ production cross section or average multiplicity, we have used both experimental data and calculations from MC codes, which are themselves tuned to describe experimental data over a given energy range.

For proton kinetic energies $T_p\le 2$~GeV, $\pi^0$ production is dominated by baryon resonance production. The leading $\pi^0$ production channels for these energies are $pp\to pp \pi^0$, $pp\to pp 2\pi^0$ and $pp\to \{pn, D\} \pi^+\pi^0$. Other two-pion and three-pion channels are negligible. Good quality data exist for these channels in this energy region, which we have here compiled. The references to these data are shown in tables~\ref{tab:XS1pi0} to \ref{tab:XSpippi0}.

%==============================================
For energies $T_p>2$~GeV, we calculated the average $\pi^0$ production multiplicity using Geant~4.10.0, Pythia~8.18, SIBYLL~2.1 and QGSJET-I. Note that the latest version of the QGSJET description presently available is QGSJET-II. We have not run either SIBYLL or QGSJET codes here, we instead rely on the fits of $\pi^0$ and $\eta$ mesons spectra for both SIBYLL~2.1 and QGSJET-I provided in \cite{Kelner2006}, valid for $T_p> 100$~GeV.

The numerical descriptions we run are those provided in Geant~4.10.0 and Pythia~8.18. Geant~4.10.0 is itself actually a toolkit for the simulation of the passage of particles through matter. Thus it does not stand in any of the MC code classes that we described previously. However, Geant~4.10.0 contains sophisticate hadronic interaction models such as ``\textit{FTFP\_BERT}'', that we use here, which implements Bertini-style cascade at low energies $T_p\leq5$~GeV. For $4<T_p\leq 10^5$~GeV it includes the FRITIOF string model, see \citep{Geant42003,Geant42006}. Simulations with Pythia~8.18 are done by using the default Pythia 8.18 tune and selecting ``\textit{SoftQCD}'' processes.

In this way, we adopt the Geant~4.10.0 average $\pi^0$ production multiplicity to fill in the energy gap between the experimental data $T_p\leq 2$~GeV and the high energy models adopted. At high energies, different hadronic models predictions start to diverge. Thus, at these high energies we provide the option to adopt the multiplicity from one of these hadronic models: namely Geant~4.10.0; Pythia~8.18; SIBYLL~2.1 and QGSJET-I.

We have fitted the inclusive $\pi^0$ production cross sections and multiplicities from kinematic threshold to very high energies ($T_p\sim 1$~PeV) and parametrized their descriptions. 

%==============================================
% 1 pi0 production XS
%==============================================
\subsubsection{Parametrization of the $pp\to pp \pi^0$ cross section}
An accurate parametrization of the $pp\to pp\pi^0$ cross section data from the kinematic threshold to several GeV, is given by:
\begin{equation}\label{eq:XS1pi}
\sigma_{1\pi}= \sigma_0\times \eta^{1.95}(1+\eta+\eta^5)\times \left[ f_{BW}(\sqrt{s}) \right]^{1.86}.
\end{equation}

\noindent Here, $\sigma_0=7.66\times 10^{-3}$ mb; $\eta=P_\pi^*/m_\pi$, where $P_\pi^*$ is the maximum pion momentum in the center-of-mass system. $\eta$ is given by:
\begin{equation}
\eta= \frac{\sqrt{(s-m_\pi^2 - 4m_p^2)^2 -16m_\pi^2 m_p^2 }}{2m_\pi\sqrt{s}}~.
\end{equation}

\begin{figure}[h]
\includegraphics[scale=0.25]{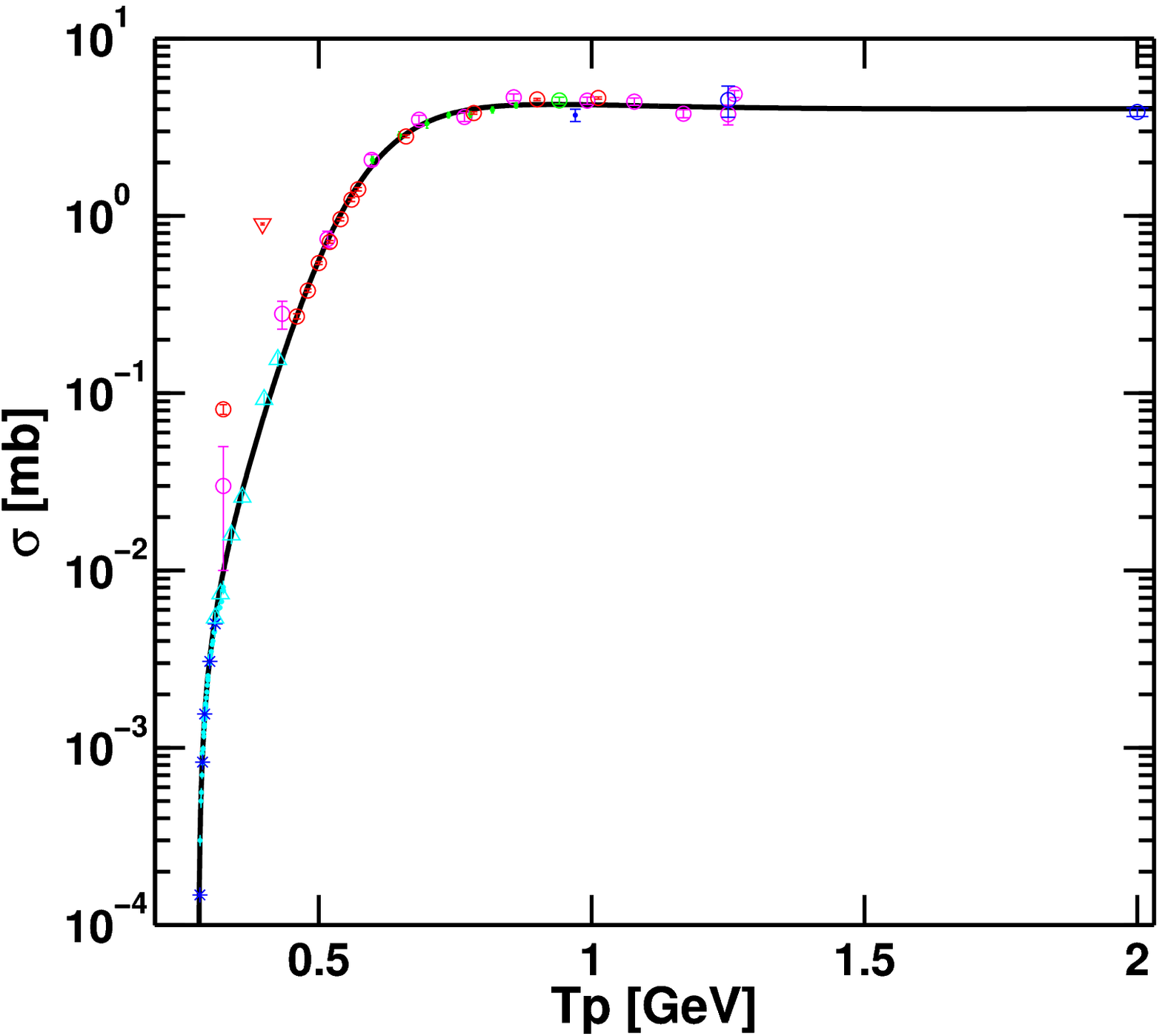} 
\includegraphics[scale=0.25]{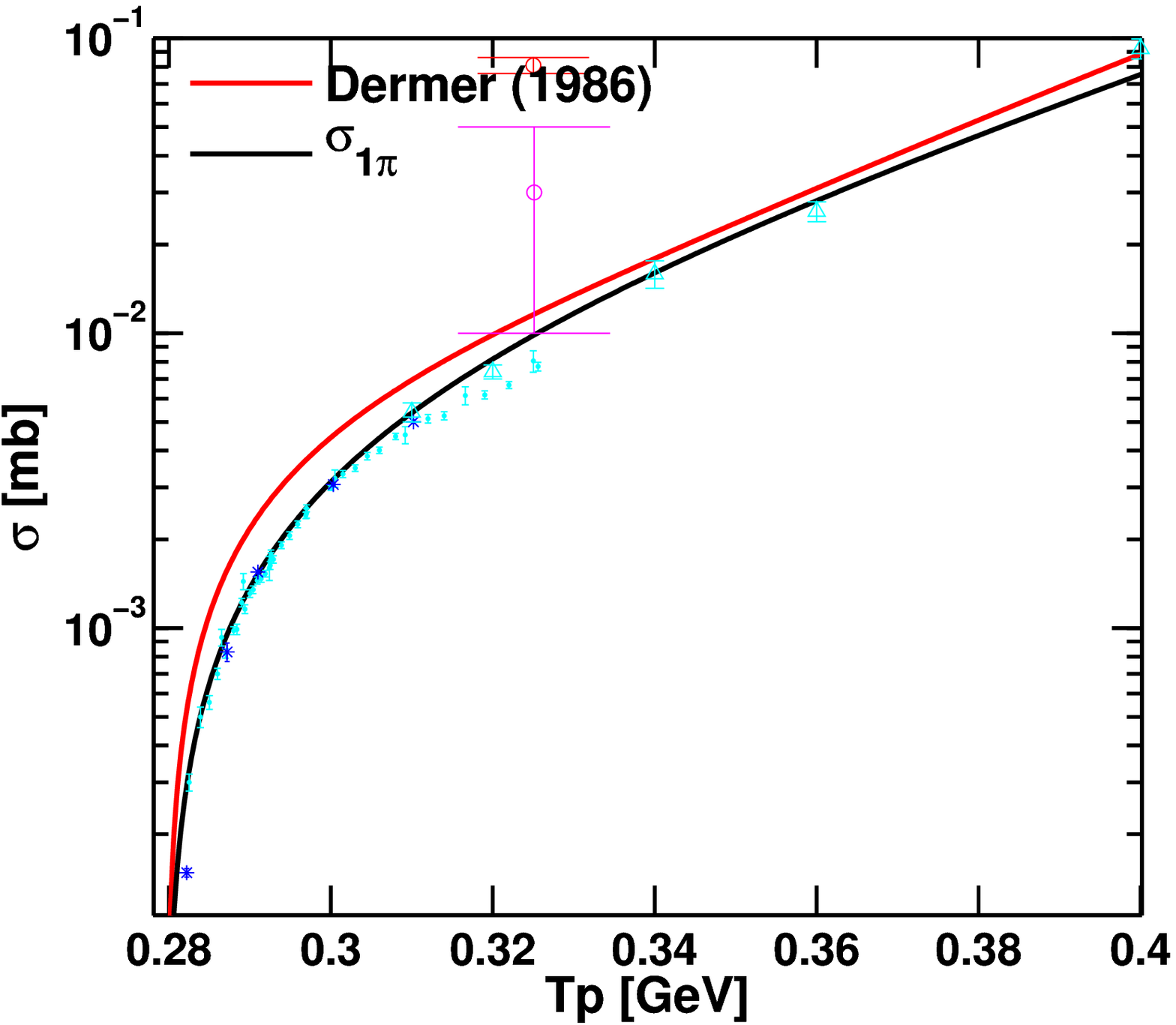}
\caption{Cross section data (see table~\ref{tab:XS1pi0}) for the $pp\to pp \pi^0$ channel compared  with eq.~(\ref{eq:XS1pi}) (left panel). In the right panel we compare data with eq.~(\ref{eq:XS1pi}) and the parametrization given in \cite{Dermer1986b}. Notice that near the kinematic threshold the parametrization of \cite{Dermer1986b} can be up to 70 \% larger than the data. \label{fig:XS1pi0}}
\end{figure}

\noindent $f_{BW}(\sqrt{s})$ is the unit-less relativistic Breit--Wigner distribution,
\begin{equation}
\label{eq:XS1pi01}
\begin{split}
f_{BW}(\sqrt{s}) &= \frac{m_p \times K}{\left( (\sqrt{s}-m_p)^2 - M_{res}^2\right)^2 + M_{res}^2\Gamma_{res}^2},\\
K&=\frac{\sqrt{8}M_{res}\Gamma_{res} \gamma }{\pi\sqrt{M_{res}^2 + \gamma}},\\
\gamma &=\sqrt{M_{res}^2(M_{res}^2 + \Gamma_{res}^2)}~~,
\end{split}
\end{equation}

\noindent where $s=2m_p(T_p+2m_p)$ is the center-of-mass energy squared. The values of resonance mass and width are $M_{res}=1.1883$~GeV and $\Gamma_{res}=0.2264$~GeV. In figure~\ref{fig:XS1pi0} the parametrization formula of eq.~(\ref{eq:XS1pi}) is compared with the experimental data in table~\ref{tab:XS1pi0}. 

It is clear from figure~\ref{fig:XS1pi0} that the parametrization given in eq.~(\ref{eq:XS1pi}) represents better the experimental data especially near the kinematic threshold. The widely used parametrization given in \citep{Dermer1986b} around $T_p=0.3$~GeV  can be about 70 \% higher than the data.

\begin{table}[h]
\caption{References for $pp\to pp \pi^0$ cross section data for $T_p\in [0.28,2.2]$~GeV. \label{tab:XS1pi0}}
\begin{tabular}{cccc}
%\toprule 
\toprule
$T_p$ [GeV] & & $\sigma$ [mb] & Reference\\
\toprule
0.28--0.31 & & (0.007--5)$\times10^{-3}$ & \cite{Bondar1995}\\
\hline
0.28--0.33 & & (0.3--8)$\times10^{-3}$ & \cite{Meyer1990}\\
\hline
0.29--0.33 & & (0.6--8)$\times10^{-3}$ & \cite{Meyer1992}\\
\hline
0.31--0.43 & & 0.005--0.155 & \cite{Bilger2001}\\
\hline
0.33--1.26 & & 0.03--4.87 & \cite{Shimizu1982}\\
\hline
0.33--1.01 & & 0.08--4.6 & \cite{Rappenecker1995}\\
\hline
0.397 & & 0.9 & \cite{ElSamad2006} \\
\hline
0.6--0.86 & & 2.1--4.2 & \cite{Andree1988} \\
\hline
0.94 & & 4.48 & \cite{Ermakov2011}\\
\hline
0.97 & & 3.7 & \cite{Bugg1964}\\
\hline
1.25, 2.2 & & 3.74, 4.15 & \cite{Agakishiev2012}\\
\toprule
\end{tabular}
\end{table}

%==============================================
% 2pi0 + pi+pi0  production XS
%==============================================
\subsubsection{Inclusive $\pi^0$ production cross section from two-pion production channels}
As already mentioned above, for $T_p\leq 2$~GeV, the dominant two-pion channels are $pp\to pp 2\pi^0$ and $pp\to \{pn, D\} \pi^+\pi^0$. For our purposes, only the sum of these two cross section channels is important. Equation~{eq:XS2pi} provides a fit formula for the inclusive cross section for the sum of the two-pion production channels.
\begin{equation}
\label{eq:XS2pi}
\sigma_{2\pi}= \frac{5.7~{\rm mb} }{1+\exp\left(-9.3(T_p-1.4)\right)}.
\end{equation}

\noindent Here, $T_p$ is in GeV and the formula is valid for $0.56~{\rm GeV}\leq T_p \leq2$~GeV. For $T_p< 0.56$~GeV, $\sigma_{2\pi}=0$.

\begin{figure}[h]
\includegraphics[width=0.235\textwidth]{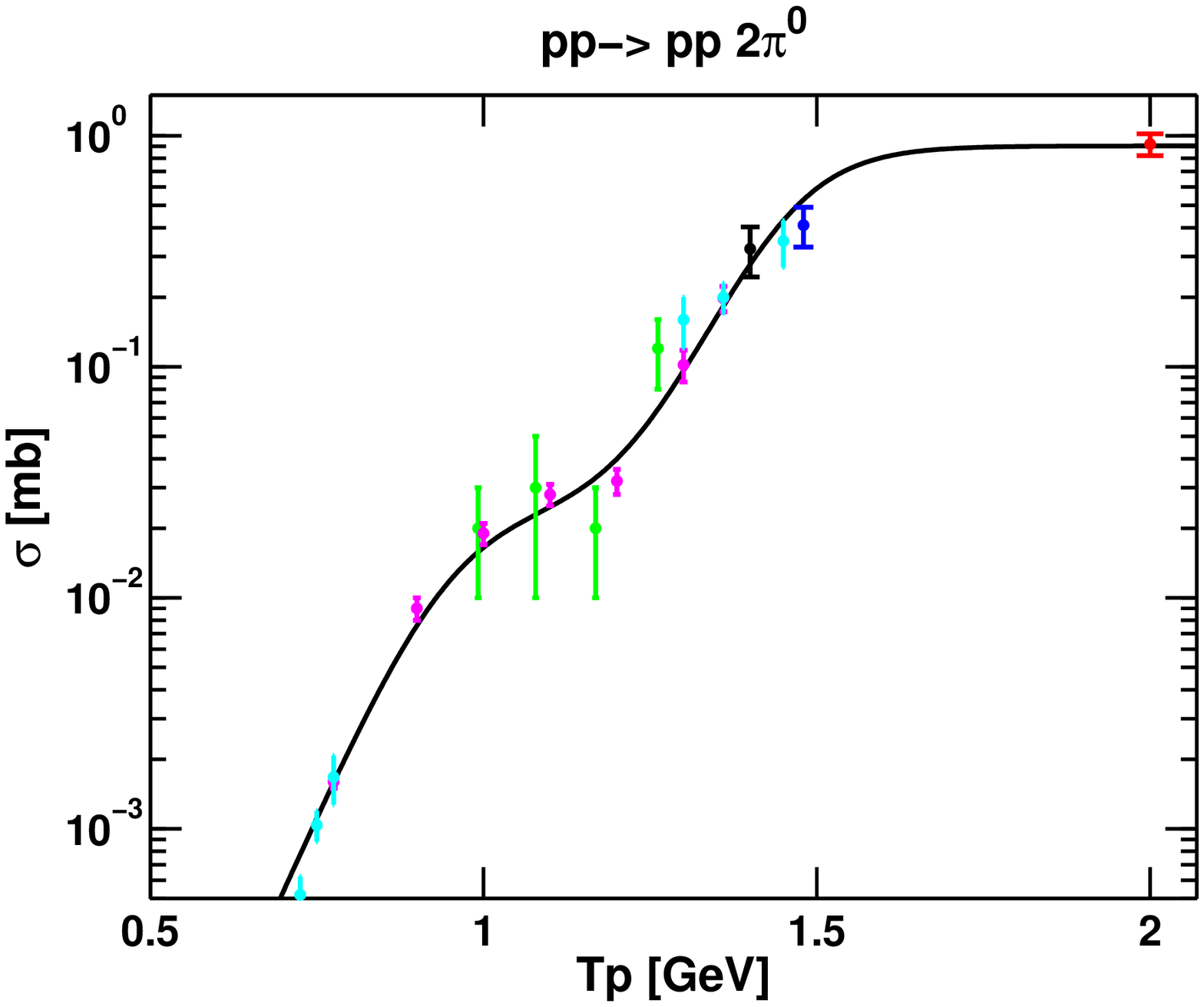}
\includegraphics[width=0.235\textwidth]{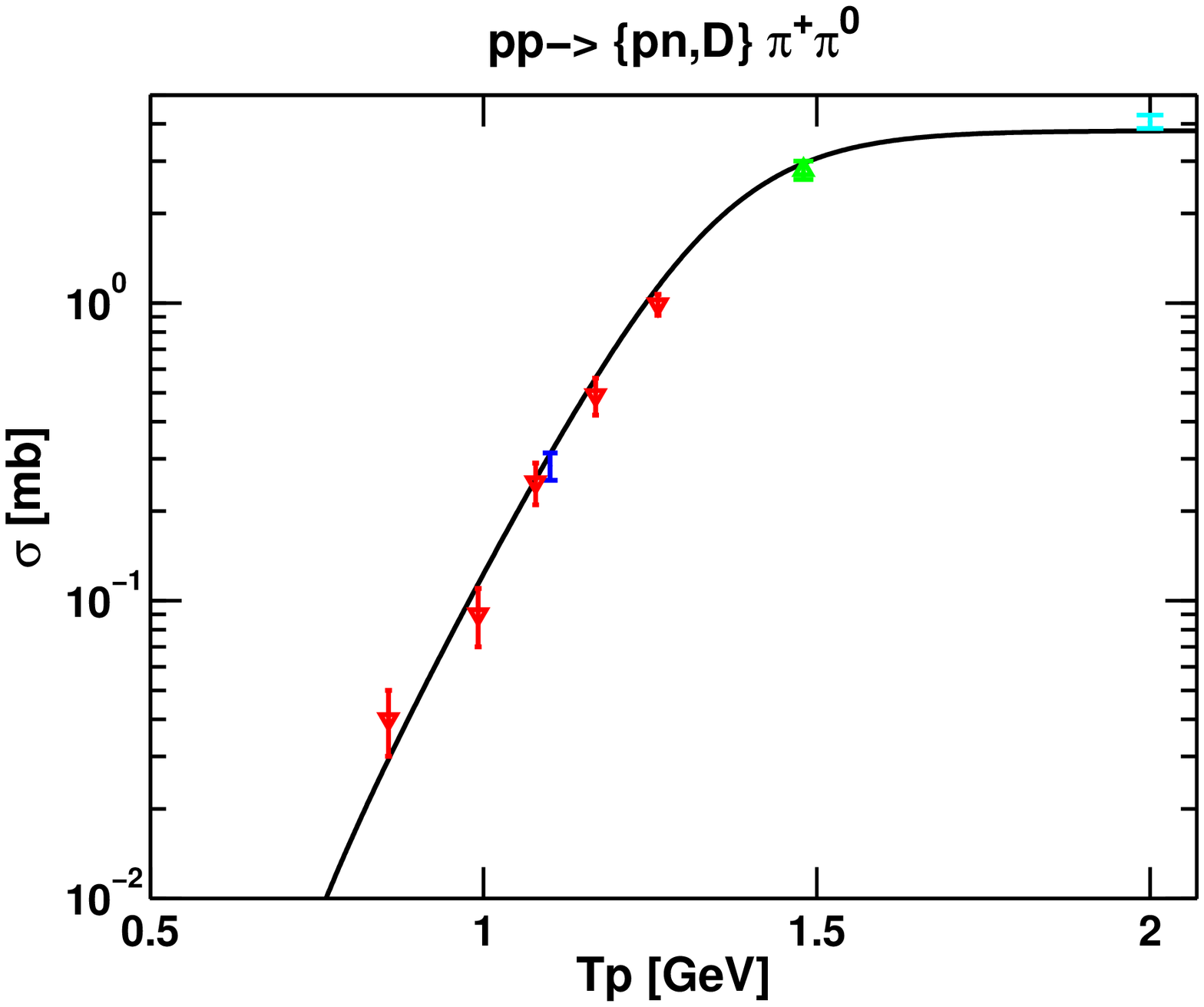}\\
\hspace*{-0.1cm}\includegraphics[width=0.49\textwidth]{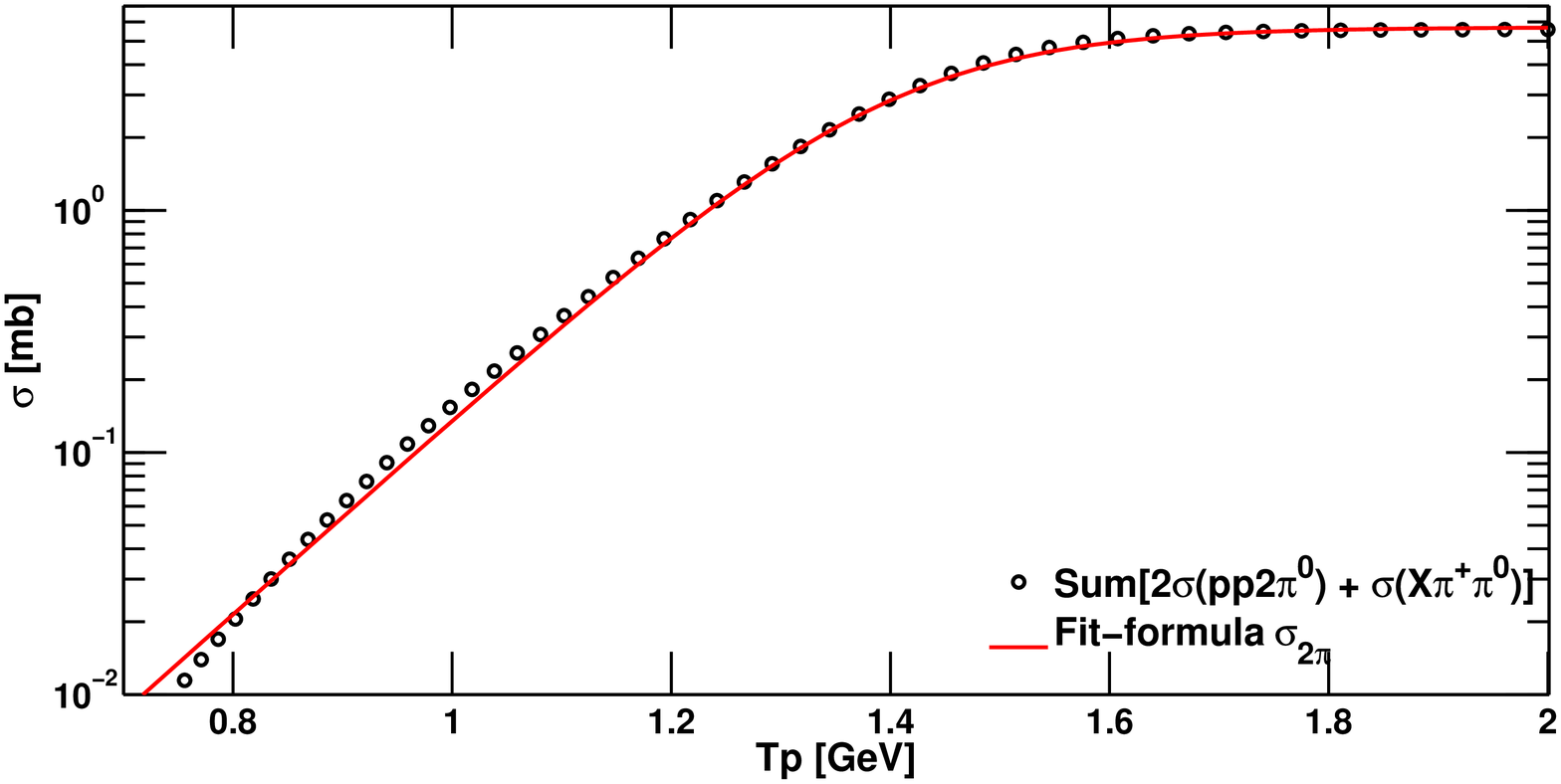}
\caption{The two panels on the top show our fit to the experimental data given in tables~\ref{tab:XS2pi0} and \ref{tab:XSpippi0}. Left panel on the top corresponds to the $pp\to pp 2\pi^0$ channel and right panel on the top to the sum of the $pp\to pn\pi^+\pi^0$ and $pp\to D\pi^+\pi^0$ channels. Experimental data reveal that the $pp\to pp 2\pi^0$ cross section has a dip around $T_p=1$~GeV. Our parametrization of these exclusive channels cross sections is $\sigma(T_p)=\sum\limits_{i=1}^2 a_i\,\left(1+\exp\left(-b_i(T_p-c_i)\right)\right)^{-1}$. For the $pp\to pp 2\pi^0$ channel, $a_1=0.88$~mb, $a_2=0.023$~mb,  $b_1=b_2= 15~{\rm GeV^{-1}}$ and $c_1=1.46$~GeV, $c_2=0.95$~GeV. For the $pp\to X\pi^+\pi^0$ channel, $a_1=3.8$~mb, $a_2=-0.0075$~mb, $b_1=9.6~{\rm GeV^{-1}}$, $b_2=2~{\rm GeV^{-1}}$ $c_1=1.35$~GeV and $c_2=0.8$~GeV.\\
The panel on the bottom show the sum of these two-pion production channels multiplied with their respective multiplicities (i.e. $2\times\sigma(pp\to pp2\pi^0) + \sigma(pp\to X\pi^+\pi^0)$). They are denoted in the figure with open-circles. The full red line is the parametrization given in eq.~(\ref{eq:XS2pi}). The two-pion channel contributions to the total inclusive $\pi^0$ production become non-negligible for $T_p>1.2$~GeV, thus the accuracy below this energy is not necessarily important. \label{fig:XS2pi}}
\end{figure}

Figure~\ref{fig:XS2pi} compares individual two-pion production channel data against our parametrizations. The sum of the two channels against the parametrization in eq.~(\ref{eq:XS2pi}), is also shown. Tables~\ref{tab:XS2pi0} and \ref{tab:XSpippi0} provide references to the experimental data for the different two-pion production channels.

\begin{table}[h]
\caption{References for $pp\to pp 2\pi^0$ cross section data for $T_p\in [0.65,2.0]$~GeV. \label{tab:XS2pi0}}
\begin{tabular}{cccc}
%\toprule 
\toprule
$T_p$ [GeV] & & $\sigma$ [mb] & Reference\\
\toprule
0.65--0.78 & & (0.05--1.68)$\times10^{-3}$ & \cite{Johanson2002}\\
\hline
0.78--1.36 & & 0.002--0.2 & \cite{Skorodko2009}\\
\hline
0.99--1.26 & & 0.02--0.12 & \cite{Shimizu1982}\\
\hline
1.3-1.45 & & 0.16--0.35 & \cite{Pauly2007}\\
\hline
1.4 & & 0.32 & \cite{Adlarson2012}\\
\hline
1.48 & & 0.41 & \cite{Eisner1965}\\
\hline
2.0 & & 0.92 & \cite{Pickup1962}\\
\toprule
\end{tabular}
\end{table}

\begin{table}[h]
\caption{References for $pp\to \{pn, D\} \pi^+\pi^0$ cross section data for $T_p\in [0.73,2.0]$~GeV. \label{tab:XSpippi0}}
\begin{tabular}{cccc}
%\toprule 
\toprule
& & $pp\to pn \pi^+\pi^0$ & \\
\toprule
$T_p$ [GeV] & & $\sigma$ [mb] & Reference\\
\toprule
0.73--0.78 & & (0.83--2.29)$\times10^{-3}$ & \cite{Johanson2002}\\
\hline
1.1 & & 0.284 & \cite{Skorodko2009}\\
\hline
0.86--1.26 & & 0.02--0.66 & \cite{Shimizu1982}\\
\hline
1.48 & & 2.37 & \cite{Eisner1965}\\
\hline
2.0 & & 4.07 & \cite{Pickup1962}\\
\toprule
& & $pp\to D \pi^+\pi^0$ & \\
\toprule
0.86--1.26 & & 0.02--0.33 & \cite{Shimizu1982}\\
\hline
1.48 & & 0.43 & \cite{Eisner1965}\\
\toprule
\end{tabular}
\end{table}

%===================================================
% Geant4, Pythia8, SIBYLL~2.1, QGSJET-I XS for Tp>100GeV
%===================================================
\subsubsection{Neutral pion production average multiplicity from Geant~4.10.0, Pythia~8.18, SIBYLL~2.1 and QGSJET-I \label{sec:InclusPi0Mult1} }

As mentioned, for energies $T_p\geq 2$~GeV we use Geant~4.10.0 to fill the energy gap between the experimental data and the high energy models. Pythia~8.18 is applicable for $T_p>50$~GeV, whereas, SIBYLL~2.1 and QGSJET-I work for $T_p>100$~GeV. Except QGSJET-I which has an average multiplicity a factor 1.7 times larger at $T_p\sim 100$~GeV compared to other models, the rest of them agree well with each other for $T_p<1$~TeV. We fit here the average $\pi^0$ multiplicity predicted by all these models. 

If $Q_p=(T_p - T_p^{\rm th})/m_p$, then the Geant~4.10.0 average $\pi^0$ multiplicity for $1~{\rm GeV}\leq T_p <5$~GeV, may be written as:
\begin{equation}\label{eq:multp1}
\left\langle n_{\pi^0} \right\rangle = -6\times10^{-3} + 0.237\,Q_p -0.023\,Q_p^2~.
\end{equation}

Notice that we do not use Geant~4.10.0 to calculate multiplicity at energies $T_p<1$~GeV. We find that Geant~4.10.0 over predicts $\pi^0$-production multiplicity for $T_p<0.7$~GeV.

For energies $T_p \geq 5$~GeV, we find that one formula is sufficient to fit multiplicities with very high accuracy for all the models. Setting $\xi_p=(T_p-3~{\rm GeV})/m_p$, this formula takes the form:

\begin{equation}\label{eq:multp2}
\begin{split}
\left\langle n_{\pi^0} \right\rangle =  a_1\,&\xi_p^{a_4}\,\left[1+ \exp\left(-a_2\,\xi_p^{a_5}\right)\right] \times \\
& \times \left[1-\exp\left(-a_3\,\xi_p^{1/4}\right)\right] .
\end{split}
\end{equation}

\begin{figure*}
\includegraphics[scale=0.3]{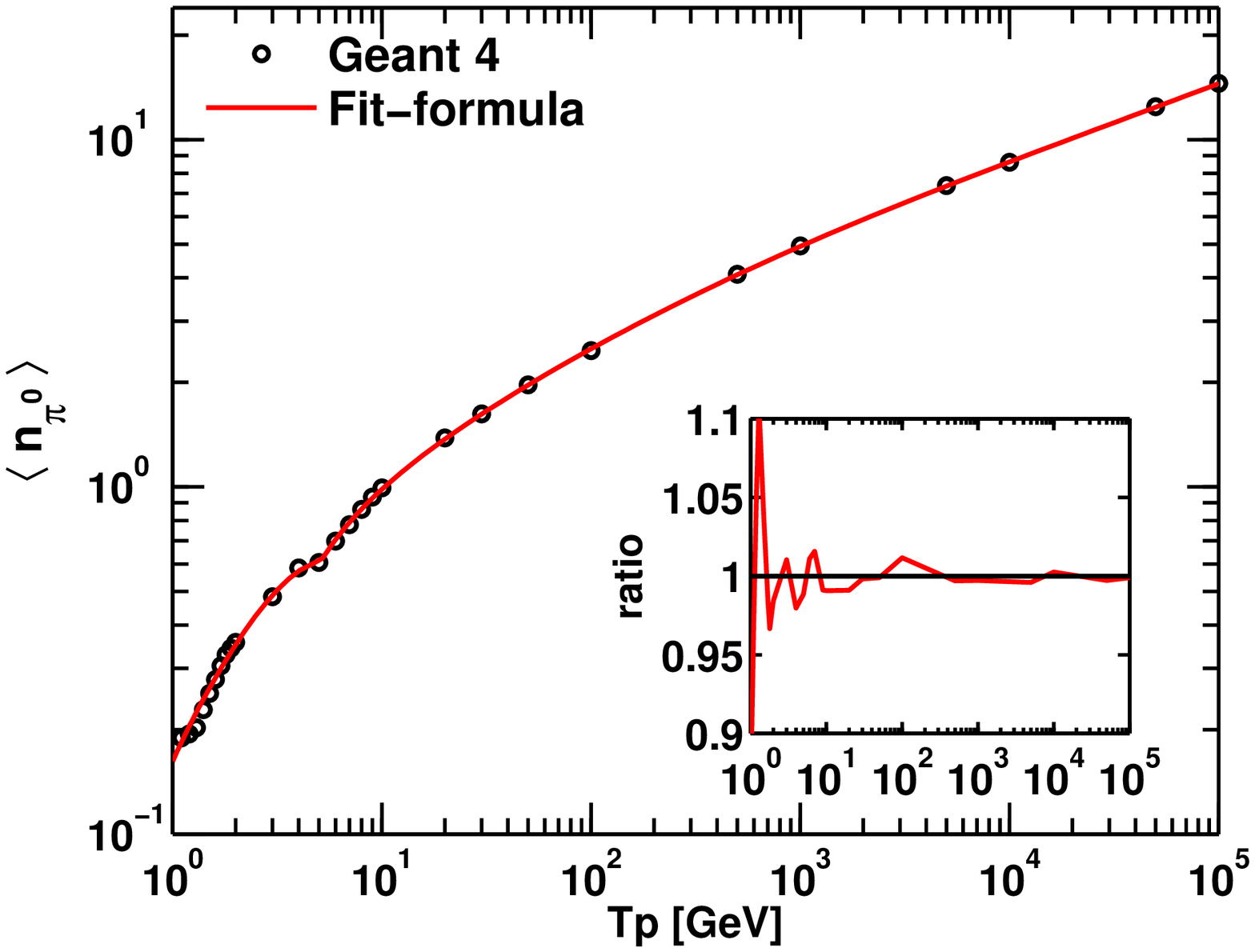}
\includegraphics[scale=0.3]{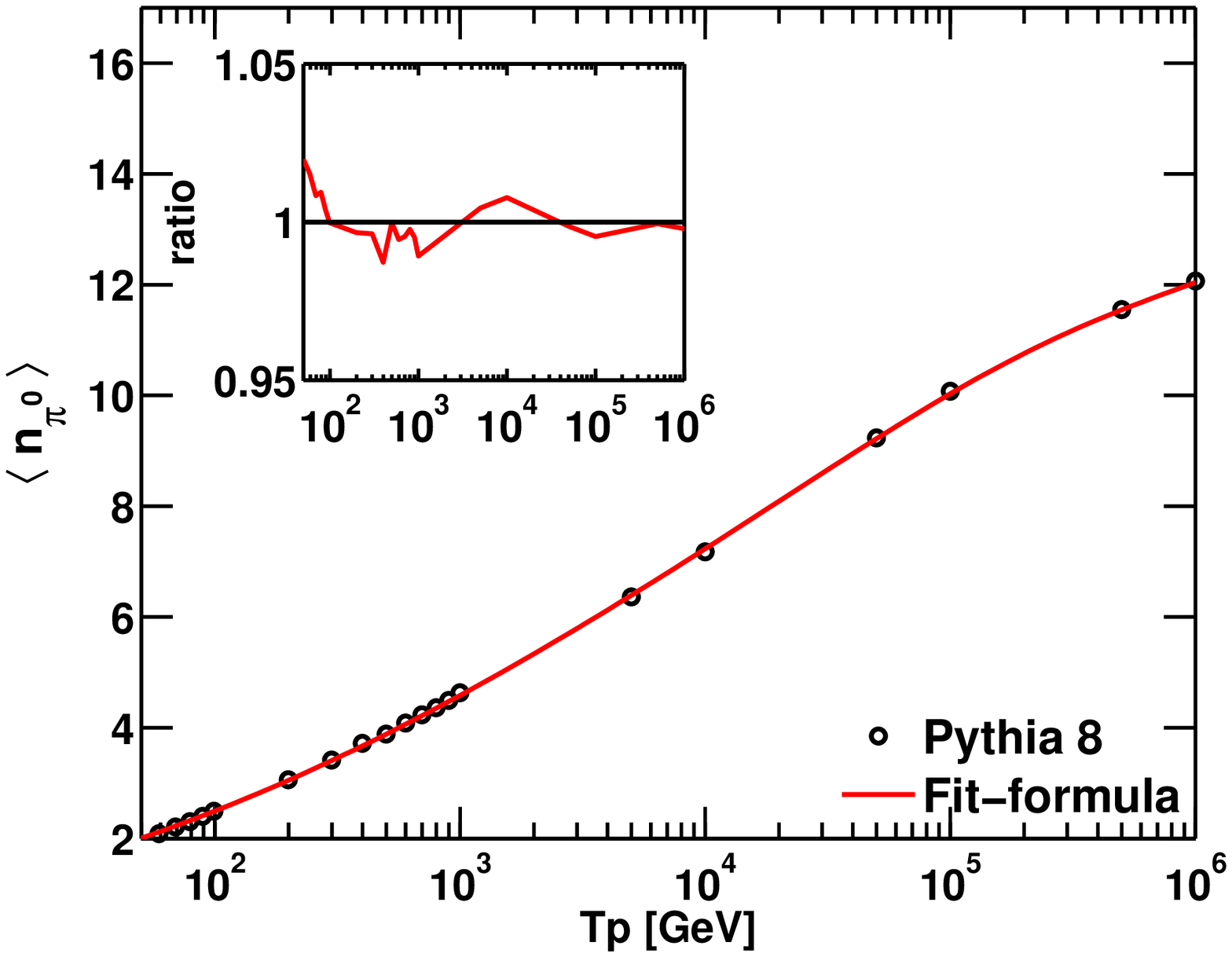}\\
\includegraphics[scale=0.3]{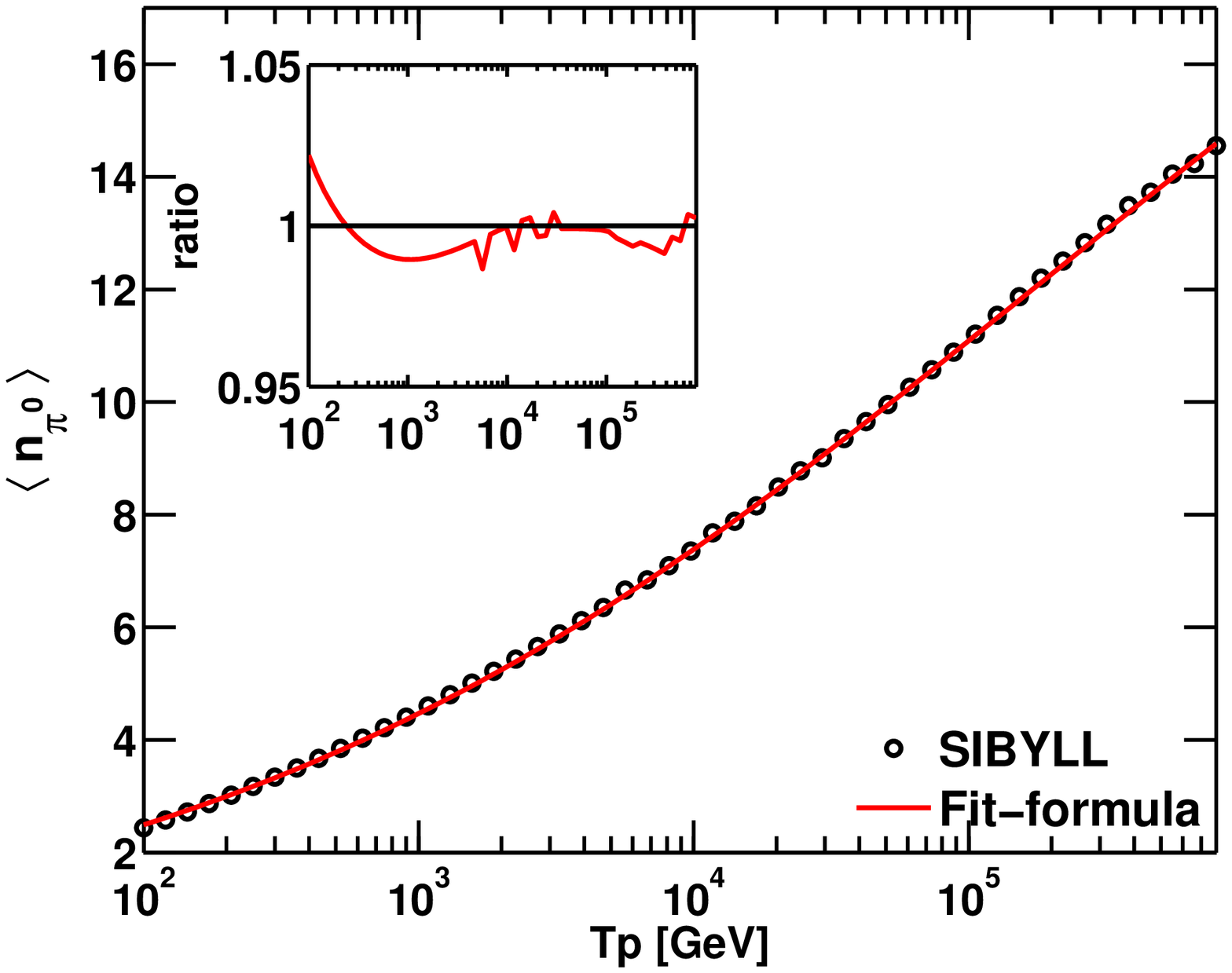}
\includegraphics[scale=0.3]{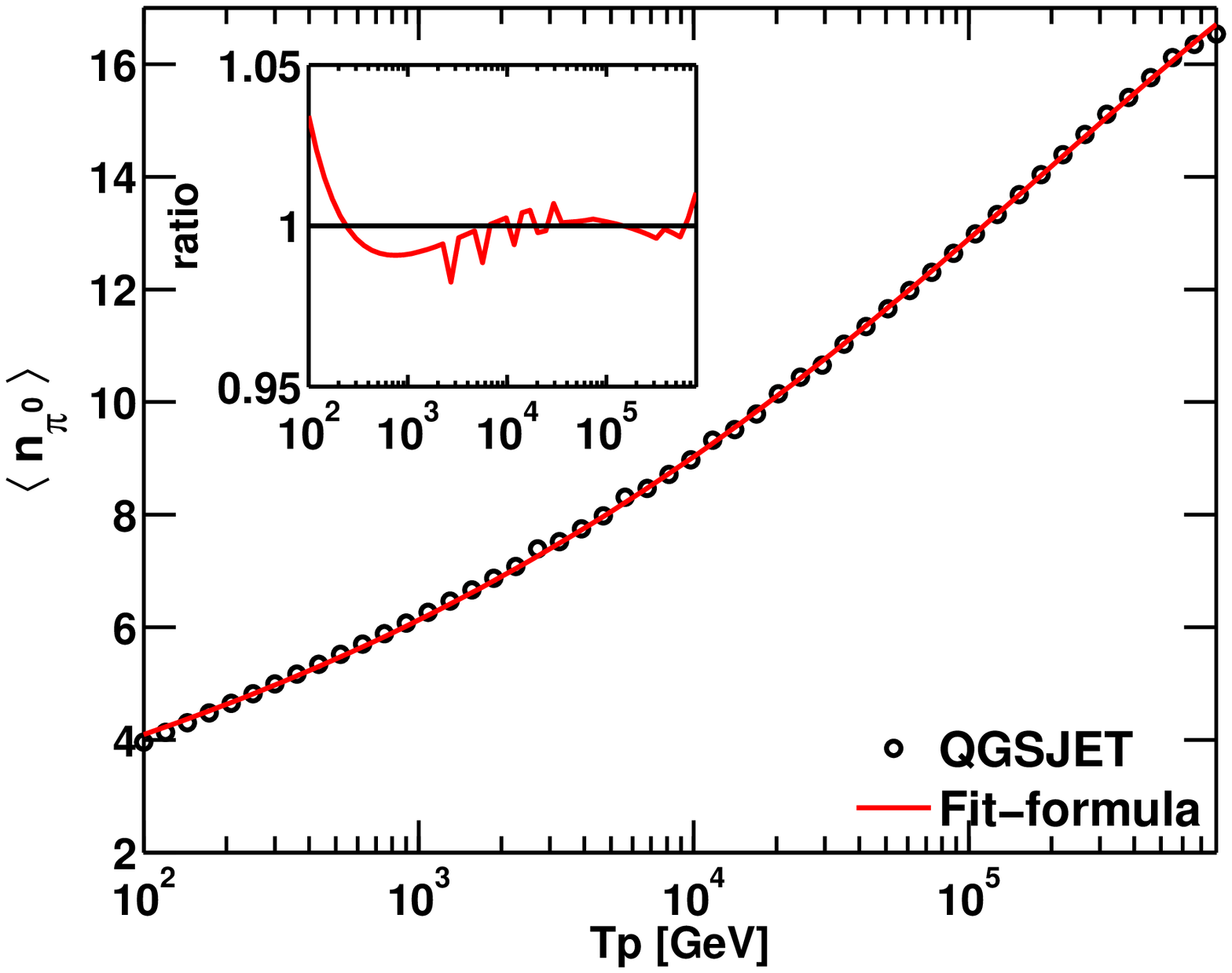}
\caption{The average $\pi^0$ production multiplicity as a function of proton kinetic energy. Geant~4.10.0 multiplicity for $1~{\rm GeV}\leq T_p <5$~GeV is calculated using the fit formula in eq.~(\ref{eq:multp1}), whereas, for $T_p \geq 5$~GeV is calculated using fit formula of eq.~(\ref{eq:multp2}) with the appropriate coefficients in table~\ref{tab:multpa1-a5}. For Pythia 8.18, SIBYLL~2.1 and QGSJET-I the fit formula is eq.~(\ref{eq:multp2}) with the corresponding coefficients presented in table~\ref{tab:multpa1-a5}. Note that SIBYLL~2.1 and QGSJET-I multiplicities were calculated using the parametrizations of the $\pi^0$ and $\eta$ spectra that are provided in ref.~\cite{Kelner2006}. The accuracy of the fit-formulas that are presented here, is better than 3~\%. An exception to this is the Geant~4.10.0 description, which for $T_p<2$~GeV has an accuracy of less than 10 \%. We, however, do not use the Geant~4.10.0 multiplicity in this energy interval, instead adopting our own fit to the experimental cross sections in this energy range. \label{fig:Multipmodels}}
\end{figure*}

\noindent The coefficients $a_1- a_5$ for each model are listed in table~\ref{tab:multpa1-a5}. Figure~\ref{fig:Multipmodels} shows different model multiplicities compared with the parametrization in eqs.~(\ref{eq:multp1}) and (\ref{eq:multp2}). The accuracy of these fits is better than 3~\% for $T_p\geq 2$~GeV.

\begin{table}[h]
\centering
\caption{Coefficients $a_1- a_5$ of eq.~(\ref{eq:multp2}) for Geant~4.10.0, Pythia~8.18, SIBYLL~2.1 and QGSJET-I. $T_p$ specifies the proton kinetic energy range for which the fit is valid.  \label{tab:multpa1-a5}}
\begin{tabular}{ccccccc}
%\toprule 
\toprule
Model & $T_p$ [GeV] & $a_1$ & $a_2$ & $a_3$ & $a_4$& $a_5$ \\
\toprule
Geant 4 & $\geq5$ & 0.728 & 0.596 & 0.491 & 0.2503 & 0.117\\
\hline
Pythia 8 & $>50$ & 0.652 & 0.0016 & 0.488 & 0.1928 & 0.483 \\
\hline
SIBYLL & $>100$ & 5.436 & 0.254 & 0.072 & 0.075 & 0.166 \\
\hline
QGSJET & $>100$ & 0.908 & 0.0009 & 6.089 & 0.176 & 0.448 \\
\toprule
\end{tabular}  
\end{table}

\subsubsection{Inclusive $\pi^0$ production cross section $\sigma_{\pi}$ \label{sec:InclusPi0XS}}

Our final parametrization for the inclusive $\pi^0$ production cross section is a combination of the experimental data cross sections at low energies, the Geant~4.10.0 cross section  at intermediate energies and at higher energies one of the four hadronic models: Geant~4.10.0, Pythia~8.18, SIBYLL~2.1 and QGSJET-I. More specifically, for proton energies between $T_p^{\rm th} \leq T_p<2~{\rm GeV}$ the $\pi^0$ production cross section is given by $\sigma_{\pi} (T_p)=\sigma_{1\pi} + \sigma_{2\pi}$; where, $\sigma_{1\pi}$ and $\sigma_{2\pi}$ are given in eqs.~(\ref{eq:XS1pi}) and (\ref{eq:XS2pi}). For energies between $2~{\rm GeV}\leq T_p \leq T_p^{\rm tran}$ the cross section is given by $\sigma_{\pi} (T_p)= \sigma_{\rm inel}(T_p) \times \left\langle n_{\pi^0} \right\rangle(T_p)$; where, $T_p^{\rm tran}$ is a transition energy between Geant~4.10.0 and other high energy models, and the average $\pi^0$ multiplicity is calculated in eqs.~(\ref{eq:multp1}) and (\ref{eq:multp2}). If we choose as our high energy model Geant~4.10.0, then $T_p^{\rm tran}=10^5$~GeV, which is the highest working energy of this model. If the chosen high energy model is instead Pythia~8.18, then $T_p^{\rm tran}=50$~GeV, and for $T_p>50$~GeV one adopts the Pythia~8.18 multiplicities. If the choice at high energies is instead SIBYLL~2.1 or QGSJET-I, then $T_p^{\rm tran}=100$~GeV, and for $T_p>100$~GeV we adopt the multiplicities of the respective model.

\begin{figure}[h]
\includegraphics[scale=0.22]{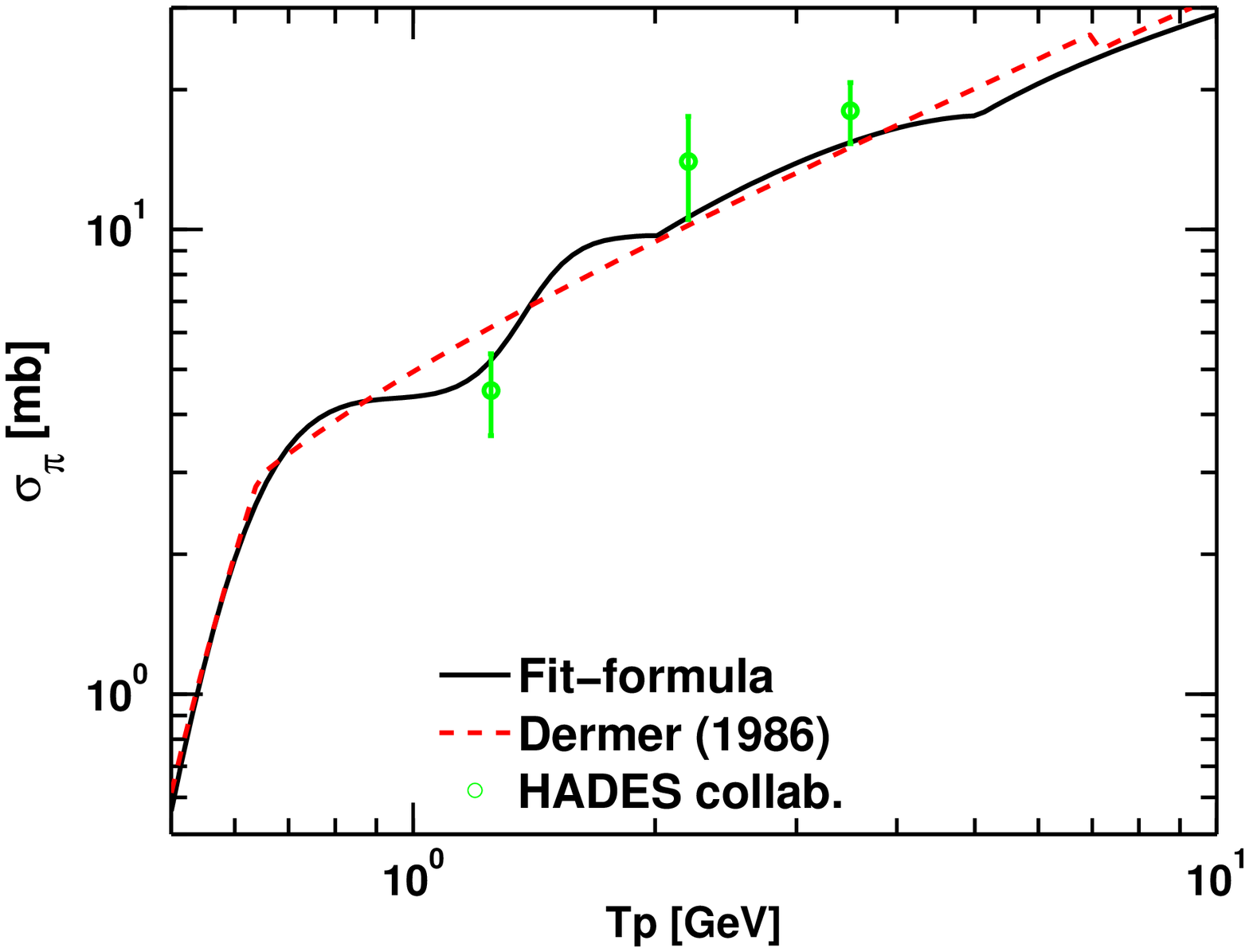}
\includegraphics[scale=0.22]{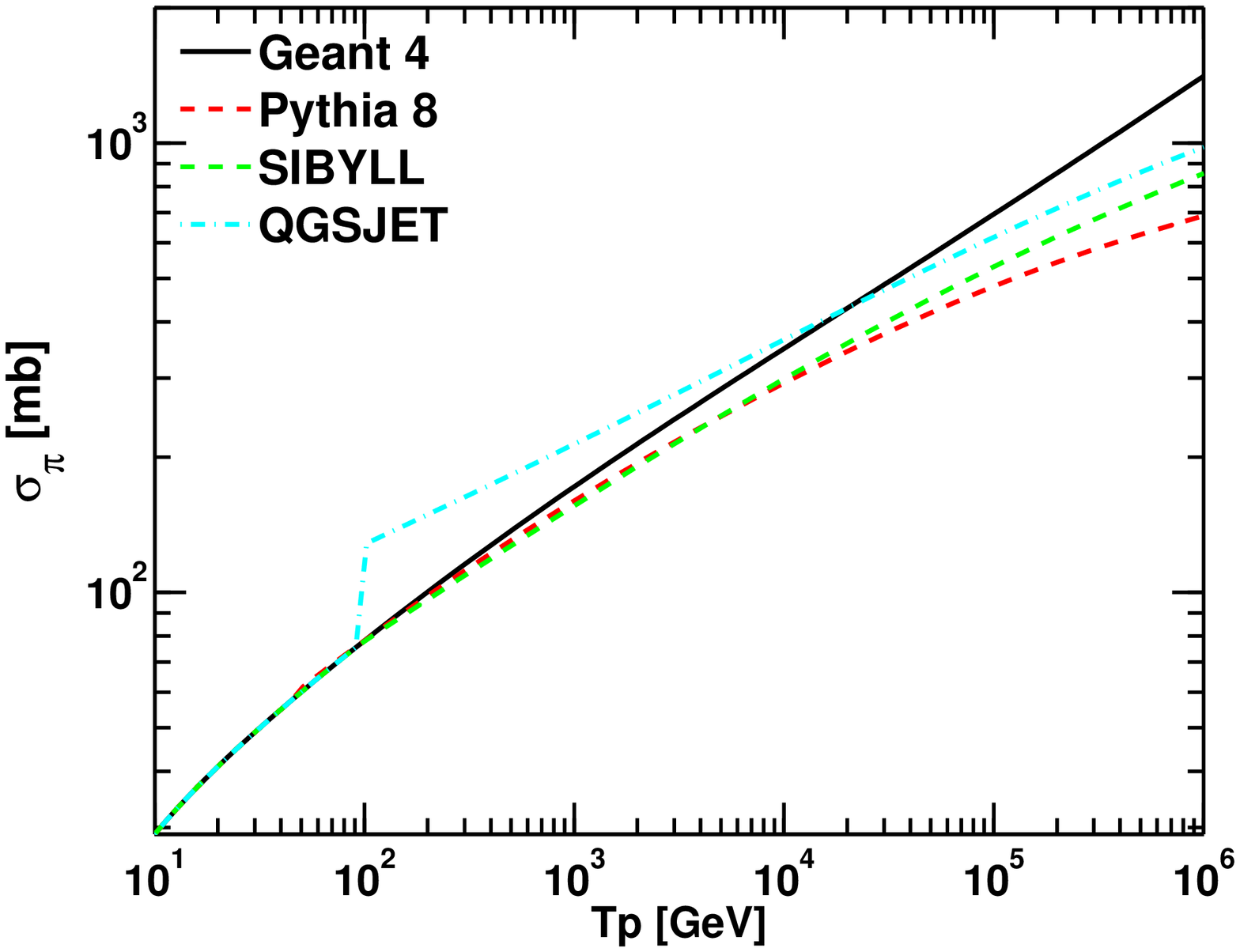}
\caption{Inclusive $\pi^0$ production cross section as a function of proton kinetic energy. The left panel compares the parametrization given in ref.~\citep{Dermer1986b}, our parametrization and the experimental data at 1.25, 2.2 and 3.5 GeV from the HADES collaboration \cite{Agakichiev2010, Agakishiev2012inc, Agakishiev2012}. The panel on the right shows the differences between high energy models using the parametrization we have introduced here. Notice that Geant~4.10.0 works for $T_p\leq 10^5$~GeV, in the plot we have extrapolated its multiplicity up to $T_p =10^6$~GeV. \label{fig:XSinclPi0} }
\end{figure}

Figure~\ref{fig:XSinclPi0} compares our low energy $\sigma_\pi$ values with the parametrization provided in ref.~\cite{Dermer1986b} and the data from the HADES collaboration \cite{Agakichiev2010, Agakishiev2012inc, Agakishiev2012}. We can conclude that the ref.~\cite{Dermer1986b} parametrization is missing some features that the inclusive $\pi^0$ production cross section has around 1--2 GeV. Differences between the high energy models that we have used here are also shown in figure~\ref{fig:XSinclPi0}. 

As we have mentioned, QGSJET-I predicts a $\pi^0$ production average multiplicity or cross section at $T_p=100$~GeV, about 1.7 times higher than the other models considered. Notice also that Geant~4.10.0 works up to $T_p=10^5$~GeV, thus, the discrepancies between models are less than 30~\% for $10^3~{\rm GeV}\leq T_p\leq 10^6$~GeV.

%===================================================
% Gamma-ray spectra
%===================================================
\section{Parametrization of the $\gamma$-ray differential cross section \label{sec:GamFit}}
We divide the $\gamma$-ray differential cross section into two parts. The fist part is the maximum value $A_{\rm max}(T_p)=\max(d\sigma/dE_\gamma)$ that depends only on the proton energy $T_p$. The second part is $F(T_p,E_\gamma)$ which describes the shape of the spectrum and is a function of $T_p$ and $\gamma$-ray energy $E_\gamma$. We therefore express the differential cross section as,
\begin{equation}
\frac{d\sigma}{dE_\gamma}\left(T_p,E_\gamma\right) = A_{\rm max}(T_p)\times F(T_p,E_\gamma).
\end{equation}

\noindent The peak $A_{\rm max}(T_p)$ is a function of the pion production cross section and is fitted separately from $F(T_p,E_\gamma)$. %From the definition, the function $F(T_p,E_\gamma)$ varies from 0 to 1.
Let us define the following variables:
\begin{equation}\label{eq:YandX}
\begin{split}
Y_\gamma = & E_\gamma + \frac{m_{\pi}^2}{4\,E_\gamma},~~~
Y_\gamma^{\rm max}= E_\gamma^{\rm max} + \frac{m_{\pi}^2}{4\,E_\gamma^{\rm max}},\\
X_\gamma = & \frac{Y_\gamma - m_{\pi}}{Y_\gamma^{\rm max} - m_{\pi}}. 
\end{split}
\end{equation}

\noindent Here, $E_\gamma^{\rm max}$ is the maximum $\gamma$-ray energy allowed by the kinematics. Let us denote with $E_\pi^{CM}$ and $E_{\pi~LAB}^{max}$ the maximum $\pi^0$ total energy in the center-of-mass and laboratory systems, respectively. Let us further denote with $\gamma_{CM}$ the Lorentz factor of the center-of-mass system and $\gamma_\pi^{LAB}$ the maximum $\pi^0$ Lorentz factor in the laboratory system. The maximum or minimum $\gamma$-ray energy $E_\gamma^{\rm max/min}$ is given by:
\begin{equation}\label{eq:EpiEgMAX}
\begin{split}
E_\pi^{CM}=& \frac{s - 4m_p^2 + m_\pi^2}{2\sqrt{s}}~,\\
E_{\pi~LAB}^{max} =& \gamma_{CM}\,(E_\pi^{CM} + P_\pi^{CM}\, \beta_{CM})~, \\
\gamma_{CM} =& \frac{T_p + 2m_p}{\sqrt{s}},~\gamma_\pi^{LAB} = \frac{E_{\pi~LAB}^{max}}{m_\pi},\\
E_\gamma^{\rm max/min} =& \frac{m_\pi}{2} \;\gamma_\pi^{LAB}\,\left(1\pm \beta_\pi^{LAB}\right)~.
\end{split}
\end{equation}

\noindent Here, $P_\pi^{CM} = \sqrt{(E_\pi^{CM})^2 - m_\pi^2}$ is the pion maximum center-of-mass momentum, $\beta_{CM} = \sqrt{1-\gamma_{CM}^{-2}}$ is the center-of-mass velocity, and $\beta_\pi^{LAB}=\sqrt{1-(\gamma_\pi^{LAB})^{-2}}$ is the pion maximum velocity in the laboratory system.

The value of $Y_\gamma^{\rm max}$ is the same if we replace the maximum $\gamma$-ray energy with the minimum one $E_\gamma^{\rm min}$. The reason being that $Y_\gamma$ is a symmetric function of $E_\gamma$ with respect to $E_\gamma=m_\pi/2$ on a logarithmic energy scale. Values of $Y_\gamma$ lie in $Y_\gamma\in\left[ m_{\pi}, Y_\gamma^{\rm max}\right]$, thus, $X_\gamma\in \left[ 0,1\right]$. It is expected that $F(T_p,E_\gamma)$ be a function of $Y_\gamma$, hence, a function of $X_\gamma$. Therefore, the functional shape of $F(T_p,X_\gamma)$, for a given proton energy, is defined in the range $\left[ 0,1\right] \times \left[ 0,1\right]$, which simplifies the expression of its parametrization.

With very good accuracy $F(T_p,E_\gamma)$ can be parametrized as

\begin{equation}\label{eq:Fgamma2}
F(T_p,E_\gamma)=\frac{\left(1-X_\gamma^{\alpha(T_p)} \right)^{\beta(T_p)}}{\left(1+\frac{X_\gamma}{C}\right)^{\gamma(T_p)}}~.
\end{equation}

\begin{table}[h]
\centering
\caption{Functions $\alpha(T_p)$, $\beta(T_p)$, $\gamma(T_p)$ and $\lambda$ coefficients in eq.~(\ref{eq:Fgamma2}) for Geant 4.10.0, Pythia 8.18, SIBYLL~2.1 and QGSJET-I. Functions $\kappa(T_p)$ and $\mu(T_p)$ are given in eqs.~(\ref{eq:dsigdEgThresh}) and (\ref{eq:mu}), respectively. \label{tab:FgamG4P8SQ}}
\begin{tabular}{cccccc}
%\toprule 
\toprule
Model  & Energy [GeV] & $\lambda$ & $\alpha(T_p)$ & $\beta(T_p)$ & $\gamma(T_p)$ \\
\toprule
Exp. Data  & $T_p^{\rm th}\le T_p \le 1$ & -- & 1.0 & $\kappa$ & 0 \\
\hline
Geant 4  & $1< T_p\le 4$ & 3.00 & 1.0 & ~$\mu+2.45$ & $\mu+1.45$ \\
\hline
Geant 4  & $4 < T_p\leq 20$ & 3.00 &  1.0 & $\frac{3}{2}\mu+4.95$ & $\mu+1.50$ \\
\hline
Geant 4  & $20 < T_p\leq 100$ & 3.00 & 0.5 & 4.2 &  1 \\
\hline
Geant 4  & $T_p > 100$ & 3.00 & 0.5 & 4.9 & 1 \\
\hline
Pythia 8 & $T_p > 50$ & 3.50 & 0.5 & 4.0 & 1 \\
\hline
SIBYLL~2.1   & $T_p > 100$ & 3.55 & 0.5 & 3.6 & 1 \\
\hline
QGSJET-I   & $T_p > 100$ & 3.55 & 0.5 & 4.5 & 1 \\
\toprule
\end{tabular}  
\end{table}

\noindent Here, $\kappa(T_p)$ and $\mu(T_p)$ are given by eq.~(\ref{eq:dsigdEgThresh}) and (\ref{eq:mu}) respectively, and  
$C=\lambda \times m_\pi/Y_\gamma^{\rm max}$. Table~\ref{tab:FgamG4P8SQ} lists the values of $\lambda$, $\alpha(T_p)$, $\beta(T_p)$ and $\gamma(T_p)$ for different energies and models.

The peak value $A_{\rm max}$ on the other hand can be expressed as follows

\begin{equation}\label{eq:Amax}
\begin{split}
A_{\rm max}(T_p) = & b_0\times\frac{\sigma_\pi(T_p)}{E_{\pi}^{\rm max}}:~{\rm for}~ T_p^{\rm th} \leq T_p < 1~{\rm GeV}, \\
A_{\rm max}(T_p) = & b_1\, \theta_p^{-b_2}\, \exp\left(b_3\, \log^2(\theta_p)\right)\times  \frac{\sigma_\pi(T_p)}{m_p}\\ 
&  ~~~~~~~~~~~~~~~:~{\rm for}~  T_p \geq 1~{\rm GeV}. \\
\end{split}
\end{equation}
\noindent Where, $\theta_p=T_p/m_p$, $E_\pi^{\rm max}$ is the maximum total $\pi^0$ energy in the laboratory frame that is allowed by the kinematics, see eq.~(\ref{eq:EpiEgMAX}). $\sigma_\pi(T_p)$ is the inclusive $\pi^0$ production cross section as was explained in section~\ref{sec:xs}, and $b_0-b_3$ are constants that depend on the hadronic model.

\subsection{Gamma-ray spectra for $T_p^{\rm th} \leq T_p <1$~GeV }
Good quality $\pi^0$--spectral data for $T_p<1$~GeV exist in the literature. We have collected these data (see table~\ref{tab:dXSdEpi}), calculated the corresponding $\gamma$-ray spectra produced, and compared these with other models. 

\begin{figure}[h]
\includegraphics[scale=0.232]{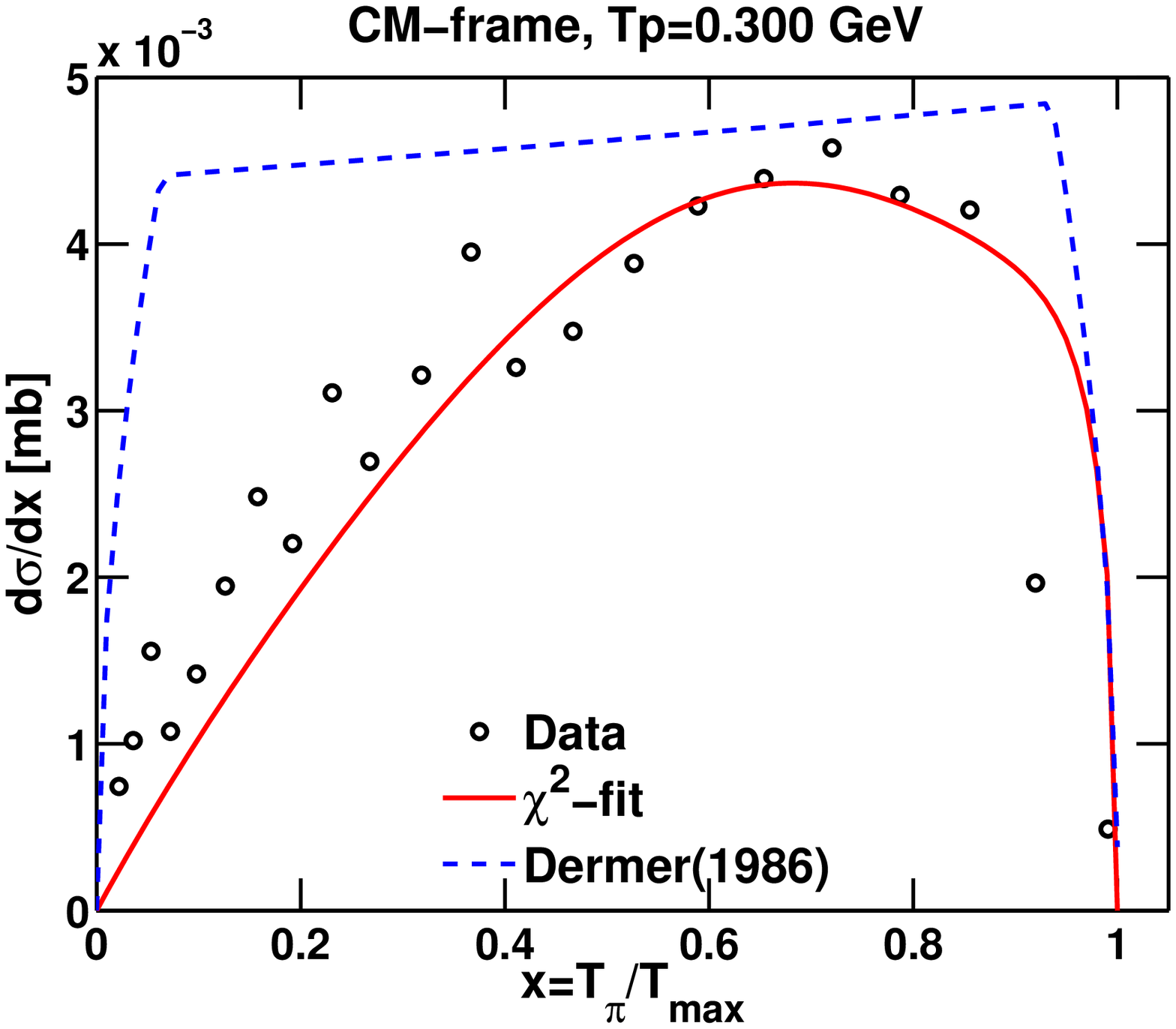}
\includegraphics[scale=0.232]{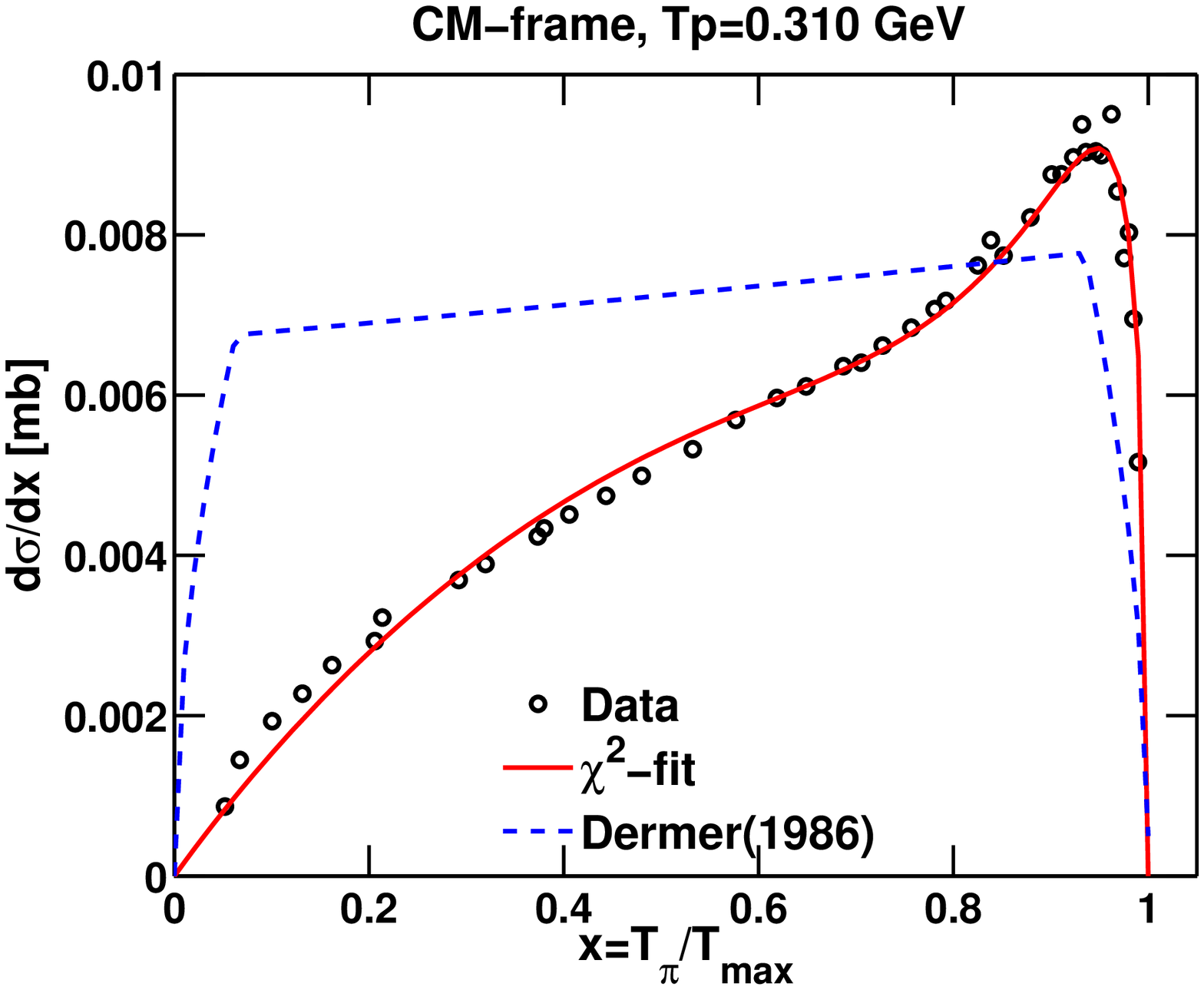}\\
\includegraphics[scale=0.233]{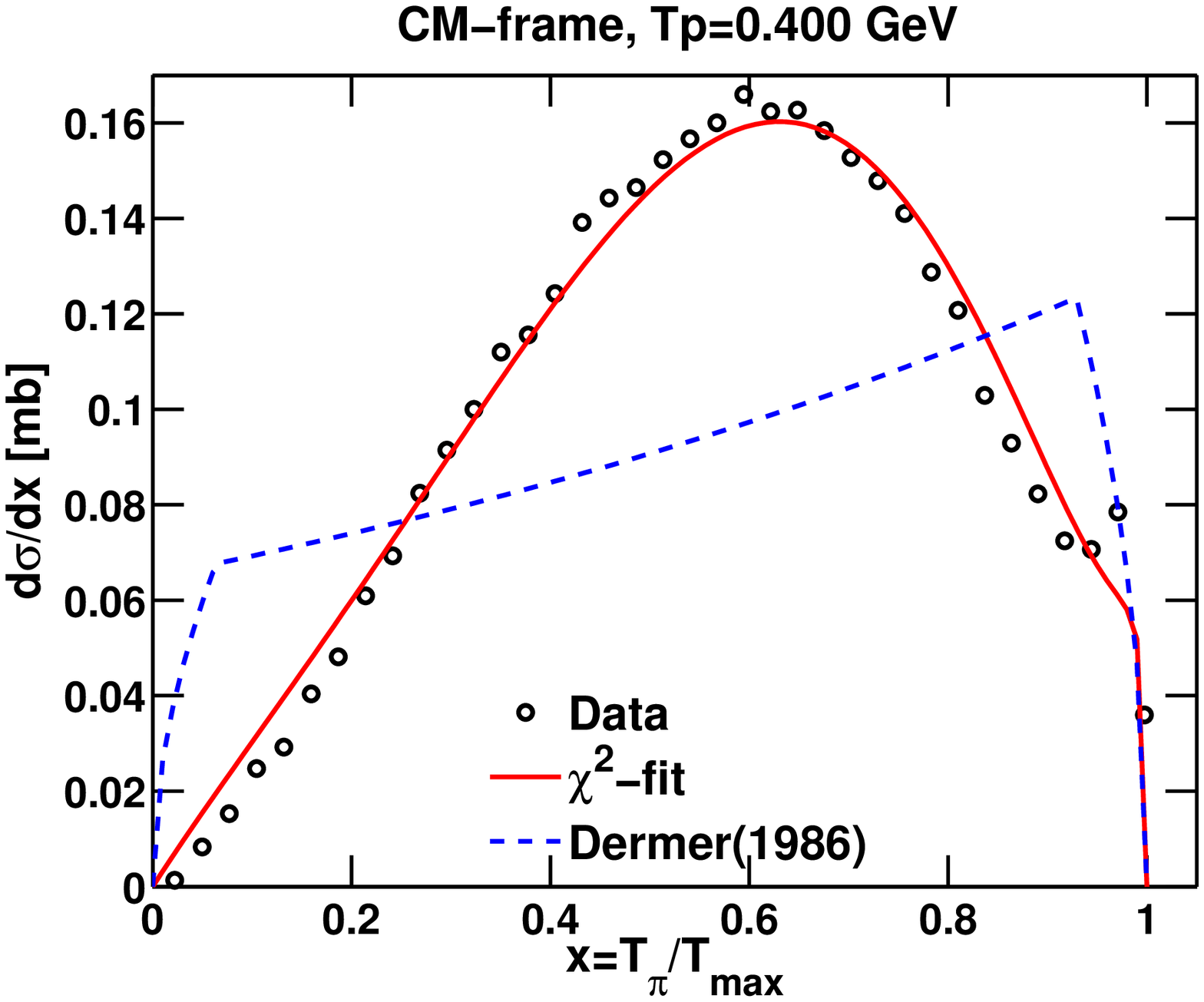}
\includegraphics[scale=0.233]{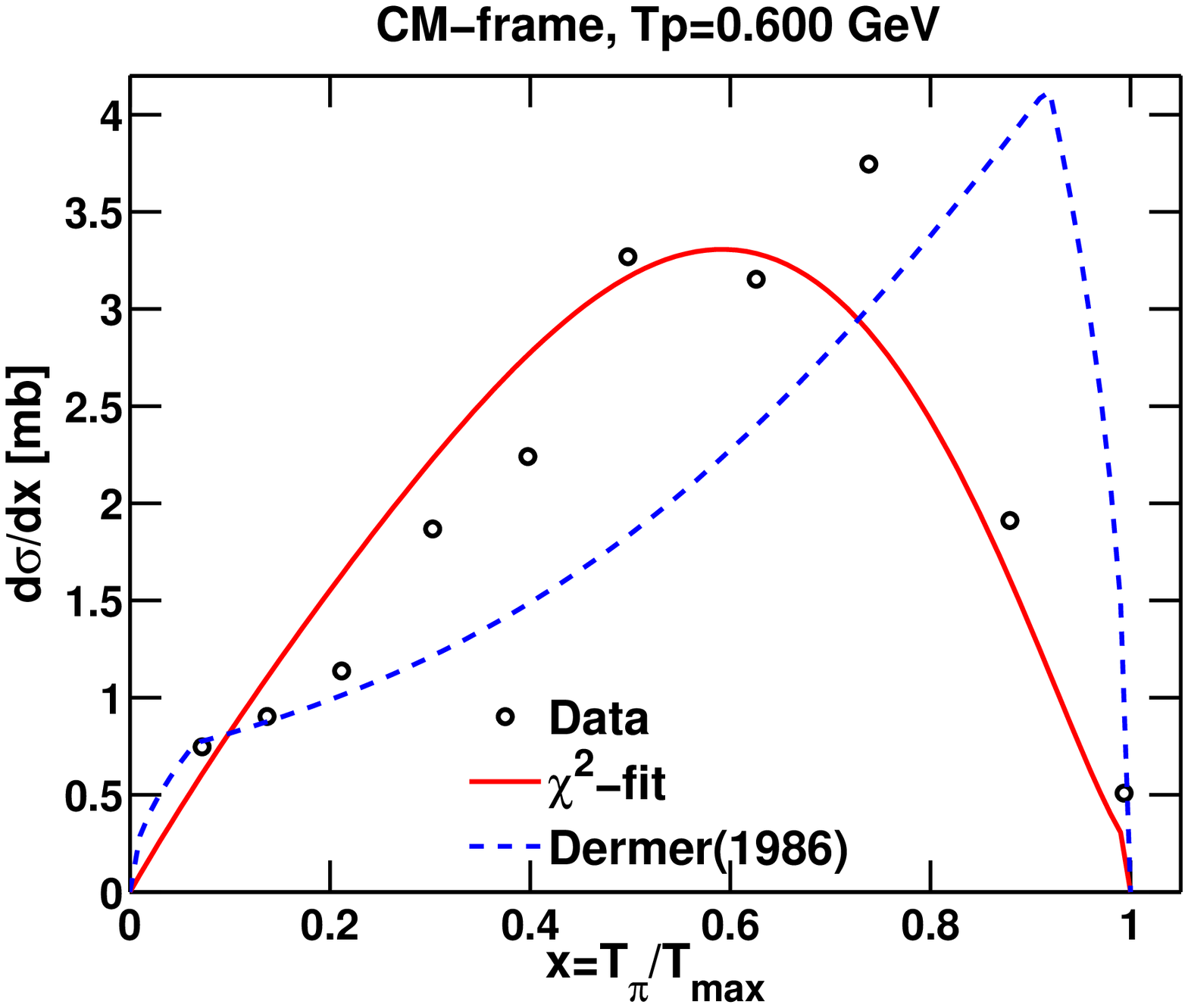}\\
\includegraphics[scale=0.237]{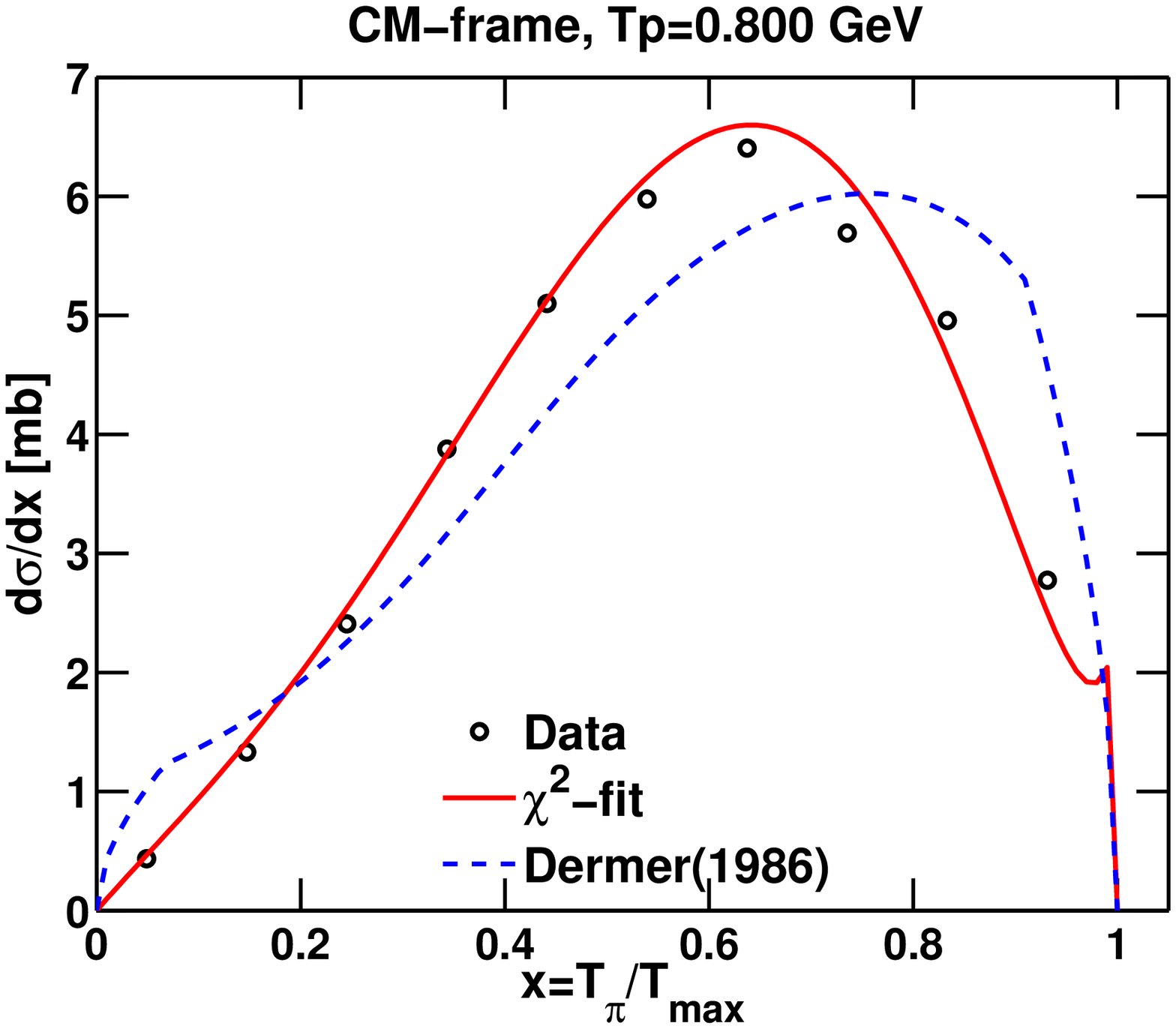}
\includegraphics[scale=0.237]{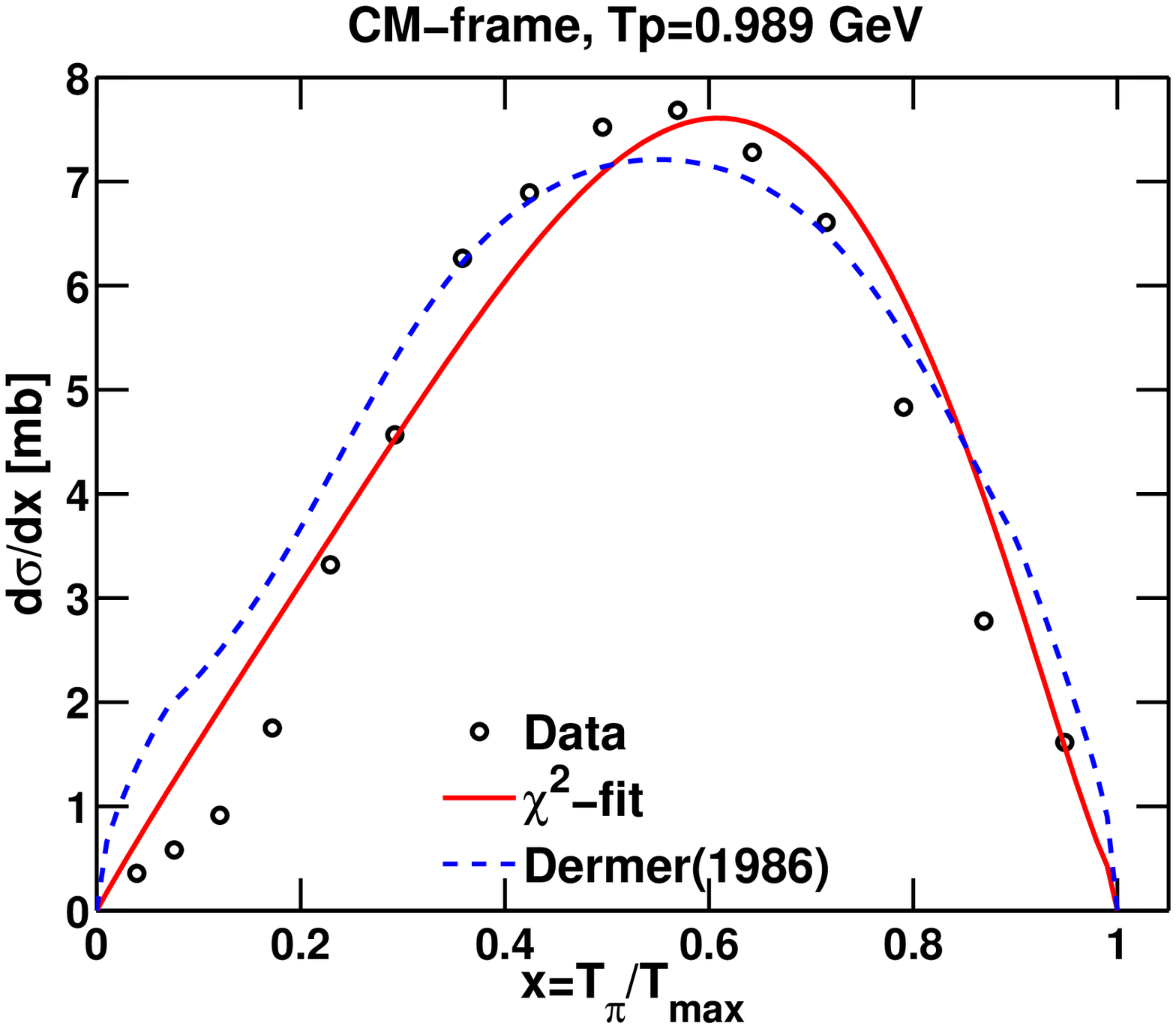}
\caption{Comparison of the center-of-mass system $\pi^0$--spectrum between data (open circles), our $\chi^2$--fit of the data (full red line) and the isobar calculations from \cite{Dermer1986b} (dash blue line). Our fit of the pion differential cross section is based on the partial wave expansion and the final state interaction (FSI). This function is a sum of three terms $f_1$, $f_2$ and $f_3$ which are a function of $x=T_\pi/T_{\rm max}$, where $T_\pi$ is the $\pi^0$ kinetic energy in the center-of-mass system, while, $T_{\rm max}$ is the maximum $\pi^0$ kinetic energy in the center-of-mass frame allowed by the kinematics, see eq.~(\ref{eq:EpiEgMAX}). The function $f_1=A_1\times x^4\frac{\sqrt{1-x}}{1-x+B}$, where $B=1{\rm MeV}/T_{\rm max}$. Function $f_2=A_2\times x\,(1-x)$ and $f_3=A_3\times x^3\,(1-x)^{3/2}$. $A_1$, $A_2$ and $A_3$ are fitted using the $\chi^2$ method. The pion differential cross section is given by $\frac{d\sigma}{dT_\pi}=\frac{\sigma_\pi(T_p)}{T_{\rm max}}\times(f_1+f_2+f_3)$, with $\sigma_\pi(T_p)$ the $\pi^0$ inclusive production cross section. \label{fig:dxsPi0}}
\end{figure}

Figure~\ref{fig:dxsPi0} shows some of these data as well as our $\chi^2$-fit of them.

\begin{table}[h]
\caption{References for $pp\to\pi^0$ differential cross section data for $T_p\in [0.293,0.989]$~GeV. \label{tab:dXSdEpi}}
\begin{tabular}{ccc} 
\toprule
$T_p$ [GeV] & & Reference\\
\toprule
0.293 & & \cite{El-Samad2003pi0Spec} \\
\hline
0.3 & & \cite{Bondar1995} \\
\hline
0.31--0.425 & & \cite{Zloma2000pi0Spec} \\
\hline
0.4 & & \cite{ElSamad2006} \\
\hline
0.6 -- 0.86 & &  \cite{Andreev1994pi0Spec} \\
\hline
0.8 & & \cite{Comptour1994pi0Spec} \\
\hline
0.97 & & \cite{Bugg1964pi0Spec} \\
\hline
0.989 & & \cite{Sarantsev2004pi0Spec} \\
\toprule
\end{tabular}
\end{table}

Assuming that the $\pi^0$ spectra in the center-of-mass system for $T_p<1$~GeV is isotropic, one can calculate the $\pi^0$ spectra in the laboratory frame ($d\sigma_\pi/dE_\pi$) using its description in the center-of-mass frame, and subsequently calculating the $\gamma$-ray spectra with the relation
\begin{equation}
\frac{d\sigma}{dE_\gamma}=2\times \int\limits_{Y_\gamma}^{E_\pi^{\rm max}} \frac{d\sigma_\pi}{dE_\pi}\,\frac{dE_\pi}{\sqrt{E_\pi^2-m_\pi^2}}~.
\end{equation}

\noindent Here, $E_\pi$ is the $\pi^0$ total energy in the laboratory frame, $E_\pi^{\rm max}$ is the maximum total $\pi^0$ energy in the laboratory frame that is allowed by the kinematics see eq.~(\ref{eq:EpiEgMAX}), $Y_\gamma$ is given in eq.~(\ref{eq:YandX}) and the factor 2 in front comes from the multiplicity of $\gamma$-rays due to $\pi^0\to2\gamma$ decay.  

Using a $\chi^2$ method, we have fitted these results with functions presented in eqs.~(\ref{eq:Fgamma2}) and (\ref{eq:Amax}). 
We find:
\begin{align}\label{eq:dsigdEgThresh}
b_0=  5.9, && \kappa(T_p)=  3.29 - \frac{1}{5}\,\theta_p^{-3/2}
\end{align}
\noindent where $\theta_p=T_p/m_p$.

\begin{figure}[h]
\includegraphics[scale=0.35]{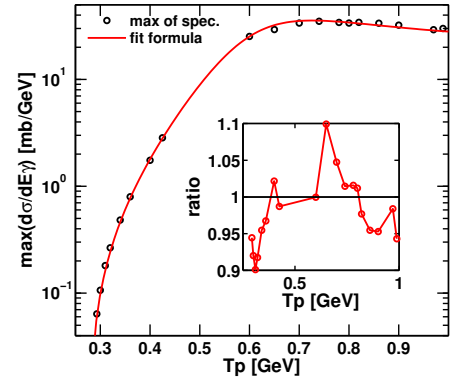}
\caption{The peak value of the $\gamma$-ray production differential cross section as a function of proton kinetic energy. Open circles represent the calculations from $\pi^0$-spectrum data; whereas, the line represents our fit function $A_{\rm max}= 5.9\times\frac{\sigma_\pi(T_p)}{E_{\pi}^{\rm max}(T_p)}$. The accuracy of the fit is better than 10 \% as it is seen from the ratio. \label{fig:ExpspecMAX} }
\end{figure}

Figure~\ref{fig:ExpspecMAX} shows how well the $A_{\rm max}$ of eq.~(\ref{eq:Amax}), fits the peak value of the $\gamma$-ray spectrum calculated from $\pi^0$ spectral data. The accuracy of the fit is better than 10 \%, as can be seen from the ratio. 

\begin{figure}[h]
\includegraphics[scale=0.2295]{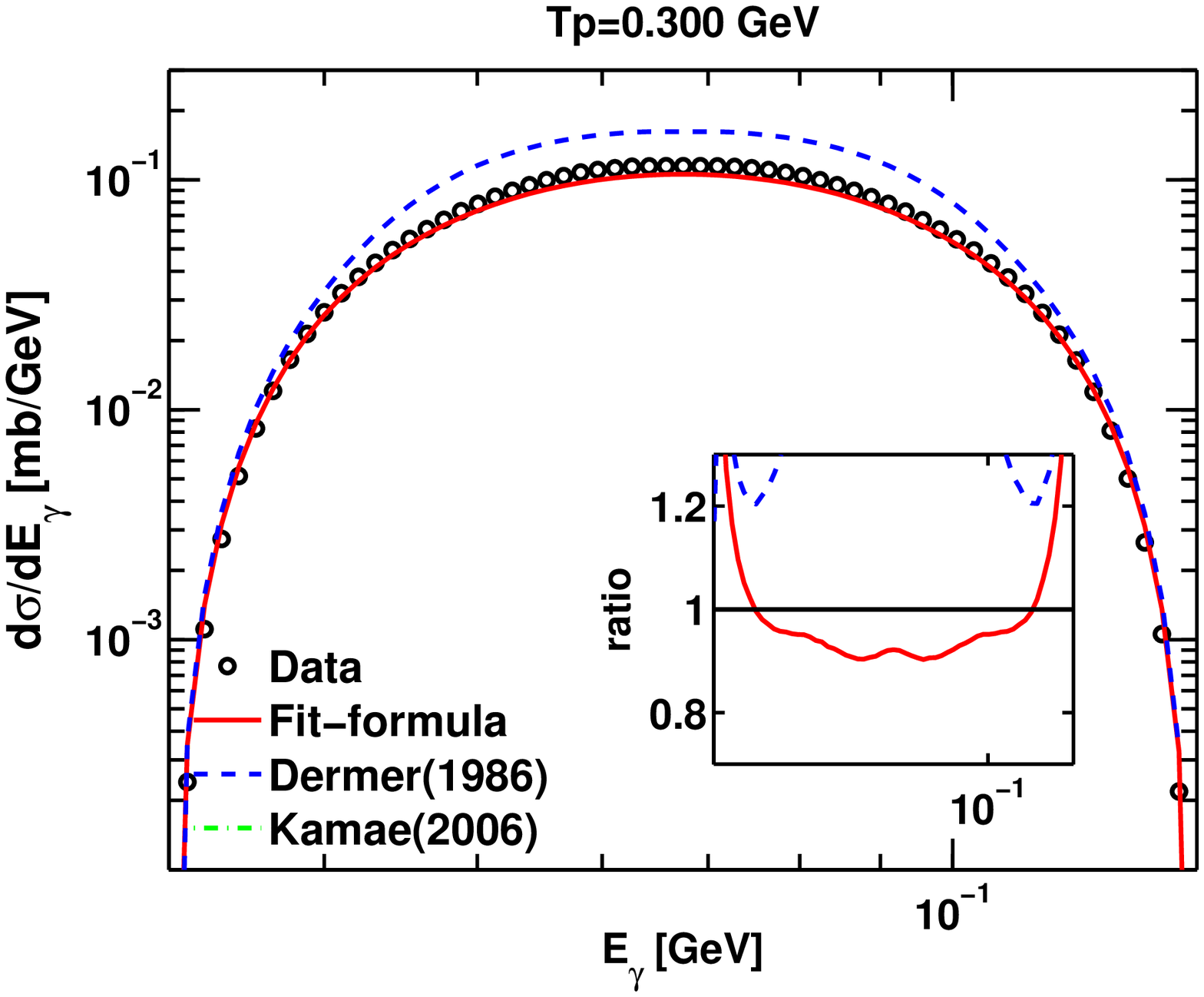}
\includegraphics[scale=0.2295]{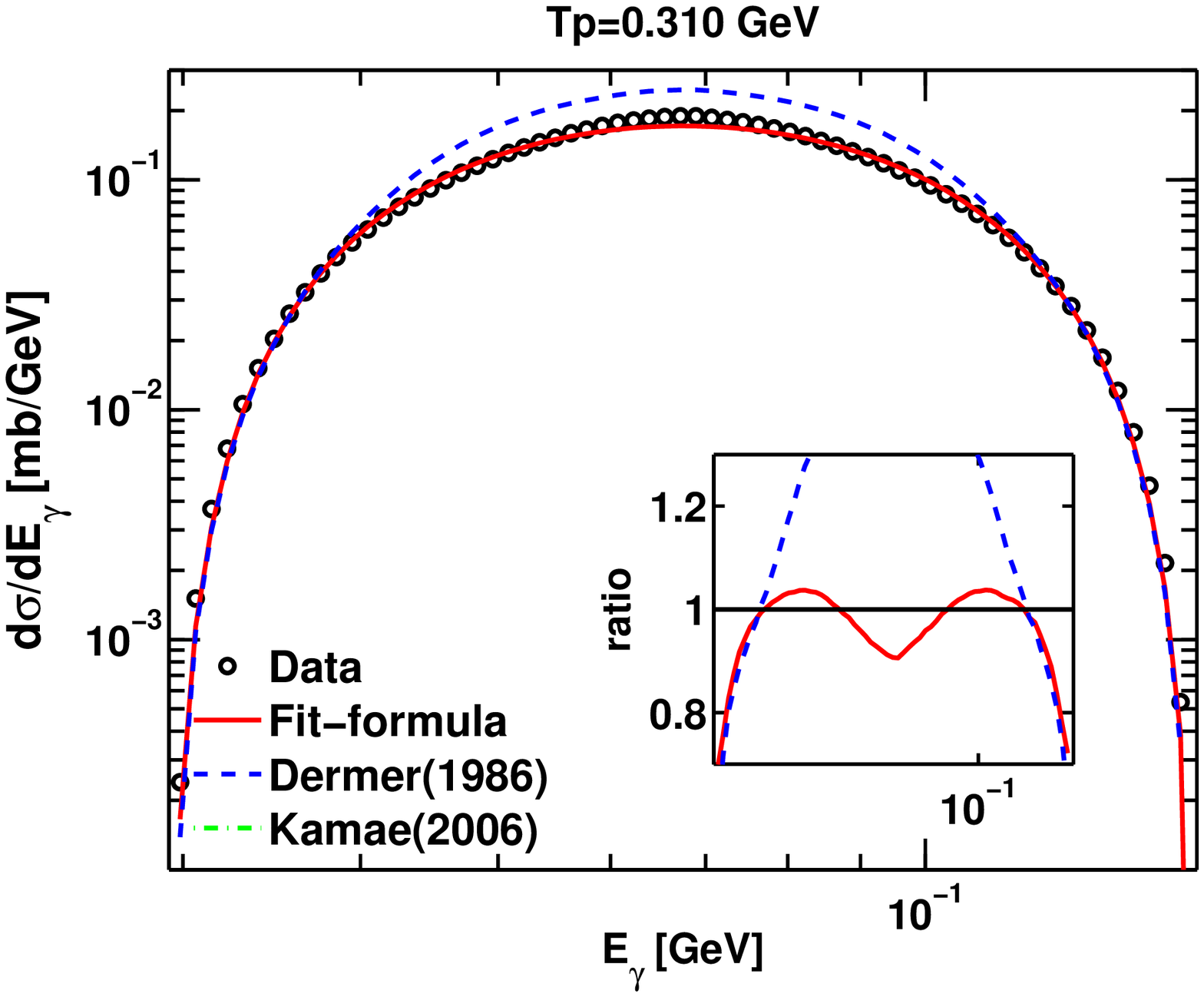}\\
\includegraphics[scale=0.2295]{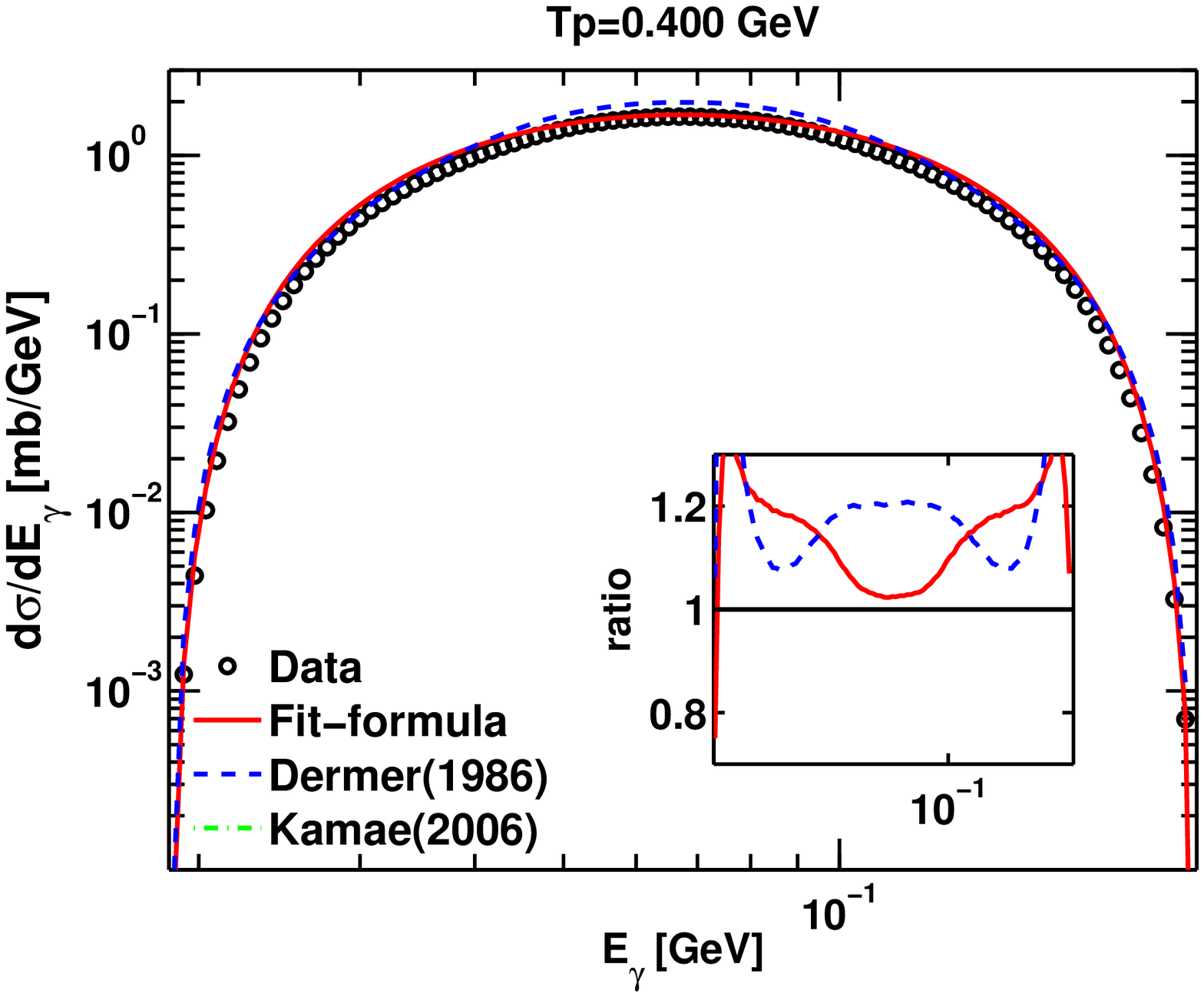}
\includegraphics[scale=0.2295]{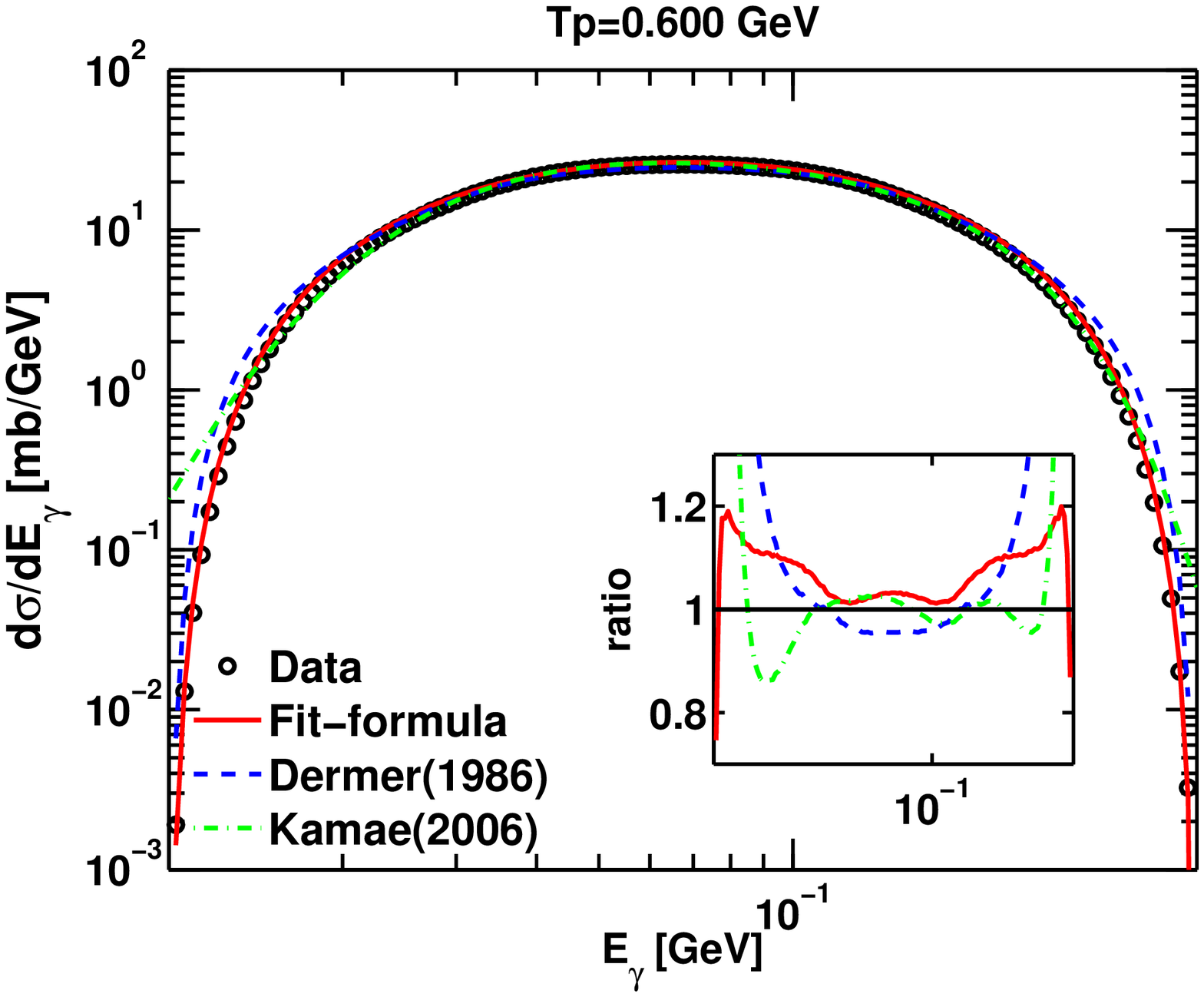}\\
\includegraphics[scale=0.2295]{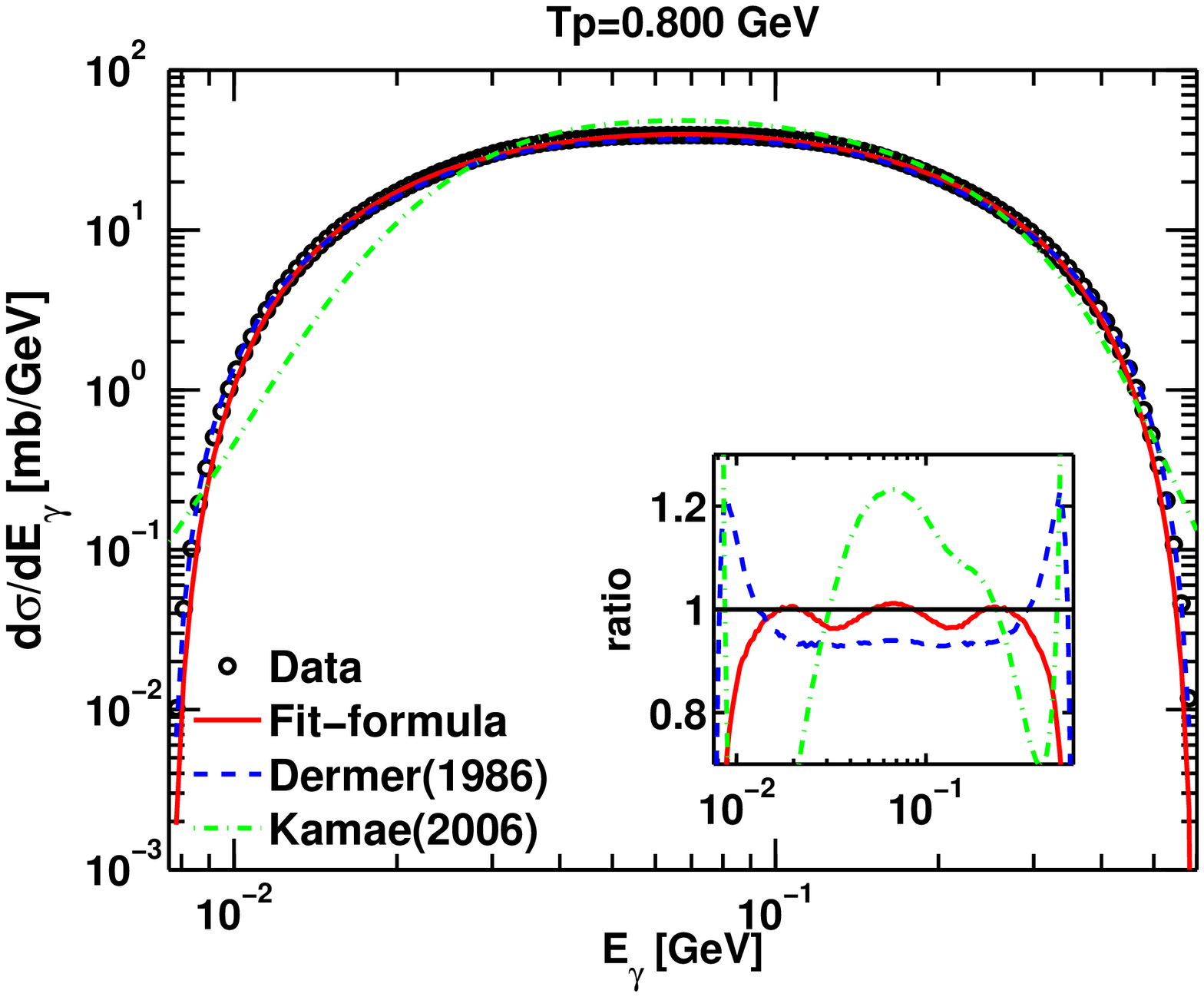}
\includegraphics[scale=0.2295]{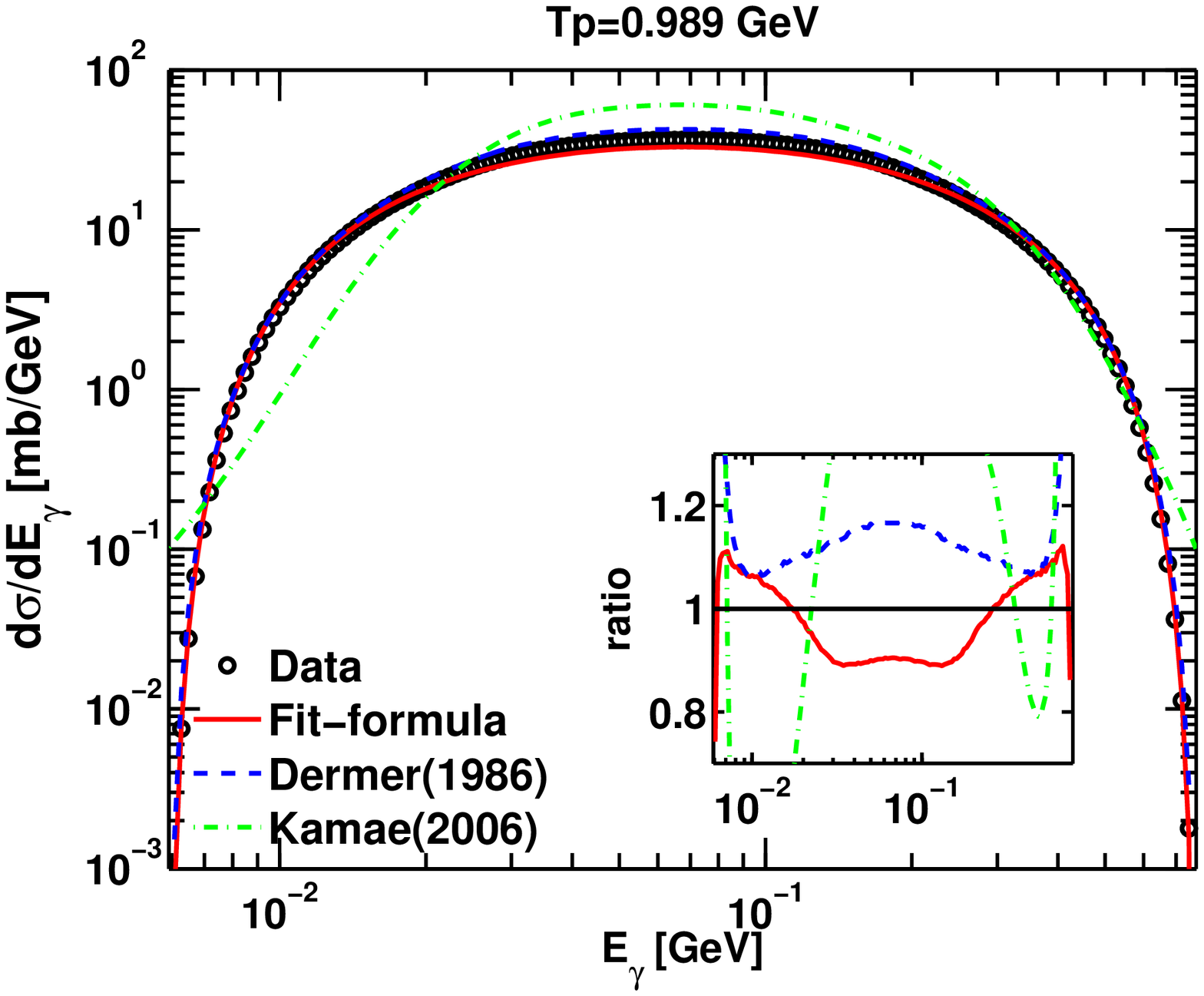}
\caption{Gamma-ray spectrum for different proton collision energies. The open circles (``Data'') are calculations using experimental $\pi^0$ spectra. Full red line is the fit formula in eq.~(\ref{eq:dsigdEgThresh}). Dash blue line are calculations from \cite{Dermer1986b}. Dash-dot green line are calculations from \cite{Kamae2006} fits. Ref.~\cite{Kamae2006} calculations are valid for $T_p\geq 0.488$~GeV and they do not respect $\gamma$-ray kinematic limit. Moreover, they do not satisfy the expected symmetry with respect to $E_\gamma=m_\pi/2$ that $\pi^0\to2\gamma$ decay create. The ratios are taken with respect to calculations from data and the accuracy of the fit is better than 20 \%. \label{fig:dxsEgam}}
\end{figure}

Figure~\ref{fig:dxsEgam} shows some examples of the $\gamma$-ray production differential cross section calculated from the experimental $\pi^0$ spectrum data and compares them with the low energy models from \cite{Dermer1986b,Kamae2006}. Although, the shape of $\pi^0$ spectra between data and the ref.~\cite{Dermer1986b} calculations are very different in the center-of-mass frame, these differences become less noticeable in the $\gamma$-ray spectra. This testifies to the fact that the $\gamma$-ray spectrum inherits from the $\pi^0$ production, the multiplicity and the kinematics of the $\pi^0\to2\gamma$ decay. Subsequently, the $\gamma$-ray spectrum washes out and is less sensitive to the actual shape of the $\pi^0$ production spectrum.

\subsection{Gamma-ray spectra from Gean~4.10.0, Pythia~8.18, SIBYLL~2.1 and QGSJET-I}
Employing the functions in eqs.~(\ref{eq:Fgamma2}) and (\ref{eq:Amax}), we have fitted the $\gamma$-ray production differential cross sections for all these models. The fitting results are summarized in tables~\ref{tab:AmaxG4P8SQ} and \ref{tab:FgamG4P8SQ}. The function $\mu(T_p)$ that is used in table~\ref{tab:FgamG4P8SQ} is given in eq.~(\ref{eq:mu}). If $q=(T_p - 1~{\rm GeV})/m_p$, then
\begin{equation}\label{eq:mu}
\mu(T_p) = \frac{5}{4}\;q^\frac{5}{4}\; \exp\left(-\frac{5}{4}\,q\right).
\end{equation}

\begin{table}[h]
%\centering
\caption{Coefficients $b_1- b_3$ in eq.~(\ref{eq:Amax}) for Geant~4.10.0, Pythia~8.18, SIBYLL~2.1 and QGSJET-I. \label{tab:AmaxG4P8SQ}}
\begin{tabular}{ccccc}
%\toprule 
\toprule
Model  & Energy range & $b_1$ & $b_2$ & $b_3$ \\
\toprule
Geant 4  & $1\leq T_p< 5$~GeV & 9.53 & 0.52 & 0.054 \\
\hline
Geant 4  & $T_p\geq 5$~GeV & 9.13 & 0.35 & 9.7e-3 \\
\hline
Pythia 8 & $T_p > 50$~GeV & 9.06 & 0.3795 & 0.01105 \\
\hline
SIBYLL   & $T_p > 100$~GeV & 10.77 & 0.412 & 0.01264 \\
\hline
QGSJET   & $T_p > 100$~GeV & 13.16 & 0.4419 & 0.01439 \\
\toprule
\end{tabular}  
\end{table}

\begin{figure*}
\includegraphics[scale=0.3]{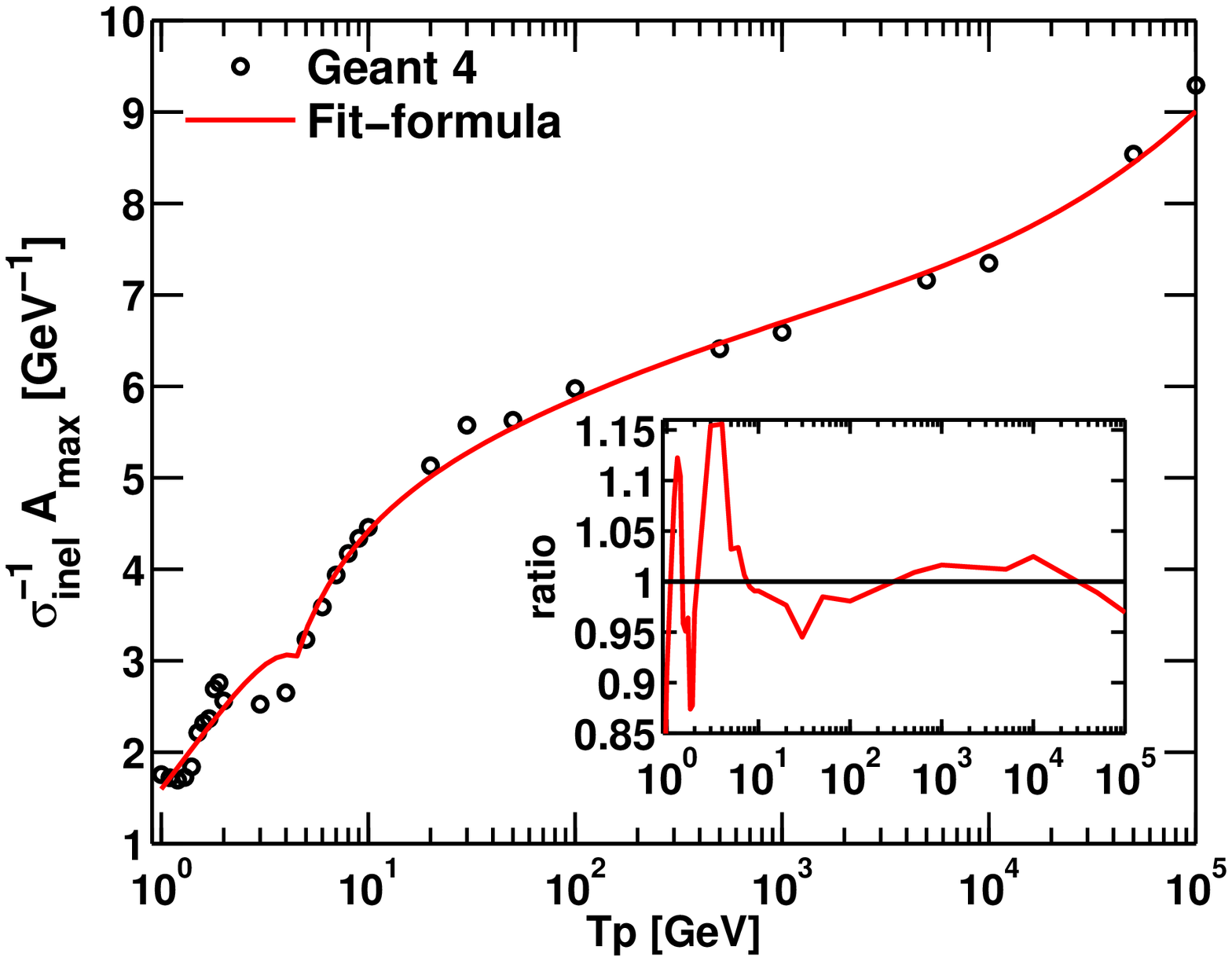}
\includegraphics[scale=0.3]{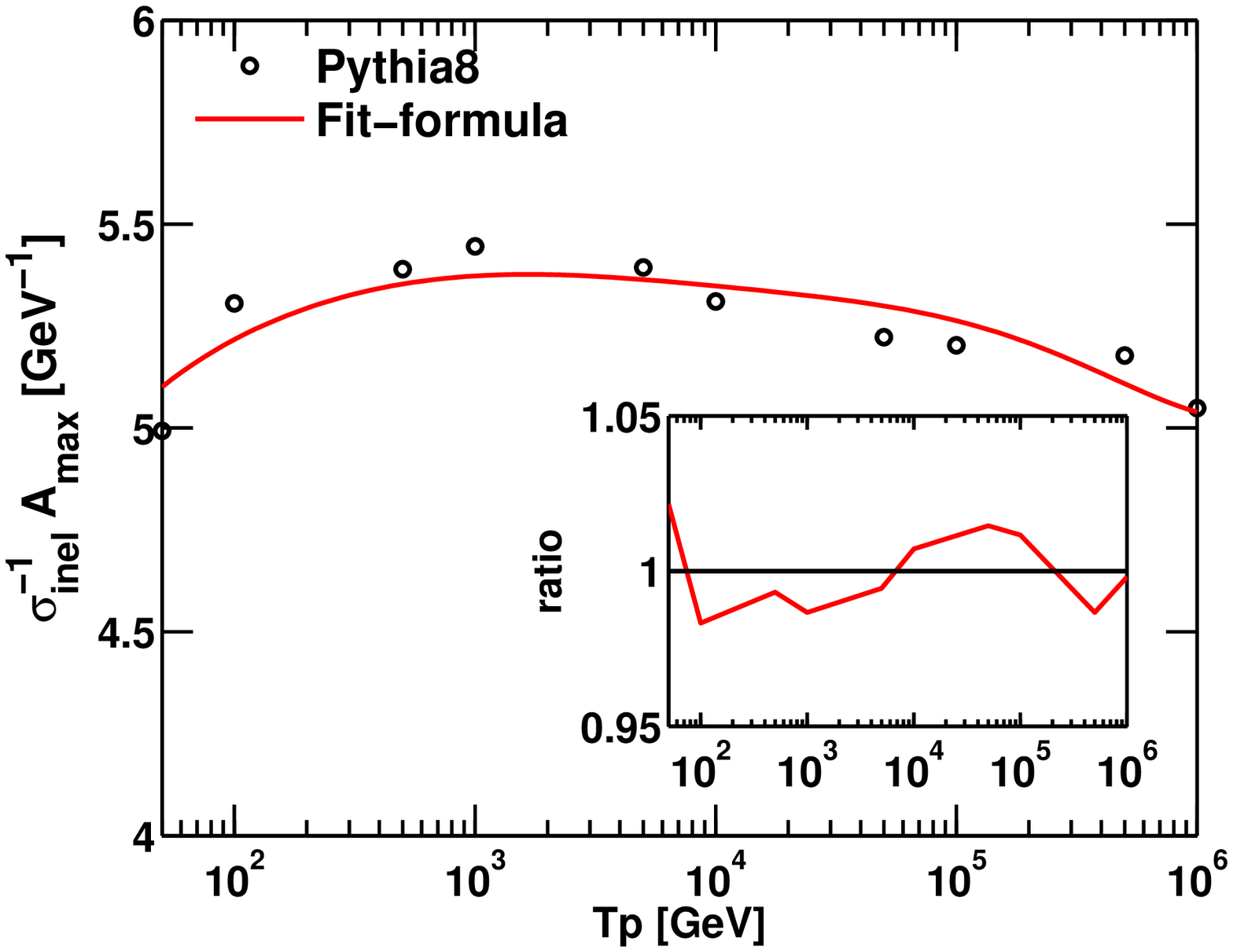}\\
\includegraphics[scale=0.3]{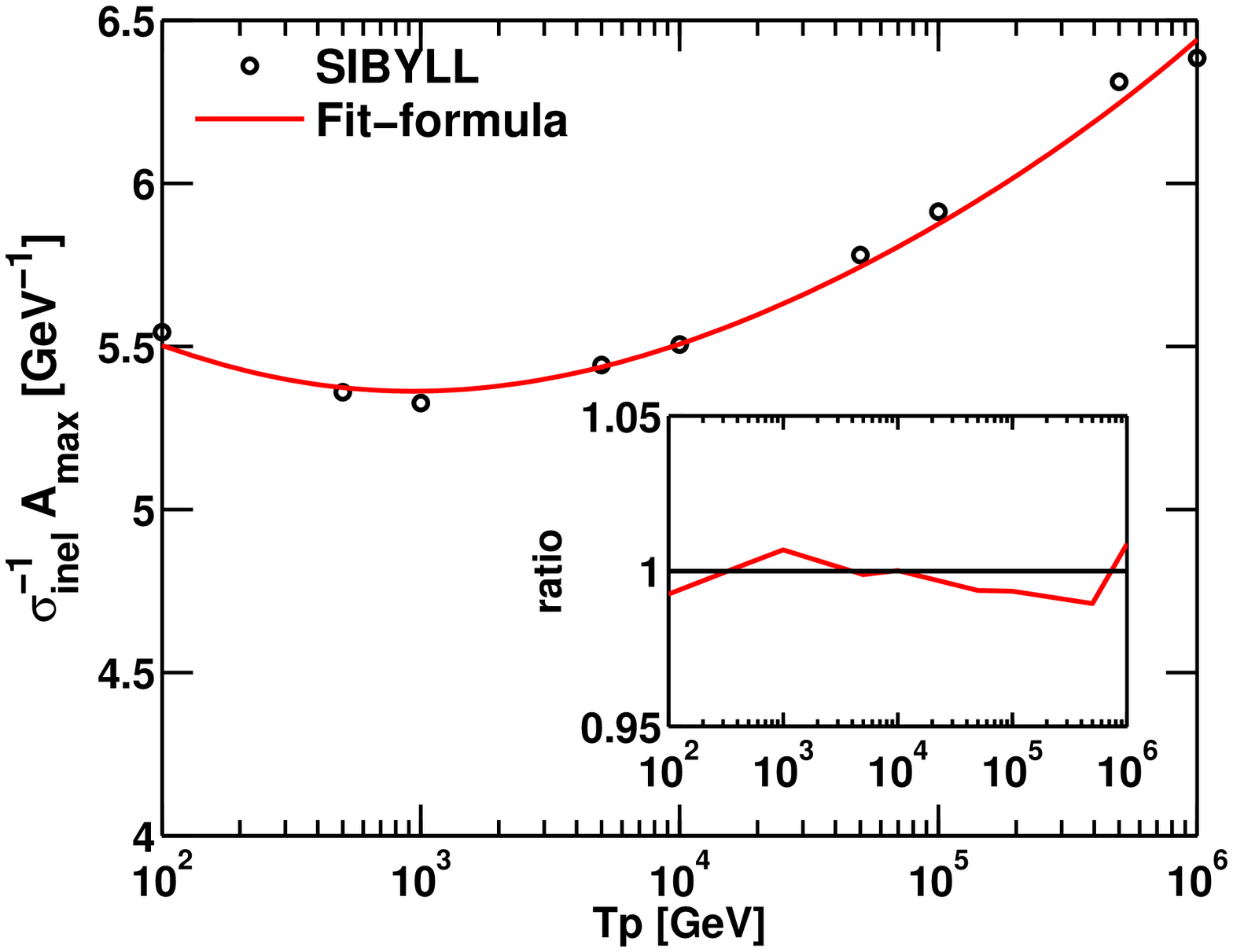}
\includegraphics[scale=0.3]{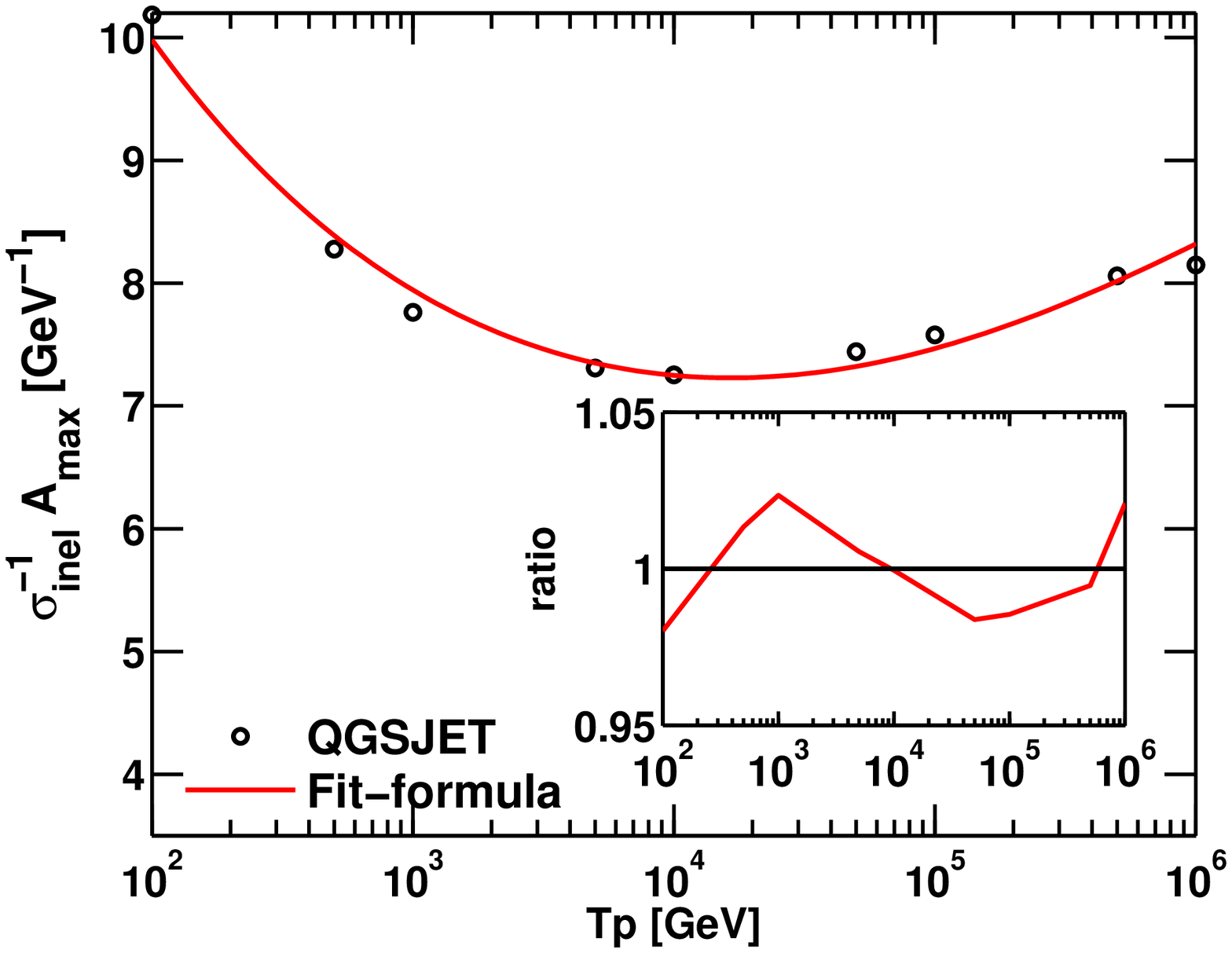}
\caption{The value of the peak of the $\gamma$-ray production differential cross section over the total inelastic cross section ($A_{\rm max}/\sigma_{\rm inel}$) as a function of proton kinetic energy. Open circles give Geant~4.10.0, Pythia~8.18, SIBYLL~2.1 and QGSJET-I calculations; whereas, the red line is the fit formula of eq.~(\ref{eq:Amax}) with coefficients listed in table~\ref{tab:AmaxG4P8SQ}. As the ratio shows, the accuracy of the fit is better than 15 \% for $1\leq T_p\leq 5$~GeV and better than 3 \% for $T_p>5$~GeV. It is remarkable that $A_{\rm max}/\sigma_{\rm inel}$, for most of the models, does not have a strong energy dependence especially for $T_p>100$~GeV. This means that $A_{\rm max}$ will have the same energy dependence, at high energies, as the total inelastic cross section. \label{fig:AmaxG4P8SQ}}
\end{figure*}

Figure~\ref{fig:AmaxG4P8SQ} shows the $A_{\rm max}$ fit calculated using eq.~(\ref{eq:Amax}) and parameters in table~\ref{tab:AmaxG4P8SQ} compared with the one calculated from Geant 4.10.0, Pythia 8.18, SIBYLL~2.1 and QGSJET-I models. As can be seen from the plots, the accuracy of the fit is better than 15 \% for $1\leq T_p\leq 5$~GeV and better than 3 \% for $T_p>5$~GeV. It is remarkable to notice that $A_{\rm max}/\sigma_{\rm inel}$, for most of the models, does not have a strong energy dependence especially for $T_p>100$~GeV. This means that $A_{\rm max}$, the peak of the $\gamma$-ray spectrum, takes the same energy dependence, at high energies, as the total inelastic cross section.

\begin{figure*}
\includegraphics[scale=0.3]{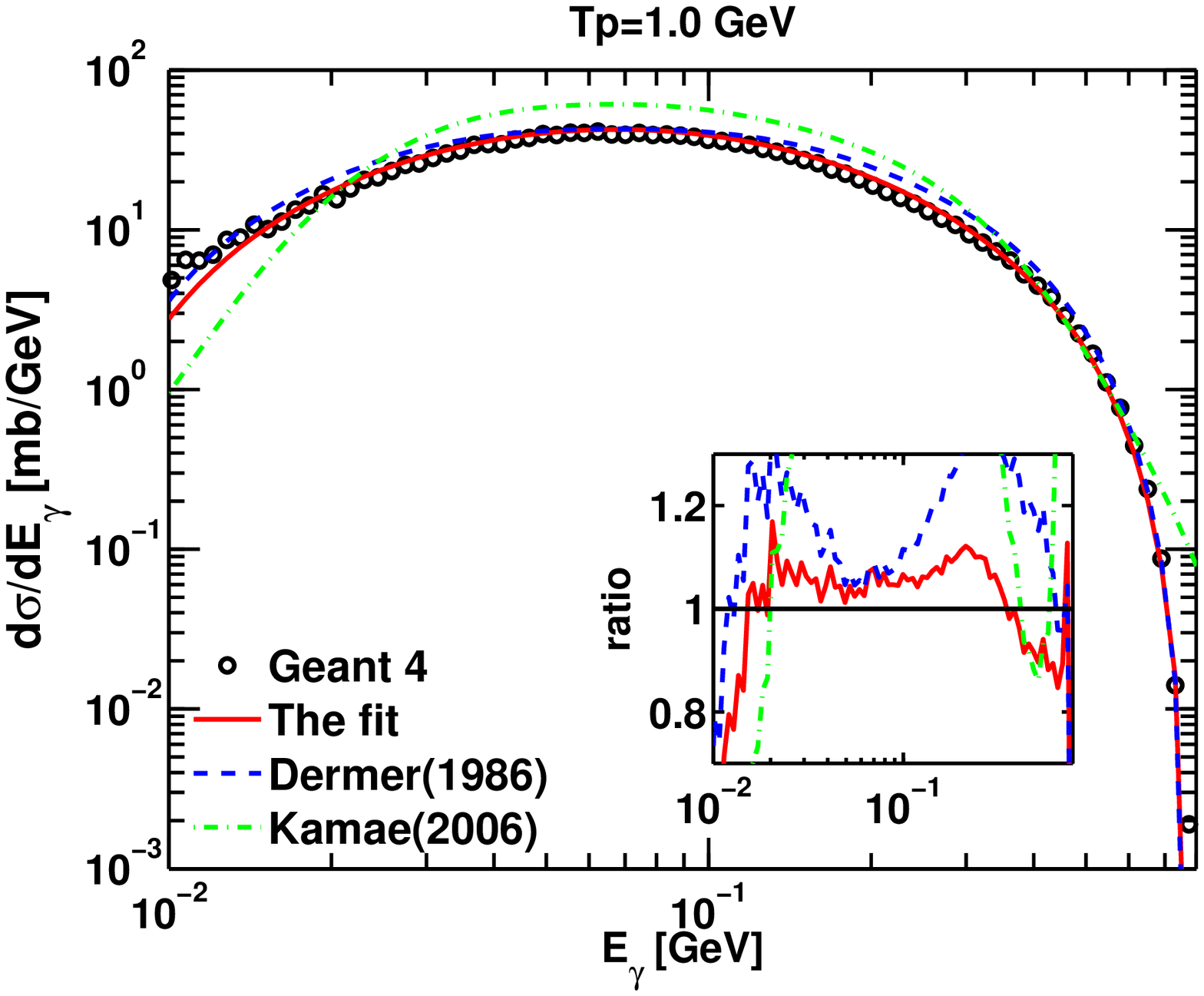}
\includegraphics[scale=0.3]{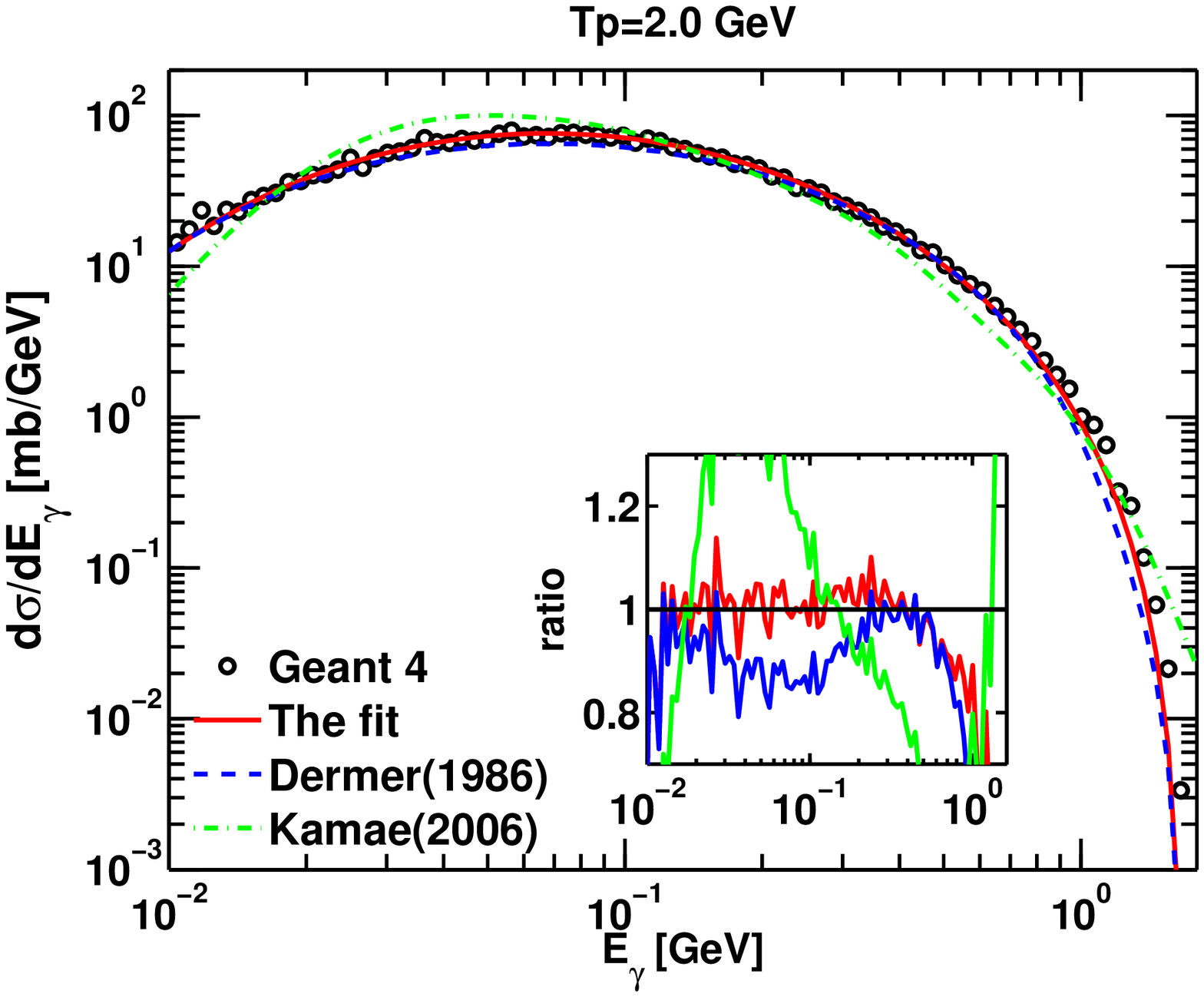}
\includegraphics[scale=0.3]{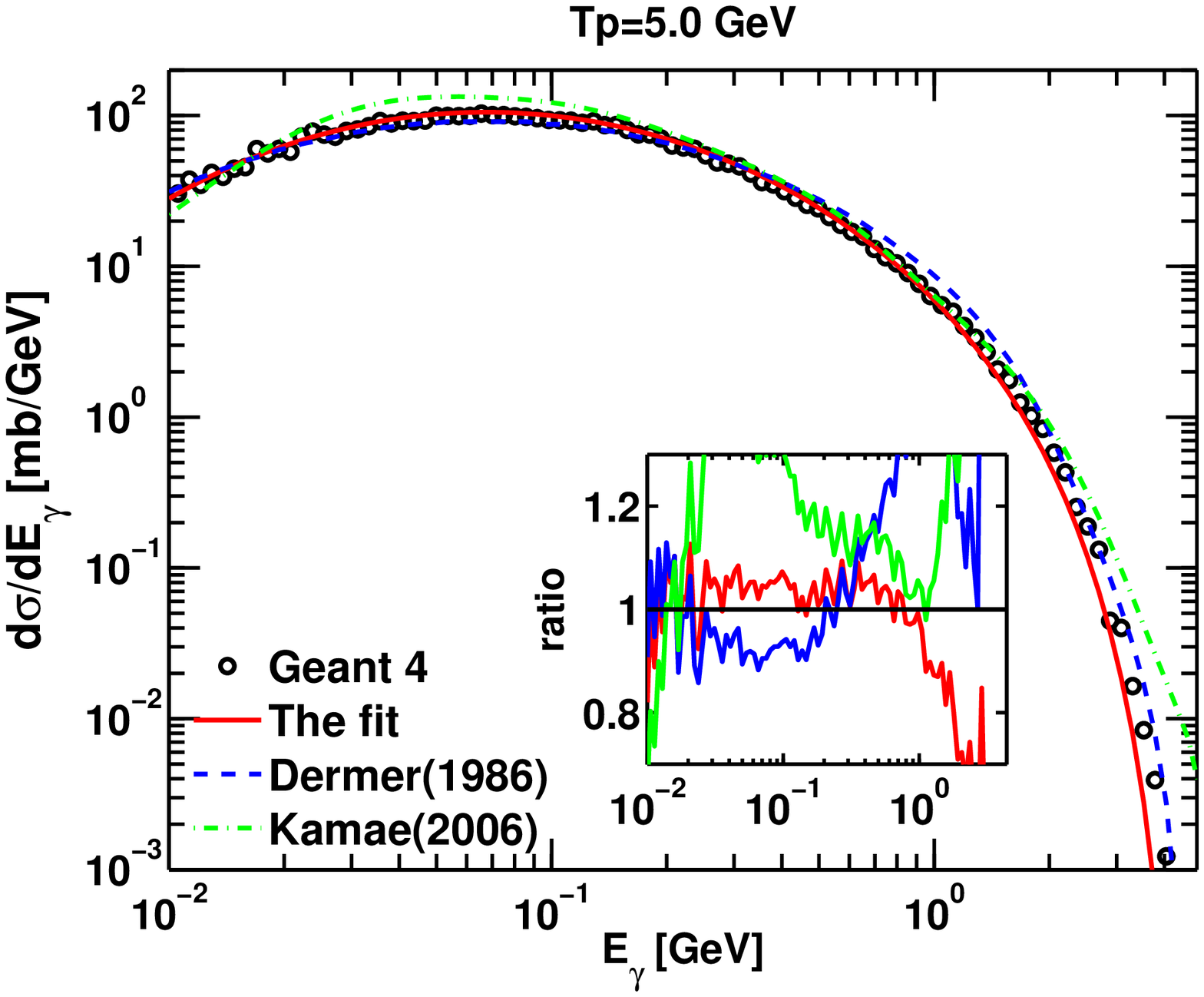}\\
\includegraphics[scale=0.3]{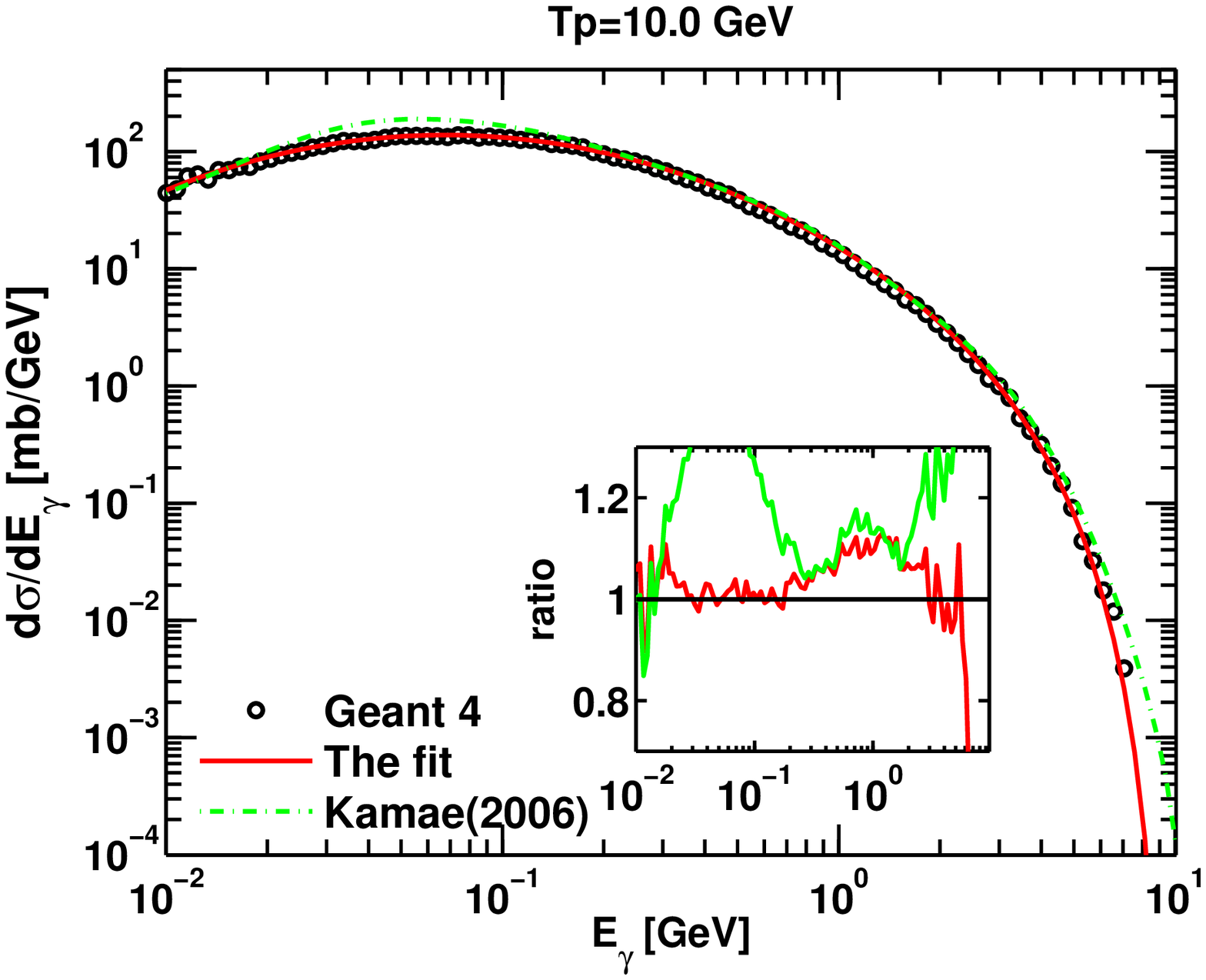}
\includegraphics[scale=0.3]{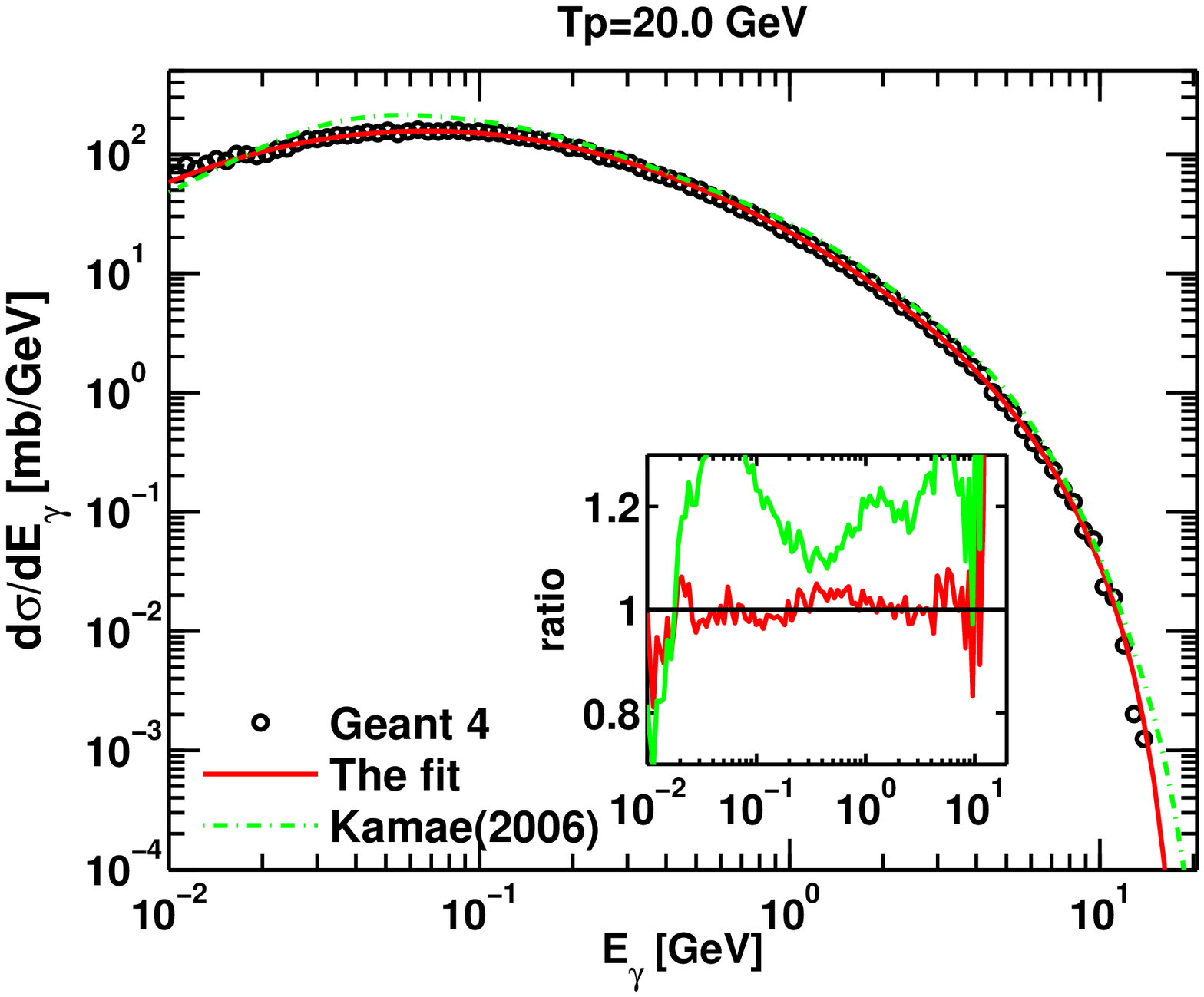}
\includegraphics[scale=0.3]{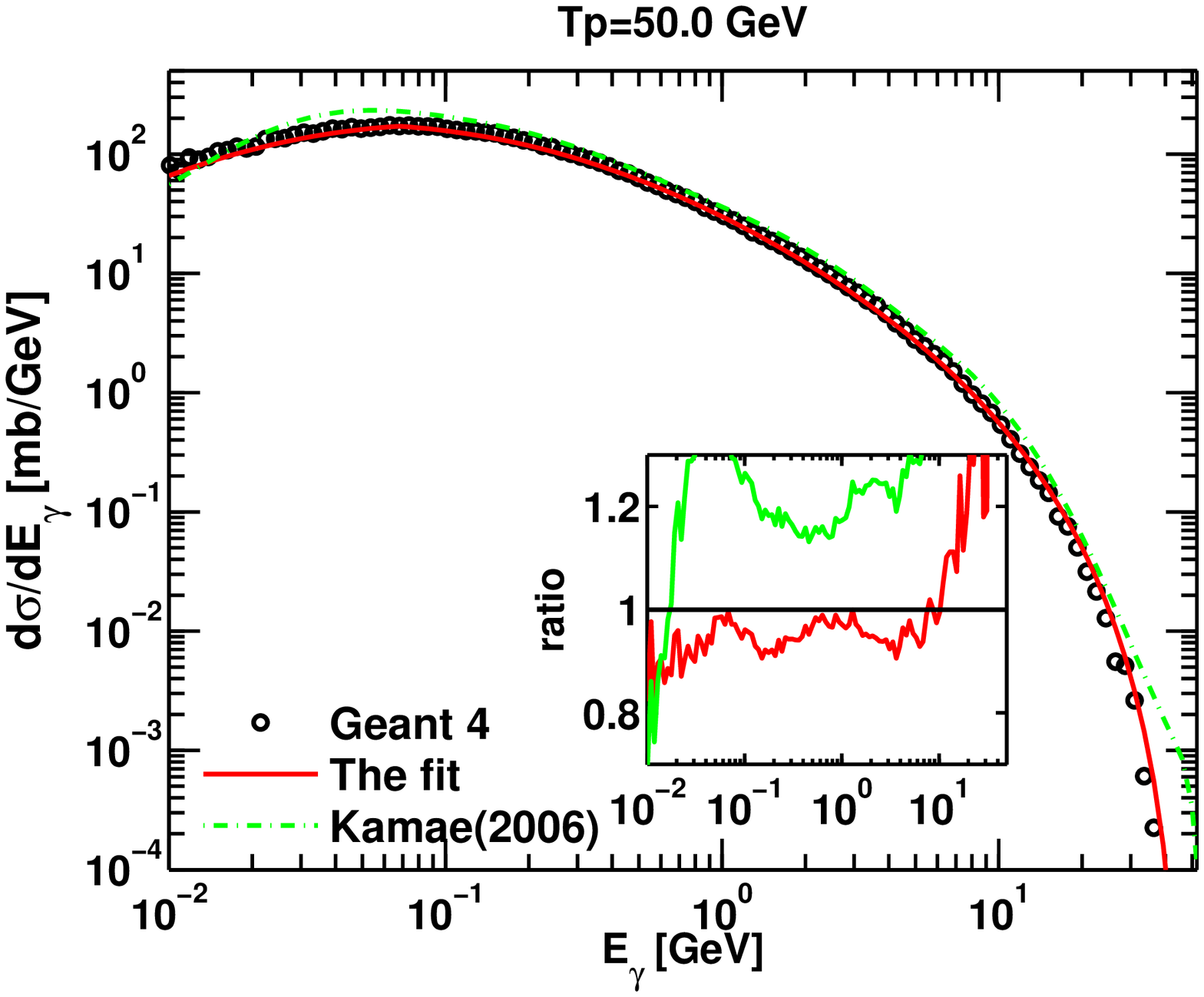}\\
\includegraphics[scale=0.3]{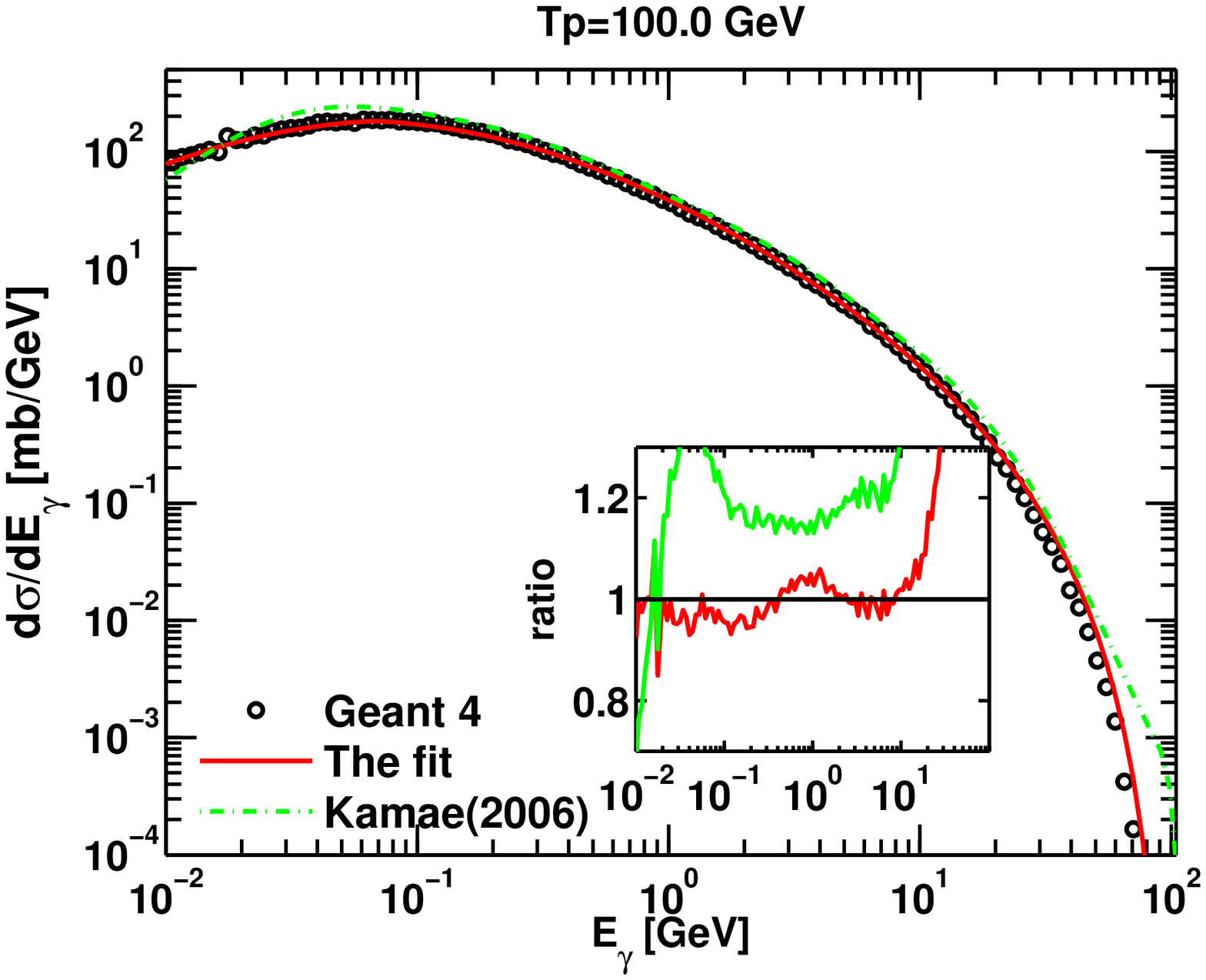}
\includegraphics[scale=0.3]{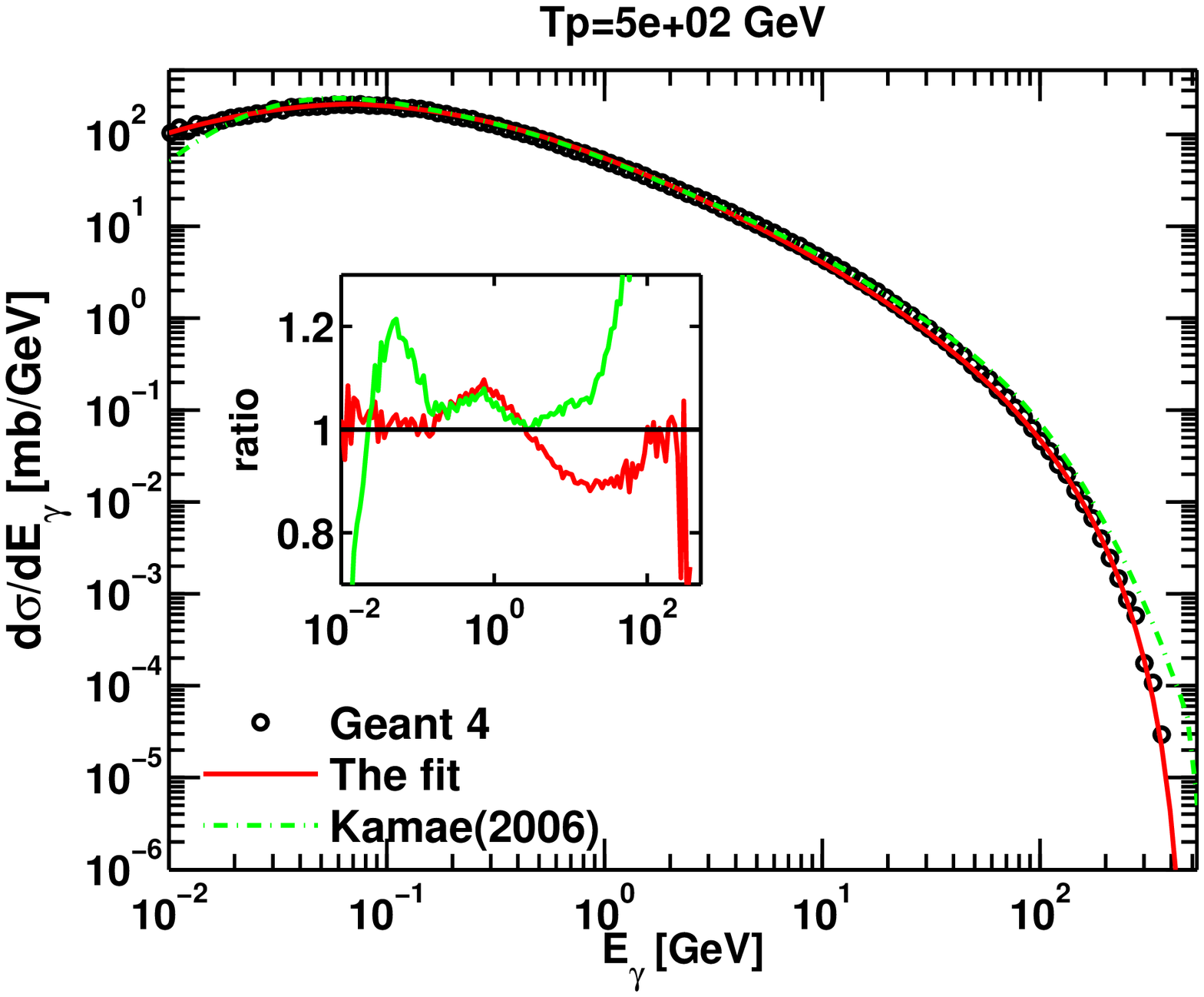}
\includegraphics[scale=0.3]{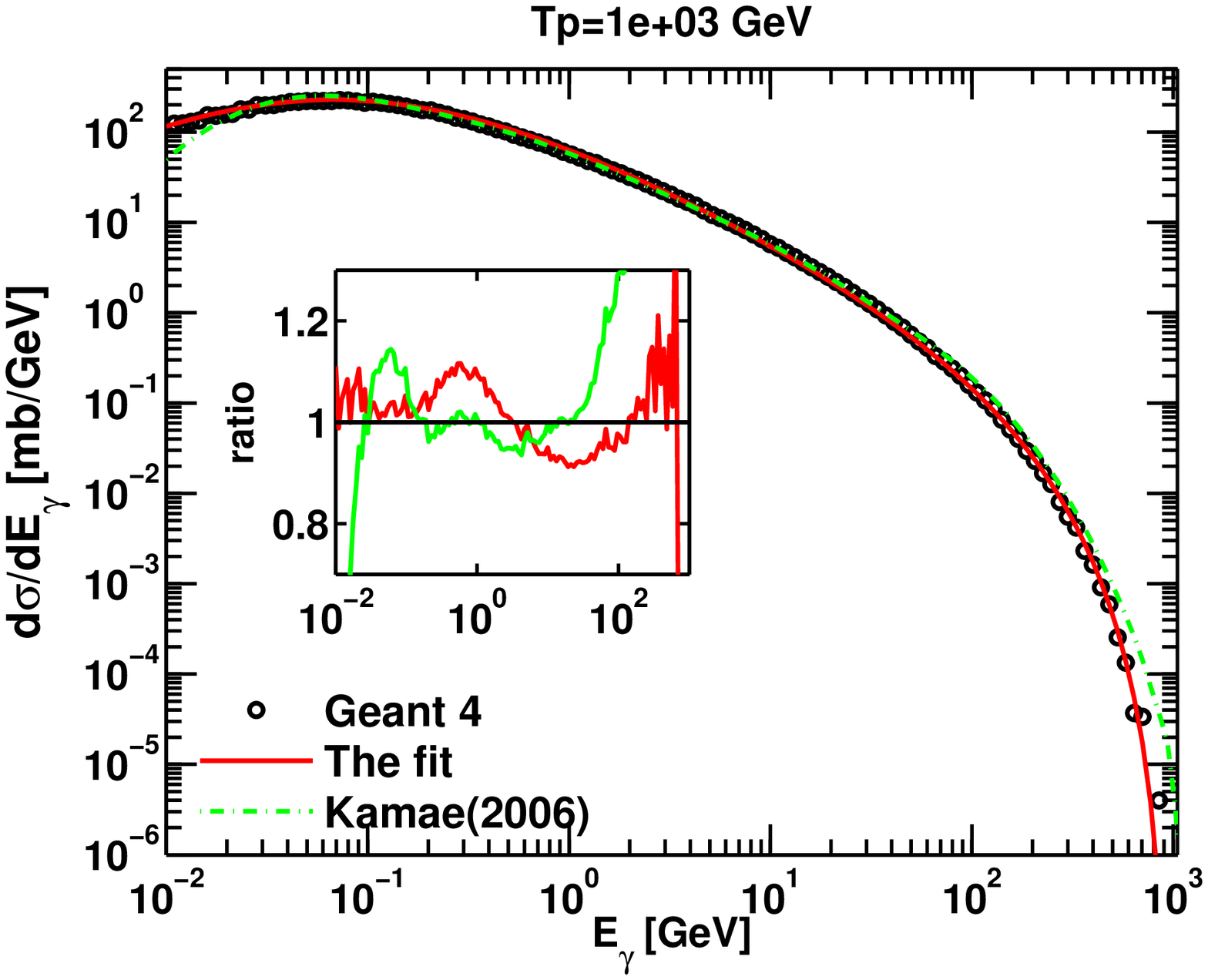}\\
\includegraphics[scale=0.3]{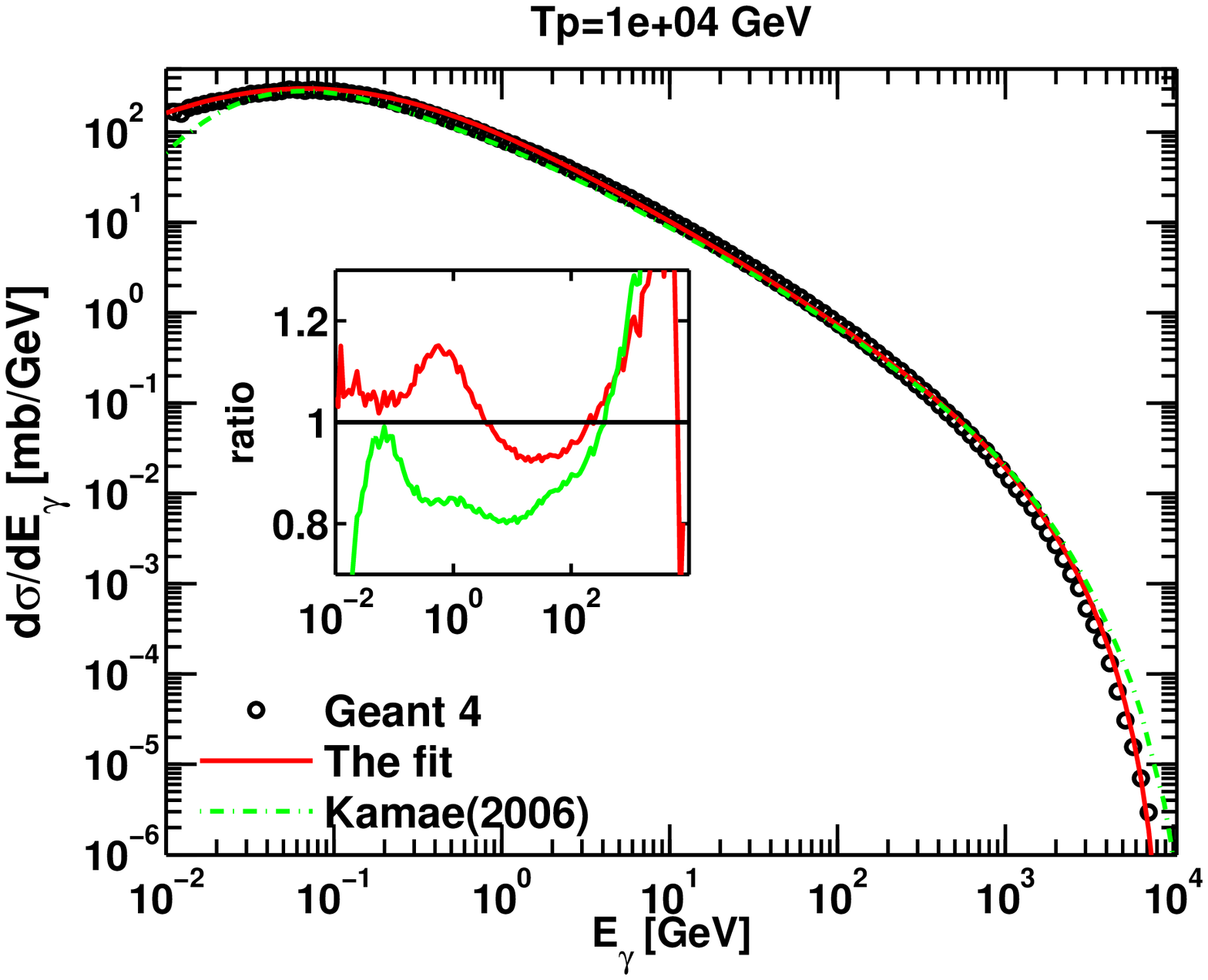}
\includegraphics[scale=0.3]{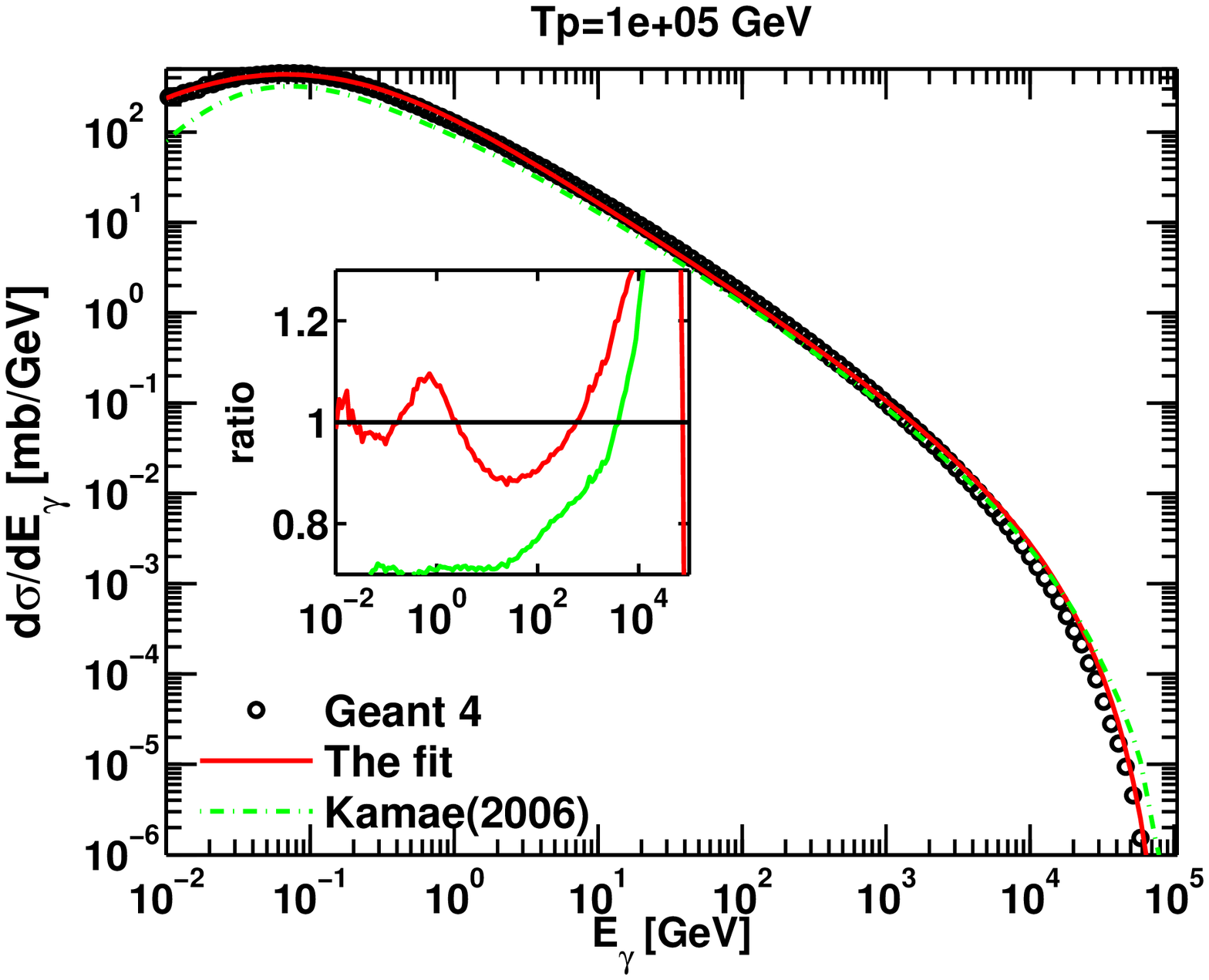}
\caption{Geant~4.10.0 $\gamma$-ray production differential cross section for some specific proton kinetic energies. The open circles are Geant~4.10.0 calculations; whereas, the full red lines are the fit formula shown in eq.~(\ref{eq:Fgamma2}) of section~\ref{sec:GamFit}, with $\alpha(T_p)$, $\beta(T_p)$ and $\gamma(T_p)$ summarized in table~\ref{tab:FgamG4P8SQ}. The dash-blue line are calculations from \cite{Dermer1986b} which are based on the isobar model. The dash-dot green line is the parametrization given by \cite{Kamae2006}. The ratio between the fit and the calculations show that the accuracy of the fit is better than 20~\%. \label{fig:G4spec} }
\end{figure*}

\begin{figure*}
\includegraphics[scale=0.3]{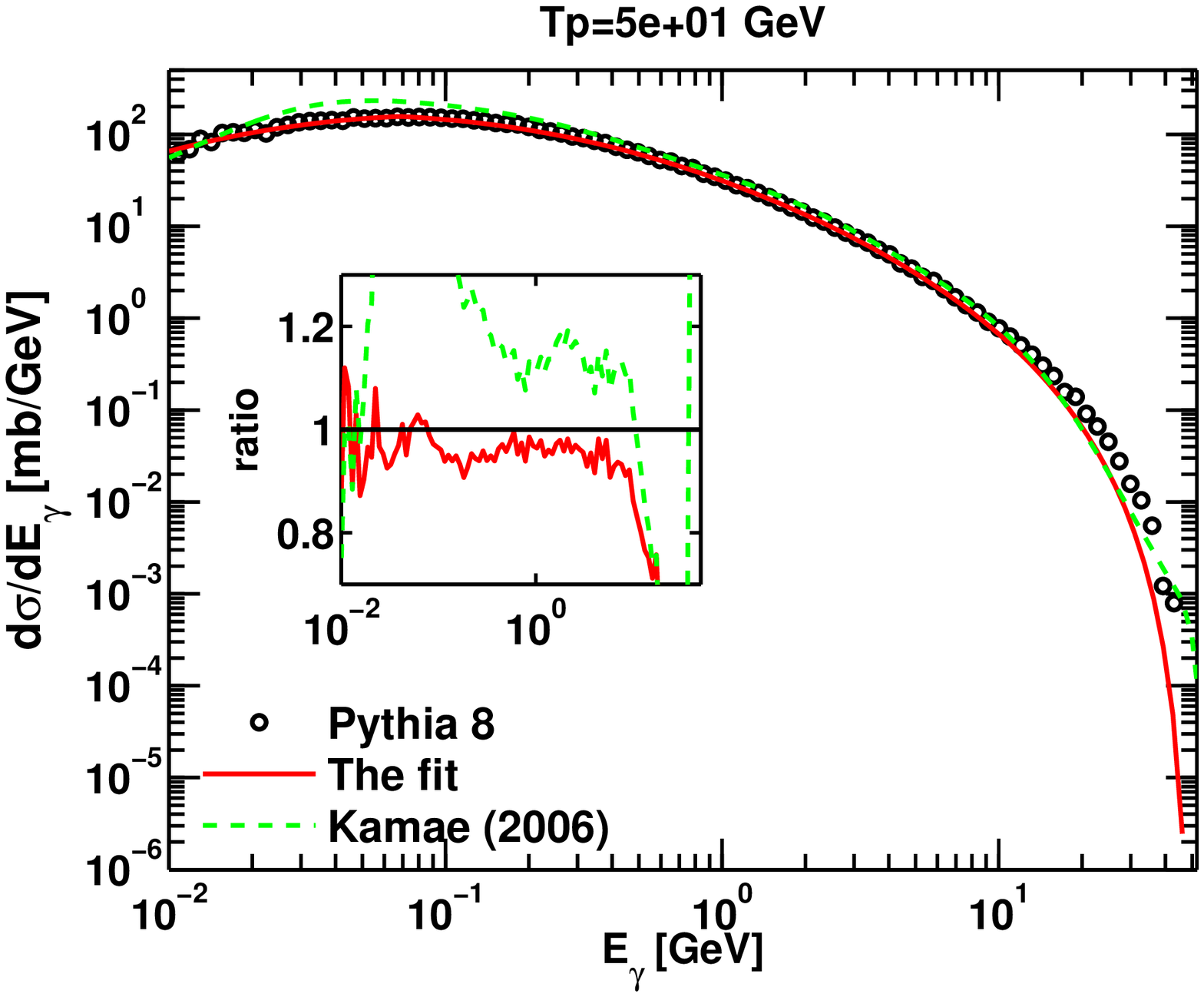}
\includegraphics[scale=0.3]{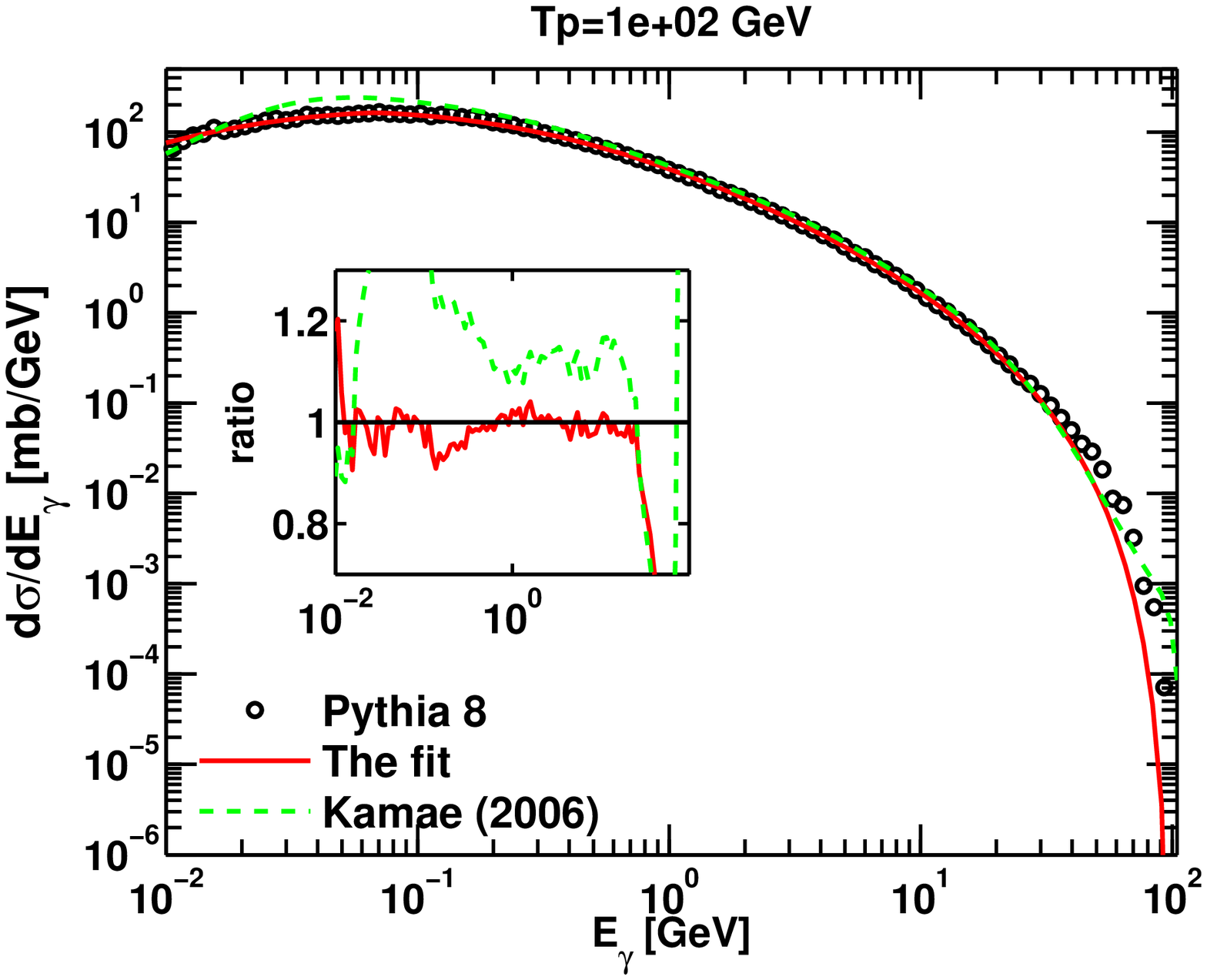}
\includegraphics[scale=0.3]{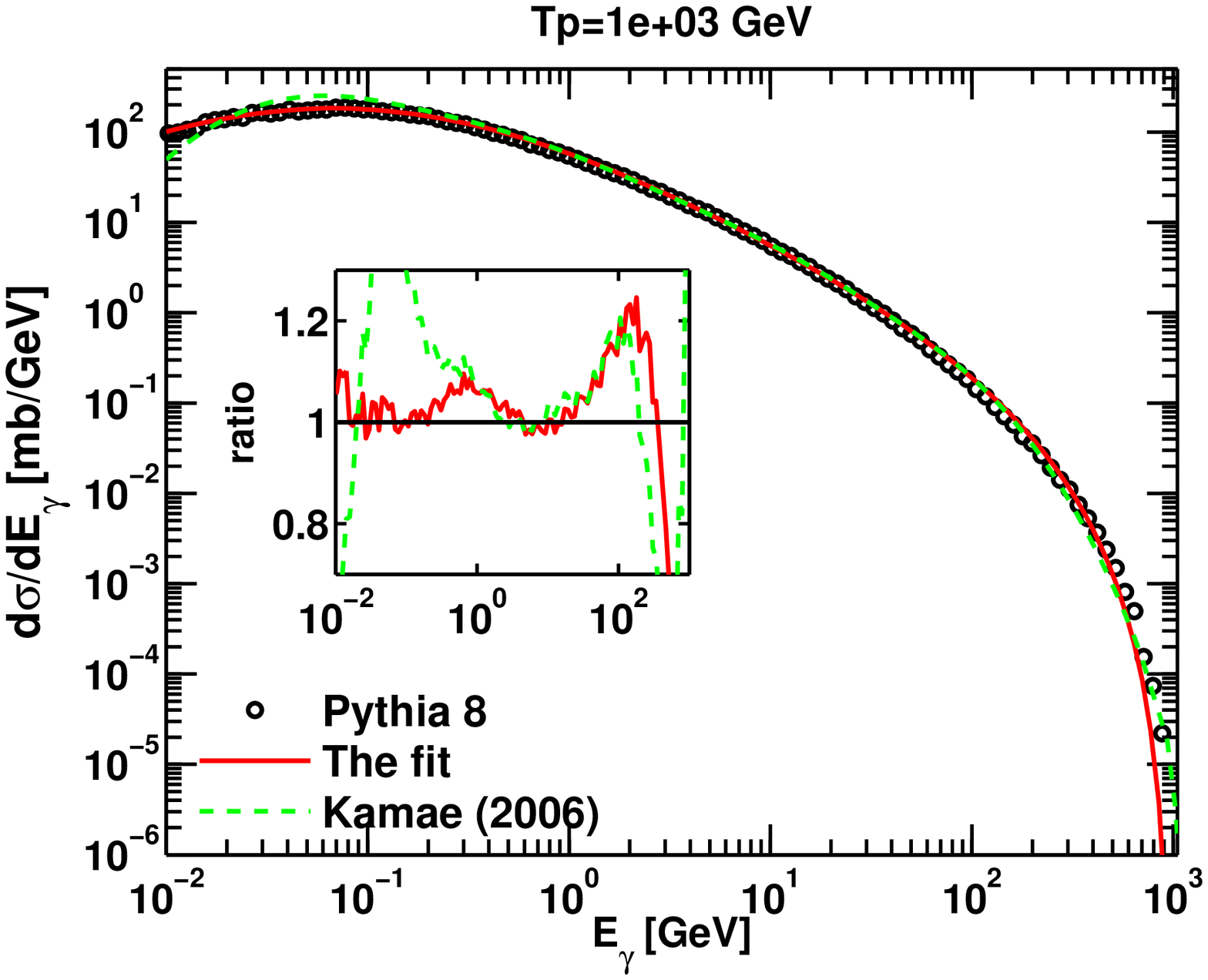}\\
\includegraphics[scale=0.3]{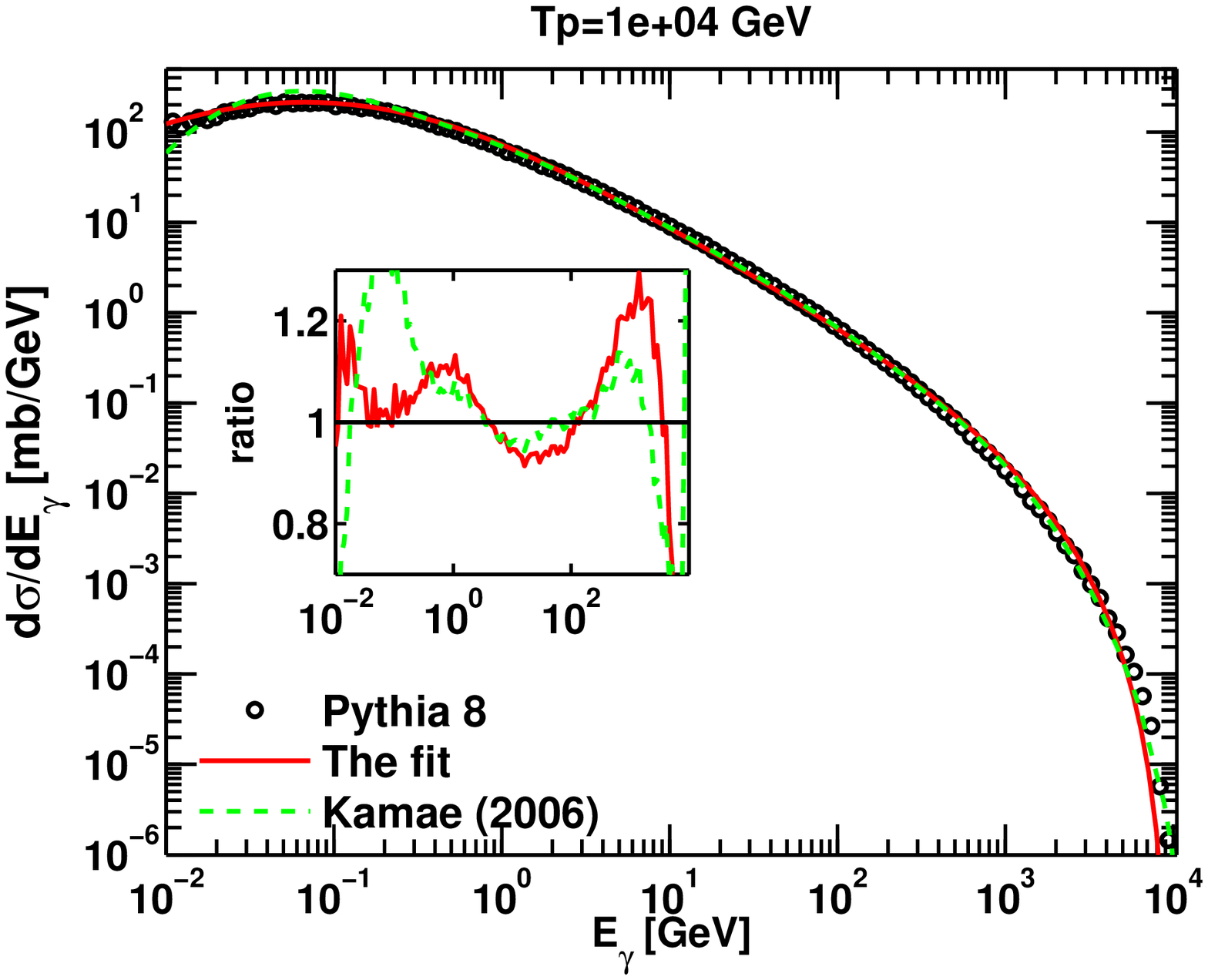}
\includegraphics[scale=0.3]{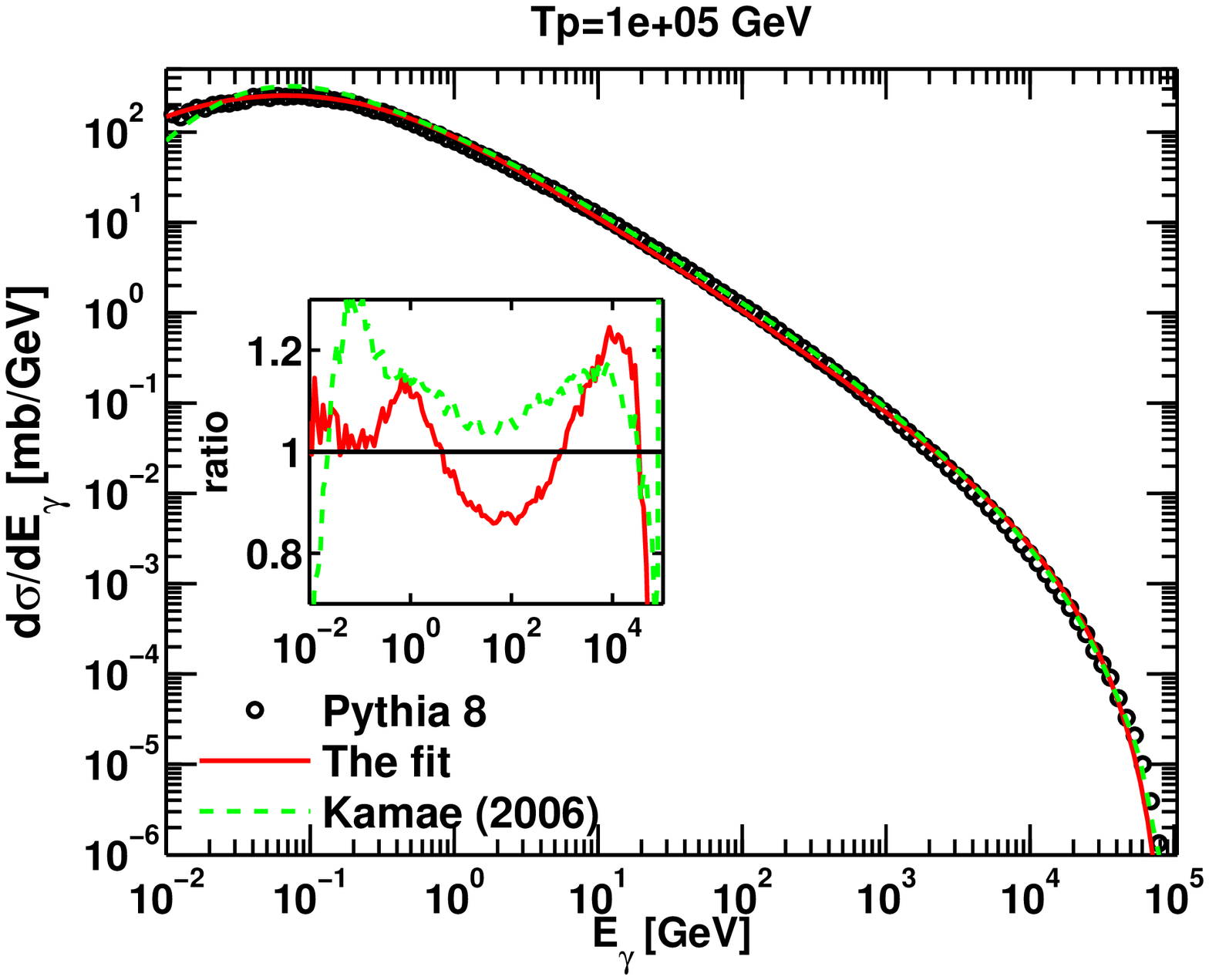}
\includegraphics[scale=0.3]{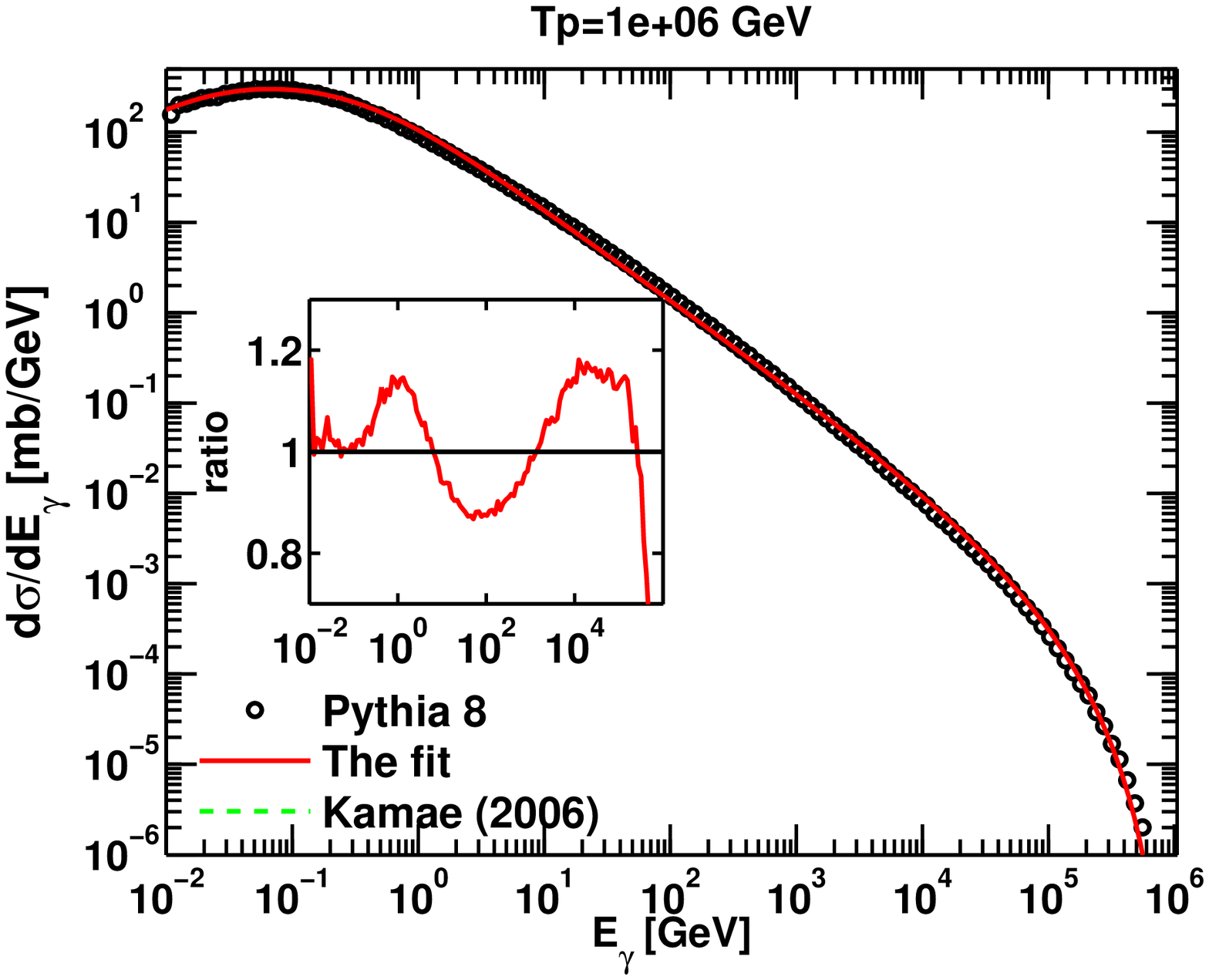}
\caption{Pythia~8.18 $\gamma$-ray production differential cross section for some specific proton kinetic energies. The open circles are Pythia~8.18 calculations, whereas, full red line is the fit formula shown in eq.~(\ref{eq:Fgamma2}) with the corresponding $\alpha(T_p)$, $\beta(T_p)$ and $\gamma(T_p)$ listed in table~\ref{tab:FgamG4P8SQ}. The dash-green line is the fit given in \cite{Kamae2006} which works for $0.448\leq T_p\leq 512$~TeV. Although not clearly shown in these plots, the fit provided by \cite{Kamae2006} violates the symmetry that $\pi^0\to2\gamma$ decay spectra has with respect to $E_\gamma=m_\pi/2$. It also violates the $E_\gamma^{\rm max}$ that is allowed by the kinematics. The ratio between the fit and the calculations shows that the accuracy of the fit is better than 20 \%. \label{fig:P8spec} }
\end{figure*}

\begin{figure*}
\includegraphics[scale=0.23]{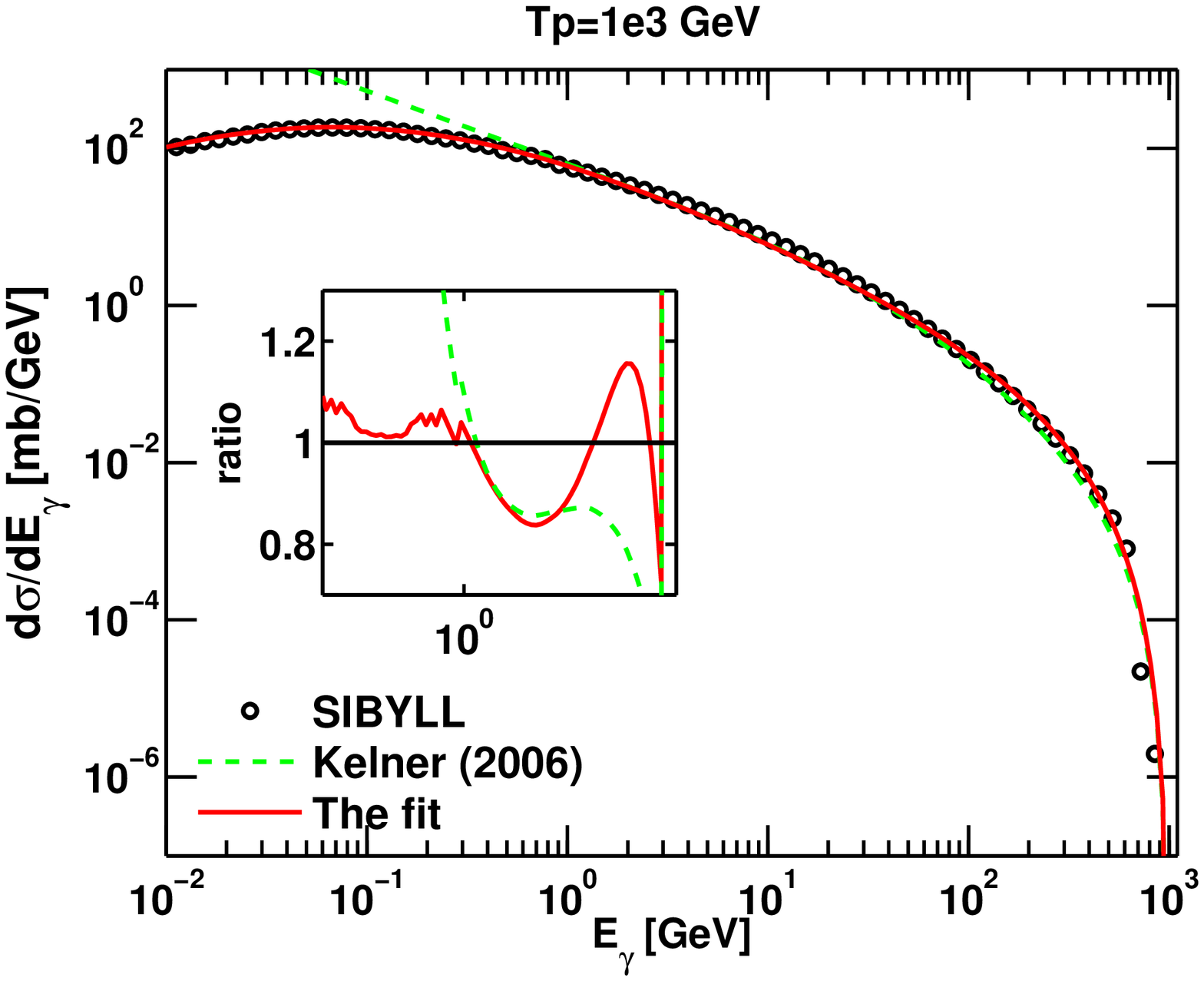}
\includegraphics[scale=0.23]{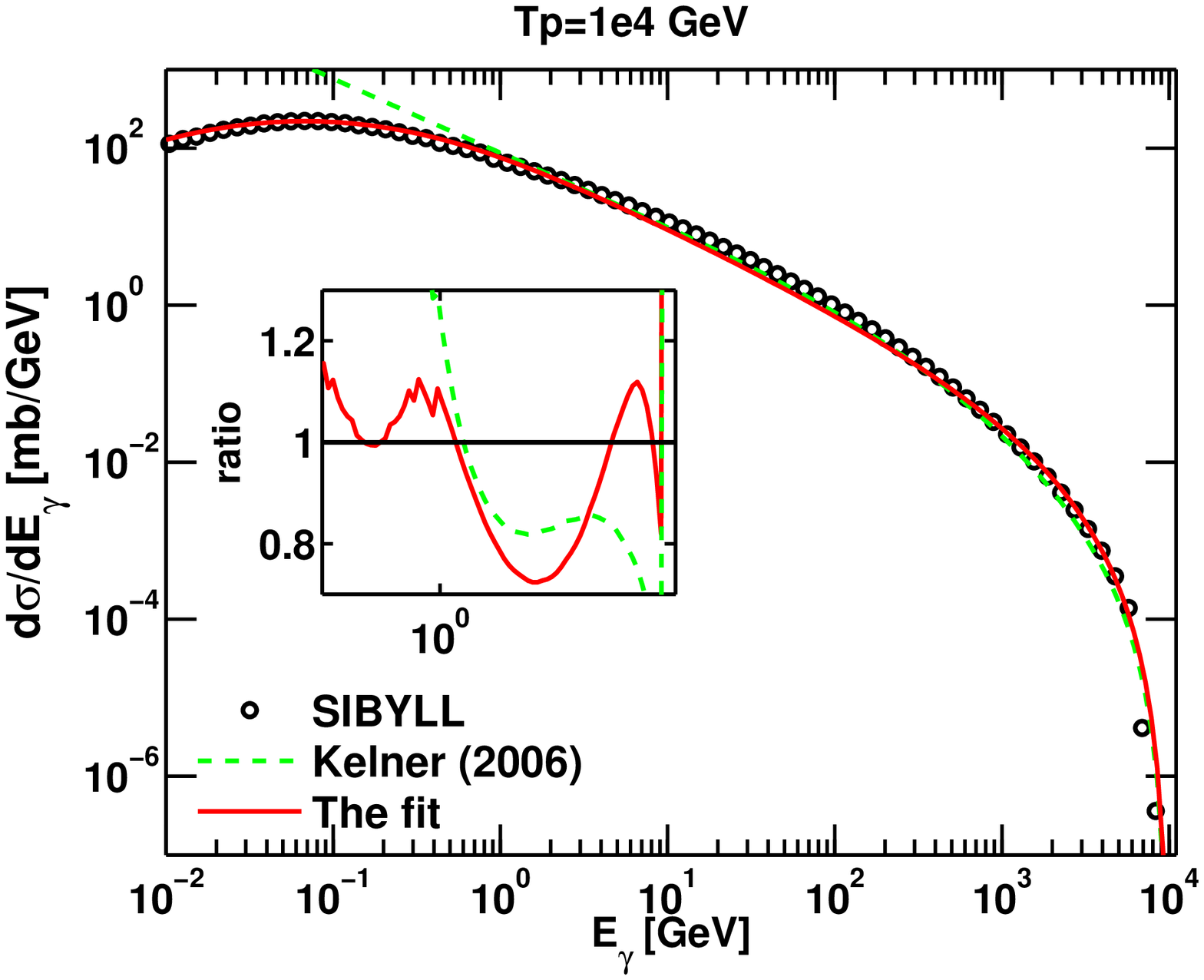}
\includegraphics[scale=0.23]{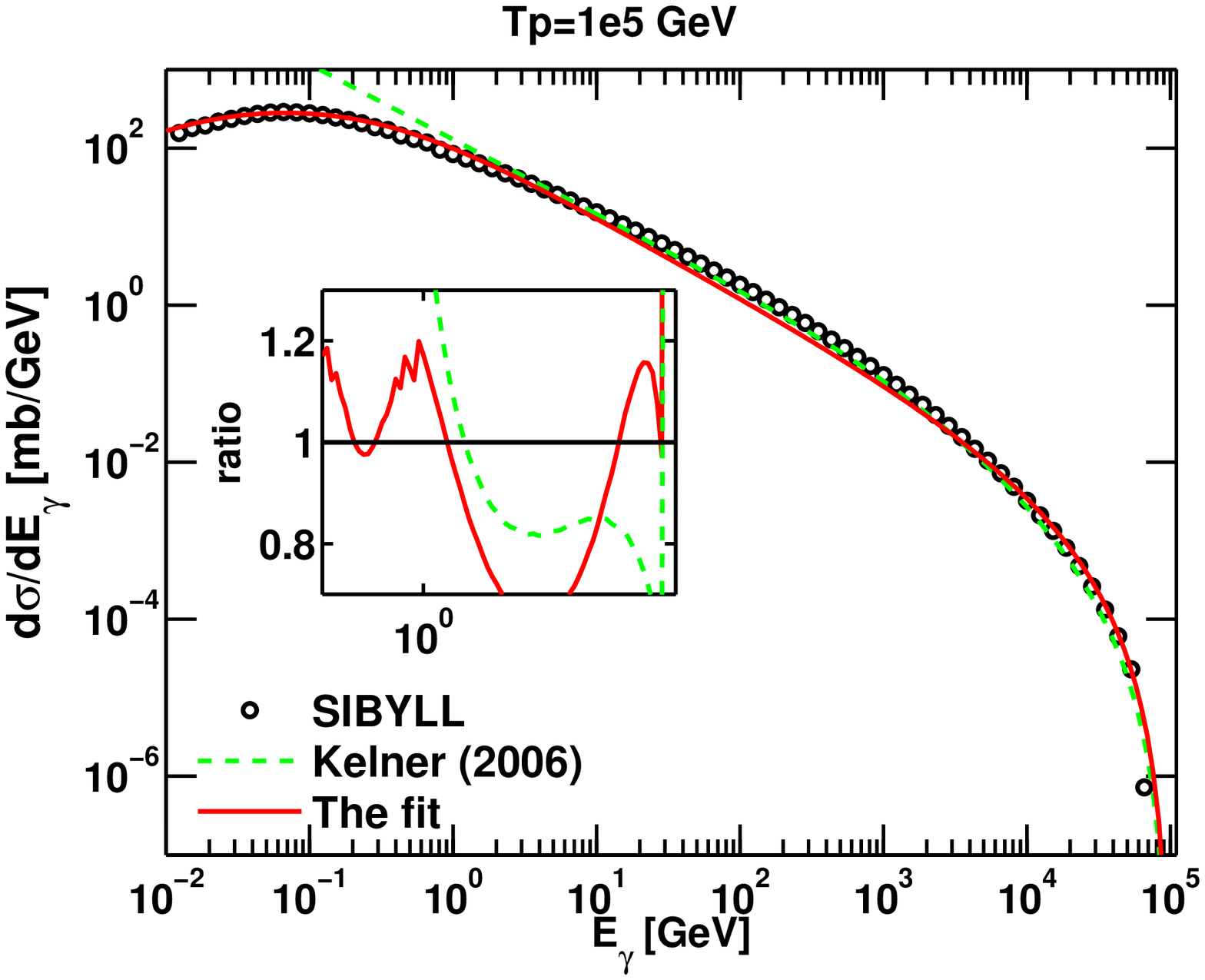}
\includegraphics[scale=0.23]{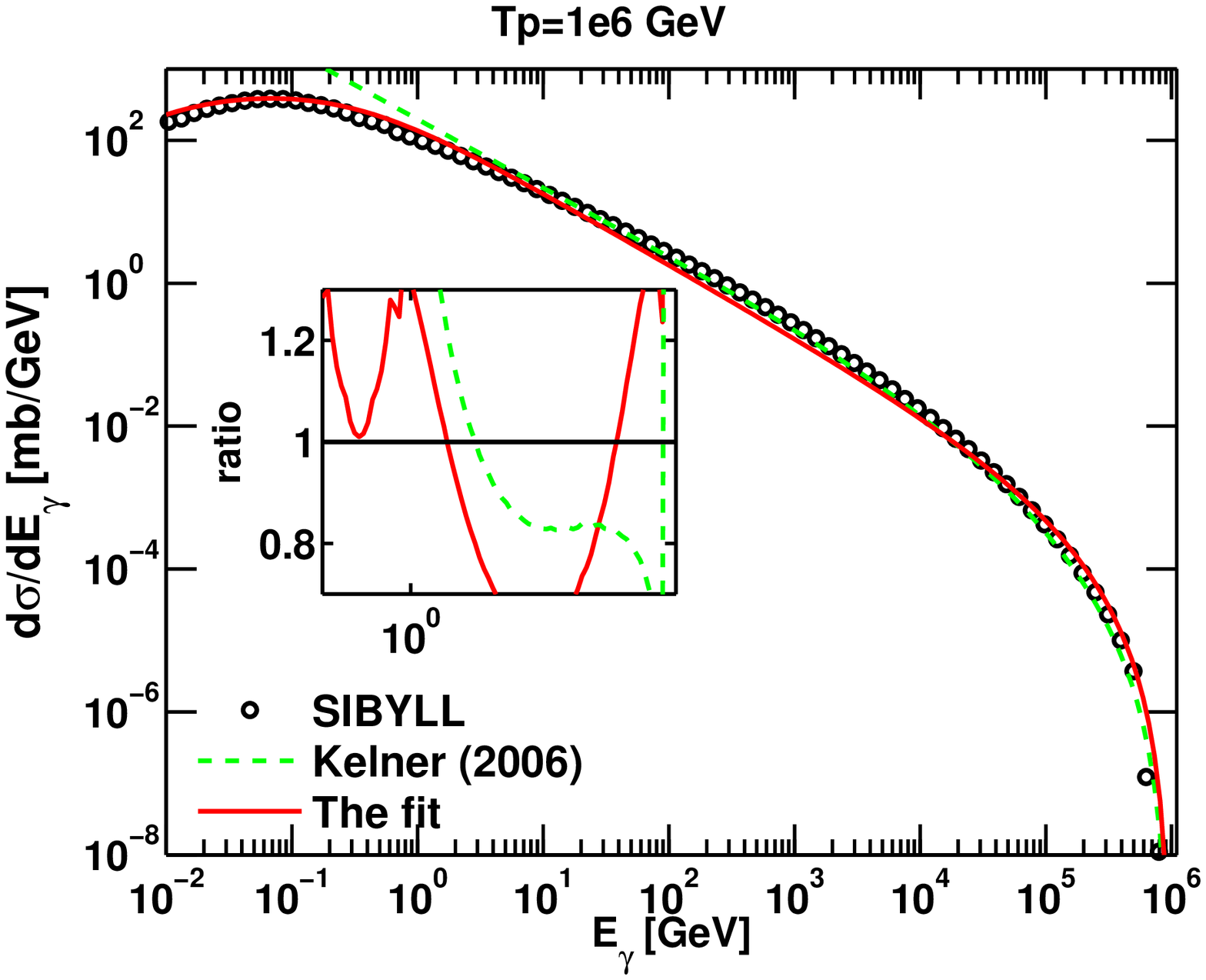}
\caption{SIBYLL~2.1 $\gamma$-ray production differential cross section for some specific proton kinetic energies. The open circles are SIBYLL~2.1 calculations, the full red line is the fit formula shown in eq.~(\ref{eq:Fgamma2}) with the corresponding $\alpha(T_p)$, $\beta(T_p)$ and $\gamma(T_p)$ listed in table~\ref{tab:FgamG4P8SQ} and the dash green line is the fit given in \cite{Kelner2006}. The ratio between the fit and the calculations shows that the accuracy of the fit is of the order 20 \%.  \label{fig:Sspec} }
\end{figure*}

\begin{figure*}
\includegraphics[scale=0.23]{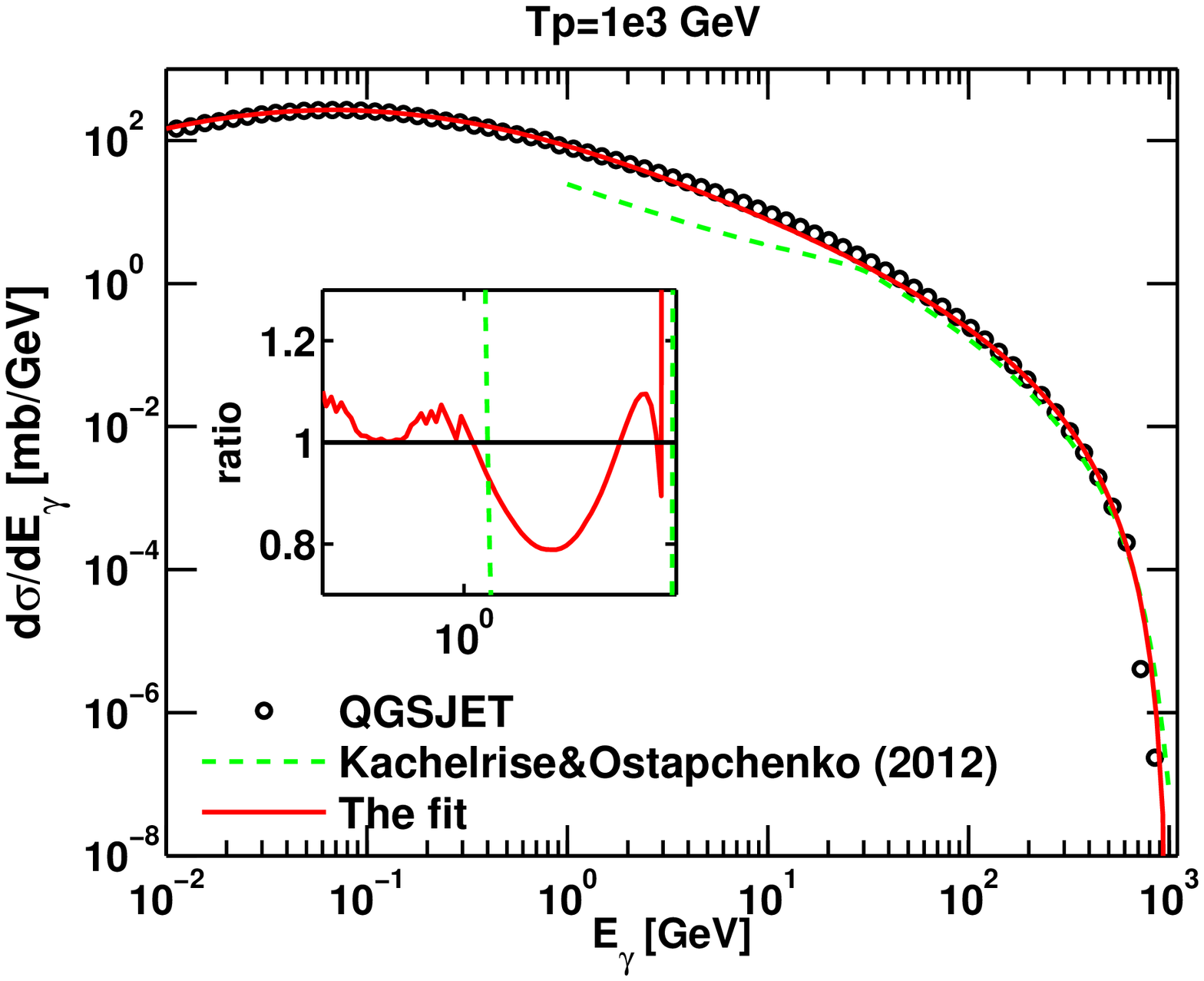}
\includegraphics[scale=0.23]{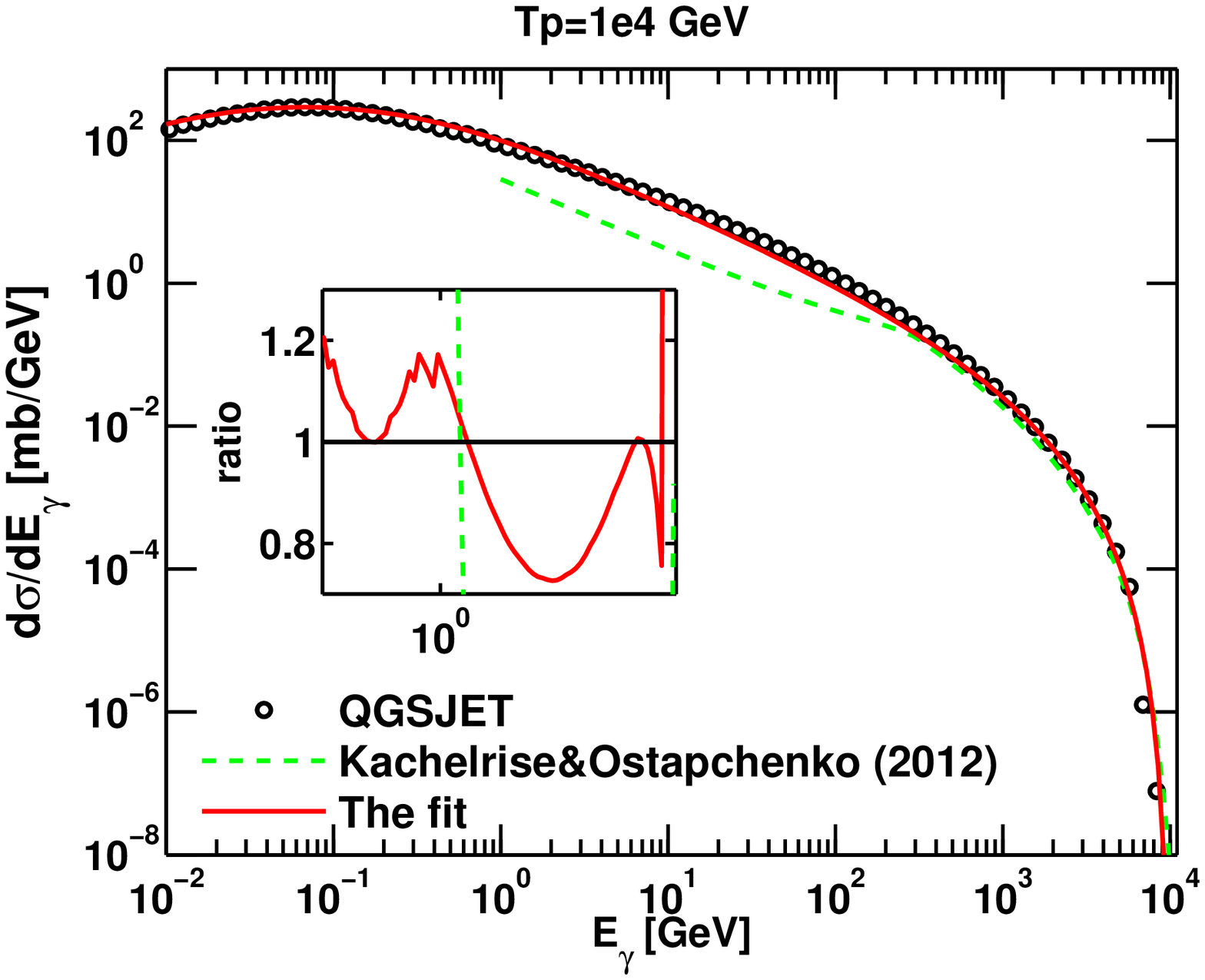}
\includegraphics[scale=0.23]{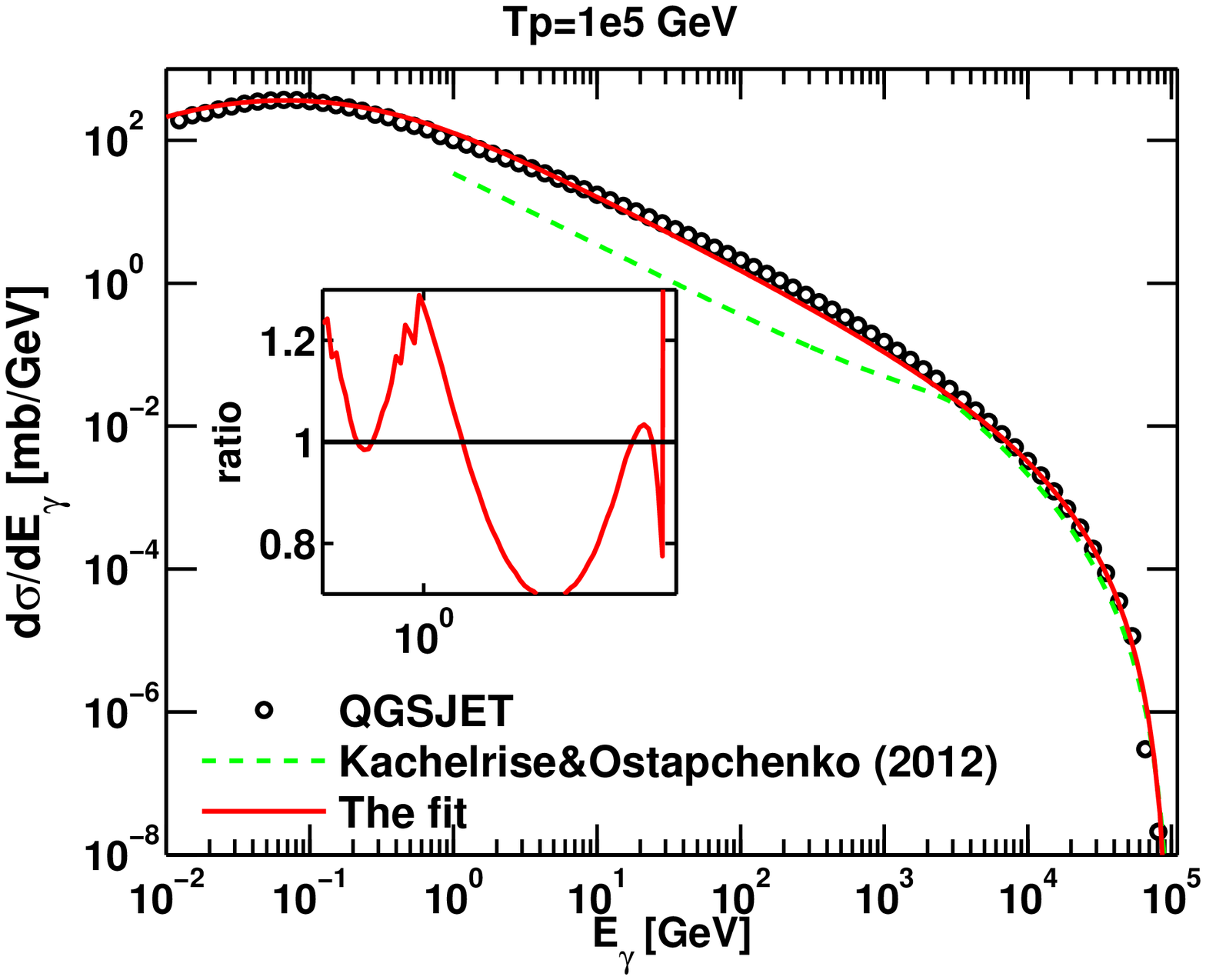}
\includegraphics[scale=0.23]{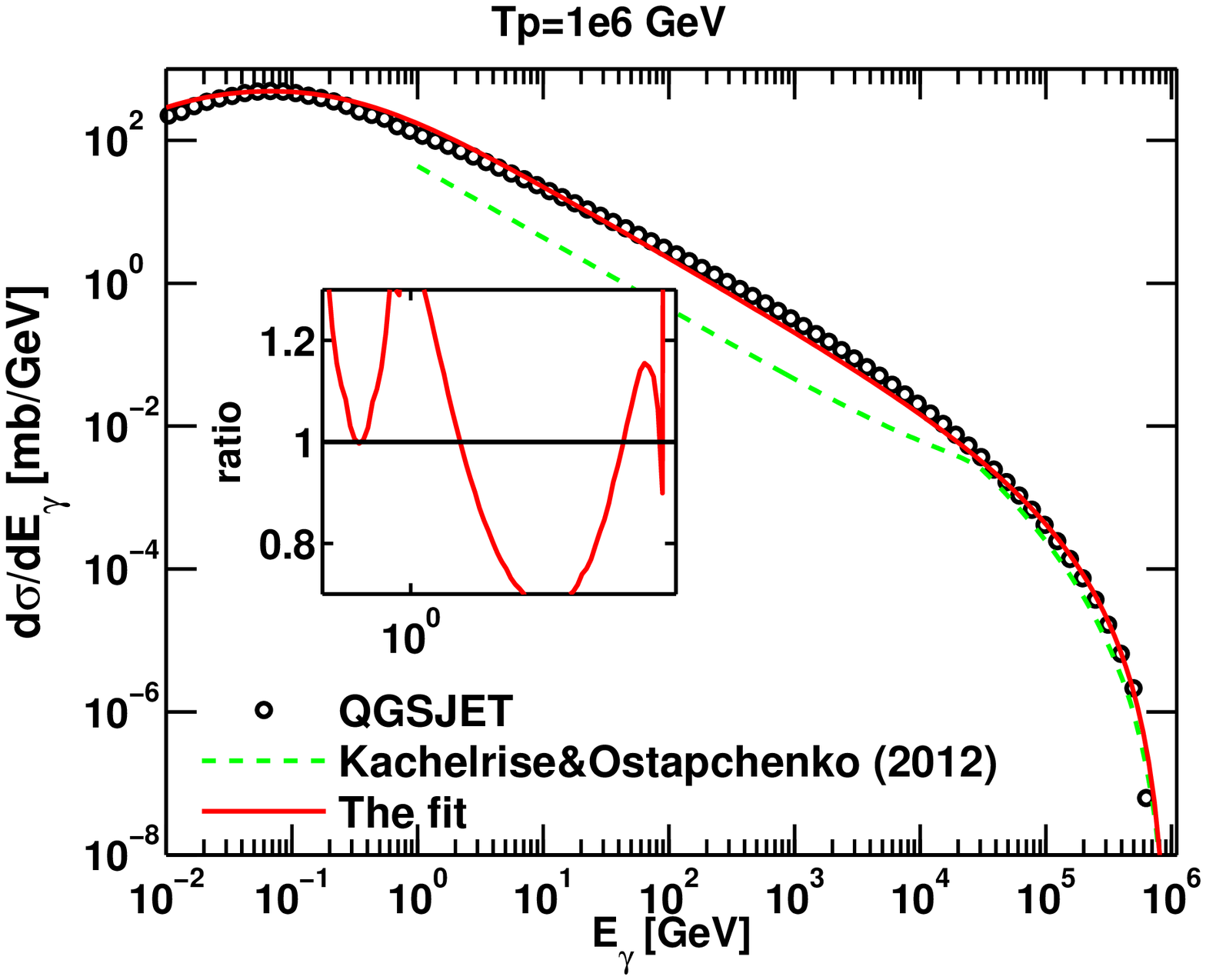}
\caption{QGSJET-I $\gamma$-ray production differential cross section for some specific proton kinetic energies. The open circles are QGSJET-I calculations, the full red line is the fit formula shown in eq.~(\ref{eq:Fgamma2}) with the corresponding $\alpha(T_p)$, $\beta(T_p)$ and $\gamma(T_p)$ listed in table~\ref{tab:FgamG4P8SQ} and the dash green line is the QGSJET-II fit calculated using the ``\textit{ppfrag}'' code given in \cite{Kachelries2012}. The ratio between the fit and the calculations shows that the accuracy of the fit is of the order 20 \%. \label{fig:Qspec} }
\end{figure*}

%Figures \ref{fig:G4spec,fig:P8spec,fig:Sspec,fig:Qspec}
Figures~\ref{fig:G4spec} to \ref{fig:Qspec} show the Geant~4.10.0, Pythia~8.18, SIBYLL~2.1 and QGSJET-I $\gamma$-ray production differential cross sections compared with their respective fit formulas given in eq.~(\ref{eq:Fgamma2}) and exponents summarized in table~\ref{tab:FgamG4P8SQ}. The accuracy of the fits is better than 20~\% for Geant~4.10.0 and Pythia~8.18. For SIBYLL~2.1 and QGSJET-I, the accuracy improves as the proton energy is reduced.

% ====================================================================
% ====================================================================
\section{The effect of Nucleus--Nucleus interaction \label{sec:NucEnhanc}}
Elements heavier than Hydrogen, are found in many astrophysical environments. As a result, it is necessary to have a formalism to describe the $\gamma$-ray production in nucleus--nucleus interactions. Generally we expect that the nucleus--nucleus contribution be small compared to the p--p one, because, the abundances of nuclei heavier than Hydrogen are small. Hence, a simplified description of the nucleus--nucleus interaction is justified.

In the context of Glauber's multiple scattering theory \cite{Glauber1955, Franco1966, Glauber1970}, high energy nucleus--nucleus collisions can be treated as a sequence of binary nucleon-nucleon scatterings. This simplifies the complicated nucleus--nucleus scattering into individual nucleon--nucleon ones and neglects the effects that come with the nuclear medium and nucleons motion inside the nucleus. Not all the nucleons, however, participate in the collisions because nucleons inside the nucleus ``eclipse'' each other. Therefore, the inelastic cross section is smaller than the sum of all binary nucleon--nucleon cross sections. In this picture, the $\gamma$-ray spectrum produced by nucleus--nucleus collisions will be essentially the same with that of p--p, but with an $A_{\rm max}$ in eq.~(\ref{eq:Amax}), scaled by the nucleus--nucleus inclusive $\pi^0$ production cross section. 

To calculate the nucleus--nucleus inclusive $\pi^0$ production cross section, we need the nucleus--nucleus inelastic cross sections and the $\pi^0$ average multiplicity. A simple and accurate model for calculating the average $\pi^0$ multiplicity, is the so called ``wounded nucleons'' model \cite{Bialas1976}. This model assumes that the average meson production multiplicity is proportional to the number of nucleons that underwent at least one inelastic collision during a nucleus--nucleus interaction. According to this model if A and B are two nuclei with mass numbers A and B, respectively, the average $\pi^0$ multiplicity is $\left\langle n_{AB\pi^0} \right\rangle = \frac{1}{2}w_{AB}\times \left\langle n_{\pi^0} \right\rangle$, where $w_{AB}$ is the number of wounded nucleons and $\left\langle n_{\pi^0} \right\rangle$ is the average $\pi^0$ multiplicity for p--p collisions. The factor $w_{AB}$ is given by:
\begin{equation}
w_{AB}=\frac{A\,\sigma_{pB} + B\,\sigma_{pA}}{\sigma_{AB}}
\end{equation}
\noindent where, $\sigma_{pA}$ and $\sigma_{pB}$ are the nucleon(proton)--nucleus A or B inelastic cross sections, and $\sigma_{AB}$ is the A+B inelastic cross section.

For the nucleus--nucleus inelastic cross sections, there exist different phenomenological models and fits of the experimental data, see e.g. \cite{Bradt1950, Karol1975, Letaw1983, Kox1987,Shen1989, Webber1990, Sihver1993, Tripathi1996, Tripathi1997, Tripathi1999}. This cross section shows an energy dependence below $T_p<0.2$~GeV/nucleon, above which it approaches the geometrical cross section and thus, thereafter, it remains constant. At very high energies though, it is expected that the nucleus--nucleus cross section increase. Soft-sphere models, for instance, predict that the nucleus--nucleus cross section will increase with energy as the logarithm of the p--p inelastic cross section, see e.g. \cite{Karol1975}.

Based on these considerations we use the parametrization of ref.~\cite{Sihver1993} for describing the nucleus--nucleus cross section. This parametrization is energy independent. For very high energies ($T_p>10^3$~GeV/nucleon) we modify this parametrization with a logarithmic term of the p--p inelastic cross section. For completeness we state here explicitly the formulas adopted for nucleus--nucleus reaction cross sections, provided in \cite{Sihver1993}:
\begin{equation}\label{eq:sigR}
\sigma_R = \sigma_{R0}\left[ A_p^{1/3}+A_t^{1/3} - \beta_0\left(  A_p^{-1/3} + A_t^{-1/3} \right)  \right]^2 ,
\end{equation} 

\noindent where, $\sigma_{R0}=\pi r_0^2\approx 58.1$~mb with $r_0=1.36$~fm. $A_p$ and $A_t$ are the projectile and the target mass numbers, respectively. The coefficient $\beta_0=2.247-0.915\,(1+A_t^{-1/3})$ if the projectile is a proton, and $\beta_0=1.581-0.876\,(A_p^{-1/3}+A_t^{-1/3})$ for a projectile different from a proton. This formula is valid for projectile energies $T_p\geq 0.2$~GeV and $T_p\geq 0.1$~GeV/nucleon, for proton and nuclei projectile species, respectively.

To account for the energy dependence of the cross section at very high energies, we slightly modify the formula 
\begin{equation}\label{eq:NucXSinel}
\sigma_{\rm inel}(A_p,A_t,T_p)=\sigma_R(A_p,A_t)\times \mathcal{G}(T_p)
\end{equation}

\noindent where, $T_p$ is the projectile kinetic energy per nucleon. For the function $\mathcal{G}(T_p)$ we assume a simple form
\begin{equation}\label{eq:GTp}
\mathcal{G}(T_p)=1 + \log\left( \max \left[ 1,\frac{\sigma_{\rm inel}(T_p)}{\sigma_{\rm inel}(T_p^0)}\right] \right) .
\end{equation}

\noindent $\sigma_{\rm inel}(T_p)$ is the p--p inelastic cross section given in eq.~(\ref{eq:XSinel}), and we choose $T_p^0=10^3$~GeV as the energy at which the nucleus--nucleus cross section growth becomes noticeable. We note that the function $\mathcal{G}(T_p)$ can increase the nucleus--nucleus cross section by about 50 \% at energies $T_p\sim 10^6$~GeV/nucleon.

If we suppose that we have $P$ projectiles and $T$ targets, the total $\gamma$-ray differential cross section will be the sum from each component. Therefore, if we scale their contributions with respect to the p--p component, we form the so called, nuclear enhancement factor $\epsilon$. Let us denote with $Y_p^i=(n_i/n_p)_{proj}$ the ratio by number of a given projectile $i$ with respect to the proton component, and let the $Y_t^j=(n_j/n_H)_{targ}$ be the ratio by number of a target $j$ with respect to Hydrogen in that medium. The ratio between nucleus--nucleus and p--p inclusive $\pi^0$ production cross sections is $\sigma_\pi^{ij}/\sigma_\pi =\frac{1}{2}\,w_{ij}\sigma_{\rm inel}(A_i,A_j,T_p)/\sigma_{\rm inel}(T_p)=\frac{1}{2}(A_i\sigma_{\rm inel}(p,A_j,T_p)+A_j\sigma_{\rm inel}(p,A_i,T_p)/\sigma_{\rm inel}(T_p)$, where $A_i$ and $A_j$ are the projectile and target mass numbers. $\sigma_{\rm inel}(T_p)$ is p--p inelastic cross section, whereas, $\sigma_{\rm inel}(p,A_i,T_p)$ and $\sigma_{\rm inel}(p,A_j,T_p)$ are proton--target and proton--projectile inelastic cross sections, respectively given by eq.~(\ref{eq:NucXSinel}). The nuclear enhancement factor is given by
\begin{equation}\label{eq:eps}
\epsilon = \sum\limits_{i,j}^{P,T} Y_p^i\, Y_t^j \left(\frac{A_i\sigma(p,A_j,T_p)+A_j\sigma(p,A_i,T_p)}{2\sigma_{\rm inel} }\right),
\end{equation}

\noindent where for simplicity we write $\sigma_{\rm inel}(p,A_i,T_p)\equiv\sigma(p,A_i,T_p)$ and $\sigma_{\rm inel}(p,A_j,T_p)\equiv \sigma(p,A_j,T_p)$. When nuclei interact with Hydrogen, the p--p inelastic cross section will appear which in principle has a different energy dependence than the nucleus--nucleus cross section. Thus we separate the contribution that comes when at least one of the interacting particles is a proton (or Hydrogen). Let us denote with
\begin{equation}\label{eq:epsC}
\epsilon_c = 1 + \frac{1}{2} \sum\limits_{i> p}^P A_i\,Y_p^i + \frac{1}{2} \sum\limits_{j> H}^T A_j\,Y_t^j~,
\end{equation}

\begin{equation}\label{eq:eps1}
\epsilon_1 = \frac{1}{2}\sum\limits_{i> p}^P \,Y_p^i\,\frac{\sigma_R(p,A_i)}{\sigma_R^{pp}} + \frac{1}{2} \sum\limits_{j> H}^T Y_t^j\,\frac{\sigma_R(p,A_j)}{\sigma_R^{pp}}~,
\end{equation}

\begin{equation}\label{eq:eps2}
\epsilon_2 = \sum\limits_{i,j>p,H}^{P,T} Y_p^i\, Y_t^j \left(\frac{A_i\sigma_R(p,A_j)+A_j\sigma_R(p,A_i)}{2\sigma_R^{pp}}\right).
\end{equation}

\noindent Where the $i>p$ and $j>H$ means that the index runs over all other elements except protons and Hydrogen as projectile and targets, respectively. We denoted with $\sigma_R^{pp}=\pi r_p^2=10\pi\approx 31.4$~mb the proton geometrical cross section, by assuming that $r_p=1$~fm. $\sigma_R(p,A)$ is the inelastic proton--nucleus cross sections given in eq.~(\ref{eq:sigR}). The nuclear enhancement factor will be expressed as:
\begin{equation}
\epsilon(T_p) = \epsilon_c + (\epsilon_1 +\epsilon_2)\times \frac{\sigma_R^{pp}\times \mathcal{G}(T_p)}{\sigma_{\rm inel}(T_p)}.
\end{equation}

The last expression of the nuclear enhancement factor is very practical. $\epsilon_c$ does not depend on energy and is not model dependent, i.e. the nucleus--nucleus cross section model. $\epsilon_1$ and $\epsilon_2$ are not functions of energy but are model dependent. The only energy dependent term in the enhancement factor is due to $\mathcal{G}(T_p)/\sigma_{\rm inel}(T_p)$ which multiplies $(\epsilon_1 +\epsilon_2)$ -- if, of course, we assume that abundances (expressed in energy per nucleon) do not change with energy. The ratio $\sigma_{\rm inel}(T_p)/\sigma_R^{pp}$ is very close to unity for energies between 10--1000~GeV/nucleon, therefore, $\sigma_R^{pp}\times\mathcal{G}(T_p)/\sigma_{\rm inel}(T_p)$ is also close to unity.

The nuclear enhancement factor $\epsilon$ enters in the calculation of the $\gamma$-ray differential cross section by multiplying the maximum $A_{max}\to \epsilon \times A_{max}$, where $A_{max}$ is given in eq.~(\ref{eq:Amax}).

To illustrate the above formulas, we consider the interaction of the primary cosmic rays with the local galactic interstellar medium. By following \cite{Gaisser2002}, the ratios of primary cosmic ray fluxes at $T_p=10$~GeV/nucleon for the H($A=1$), He($A=4$), CNO($A=14$), Mg-Si($A=25$) and Fe($A=56$) are $Y_p=1:5.51\times10^{-2}: 3.25\times 10^{-3} :1.61\times 10^{-3}:3.68 \times 10^{-4}$. The local Galactic interstellar medium abundances for H($A=1$), He($A=4$), C($A=12$), N($A=14$), O($A=16$), Ne($A=20$), Mg($A=24$), Si($A=28$), S($A=32$), Fe($A=56$) have $Y_t=1:9.59\times10^{-2}: 4.65\times10^{-4} :8.3\times 10^{-5}: 8.3 \times 10^{-4} :1.2 \times 10^{-4} :3.87 \times 10^{-5} :3.69 \times 10^{-5}: 1.59 \times 10^{-5} :3.25 \times 10^{-5}$, see e.g. \cite{Meyer1985}

\begin{figure}[h]
\includegraphics[scale=0.4]{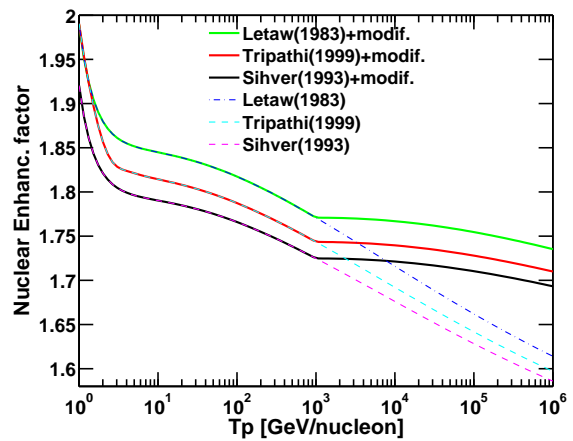}
\caption{The nuclear enhancement factor $\epsilon$ as a function of cosmic ray kinetic energy. For comparison we include ref.~ \cite{Sihver1993} cross sections and two more cross section models \cite{Letaw1983,Tripathi1999}, which take into account the energy dependence of p--nucleus inelastic cross sections at low energies. Broken lines represent $\epsilon$  in the case when $\mathcal{G}(T_p)=1$, whereas, the full lines are calculations with $\mathcal{G}(T_p)$ given in eq.~(\ref{eq:GTp}). \label{fig:Nuc.Enhanc.}}
\end{figure}

%eqs.~(\ref{eq:epsC,eq:eps1,eq:eps2}
Using eqs.~(\ref{eq:epsC}) to (\ref{eq:eps2}) we find that $\epsilon_c\approx 1.37$, $\epsilon_1\approx 0.29$ and $\epsilon_2\approx 0.1$; whereas, $\epsilon = 1.79$ if we choose $T_p=10$~GeV/nucleon. Figure~\ref{fig:Nuc.Enhanc.} shows the energy dependence of the nuclear enhancement factor $\epsilon(T_p)$. For the sake of comparison we have included two more models of p--nucleus inelastic cross sections which describe the energy dependence at low energies \cite{Letaw1983,Tripathi1999}. Different authors have estimated the nuclear enhancement factor at $T_p=10$~GeV/nucleon \cite{Cavallo1971, Stephens1981, Dermer1986a, Mori1997, Mori2009}. Ref.~\cite{Mori2009} includes in addition to p and $\alpha$ the contribution of the cosmic ray CNO, Mg-Si and Fe nuclei. The cross section of nucleon--nucleus (or nucleus--nucleus) were calculated using DPMJET-3. The value of the nuclear enhancement factor found by this author, which adopted the same abundance ratios as was done here, is $\epsilon=1.84$. For comparison, we find that $\epsilon=1.84$ using ref.~\cite{Letaw1983} cross sections, $\epsilon=1.81$ using ref.~\cite{Tripathi1999} cross sections and $\epsilon=1.79$ using ref.~\cite{Sihver1993} cross sections. 

The shape of the $\gamma$-ray spectrum due to nucleus--nucleus interactions is not exactly the same as that produced through p--p interactions. Recently ref.~\cite{Kachelriess2014} have calculated the enhancement factor using QGSJET-II-04 and EPOS-LHC. They show that the nuclear enhancement factor does not depend only on the abundances of the primary nuclei but also on the spectrally averaged photon energy. The enhancement factor that was found by these authors, agreed approximately with that of \cite{Mori2009}, however, for $E_\gamma>100$~GeV they get larger values.

The Glauber approach does not hold at low energies, because nucleons motion inside the nucleus as well as coherent interactions of many nucleons simultaneously become important. It is interesting to notice that although p--p inelastic collisions do not produce pions below $T_p<0.28$~GeV, nucleus-nucleus interactions can produce pions efficiently below $T_p<0.28$~GeV/nucleon in the so called subthreshold meson production, see e.g. \cite{BraunMunzinger1984,Noll1984,Badala1993,Badala1996,Behnke2014HADES}. This means that the enhancement factor in eq.~(\ref{eq:eps}) will go to infinity as we approach the p--p kinematic threshold, $T_p\approx 0.28$~GeV. Therefore, the enhancement factor does not have a meaning at low energies and the $\gamma$-ray spectrum should be calculated separately between p--p and nucleus--nucleus (p--nucleus) collisions.

% ====================================================================
% ====================================================================
\section{Applications \label{sec:App}}

Using the parametrized results provided in the previous sections, the determination of the resultant $\gamma$-ray spectrum produced through p--p interactions with a power-law distribution of protons is considered. To account for the energy range of the protons, and their abundance at each energy, an integral over the proton energy spectra must be carried out,
\begin{equation} \label{eq:integ_p_spec}
\Phi_{\gamma}(E_{\gamma})=4\pi n_{\rm H}\int \frac{d\sigma}{dE_{\gamma}}(T_{p},E_{\gamma}) J(T_{p})dT_{p},
\end{equation}
\noindent where $n_{\rm H}$ is the density of target protons.

For a proton spectrum of the form
\begin{equation}\label{proton_spec}
J_{p}(p_{p})=\frac{A}{p_{p}^{\alpha}}\exp{\left [-\left(\frac{p_{p}}{p_{p}^{\rm max}}\right)^{\beta}\right ]}
\end{equation}
\noindent we have investigated the corresponding functional form of the $\gamma$-ray.

Overall, the functional shape of $\gamma$-ray spectrum produced in the $0.01-1$~GeV energy region depends on the proton spectral index, $\alpha$, adopted. For sufficiently large cutoff energies ($p_{p}^{\rm max}\gg 10$~GeV), the spectrum obtained consists of a rising part from threshold to energies $\sim 0.1$~GeV, followed by a subsequent break in the spectrum at this energy giving rise to a 0.1~GeV bump feature (note- this feature sits at $\sim 0.1$~GeV in the $\Phi_{\gamma}$ representation). This bump feature remains, although becomes somewhat less pronounced, as the proton spectral slope is hardened (ie. $\alpha$ is reduced). At energies above $\sim 10$~GeV, the power-law component of the spectrum becomes well established, becoming subsequently harder above an energy of $\sim 100$~GeV, which originates from the growth in the inelastic cross-section, $\sigma_{\rm inel}$. Generally, neglecting cutoff effects, this hardening equates to the $\gamma$-ray spectral index decreasing by $\sim 0.04$ for each decadal increase in the photon energy.
 
%Hard proton spectra, however, with $\alpha<1.75$ depart somewhat from this general rule.
We next proceed to focus more closely on the specific threshold and cutoff regions in the $\gamma$-ray spectrum regions. Both of these ``end'' regions hold potential for providing key information on the underlying proton spectrum which gave rise to them.

To investigate the growth of the $\gamma$-ray spectrum produced in the region from threshold up to energies $\sim 1$~GeV, we adopt the parametrization,
\begin{equation} \label{eq:thresh_g_spec}
\Phi_{\gamma}(E_{\gamma})=B E_{\gamma}^{\eta}\left(1+\frac{2E_{\gamma}}{m_{\pi}}\right)^{-\delta}
\end{equation}
where $B$, $\eta$, and $\delta$ are fit parameters. Using this prescription, the flux in the threshold energy region, defined by the energy flux in the range $(10^{-3}-1)\times E_{\gamma}^{2}\Phi_{\gamma}(1~{\rm GeV})$, was found to be accurately described, for all hadronic models, with an accuracy of better than 5\% in all cases. Furthermore, with this prescription, the low energy power-law slope, $\eta$, was found to be closely related to the underlying proton spectral slope, $\alpha$, with the approximate expressions $\eta\approx 0.1+0.9\alpha$ and $\delta\approx 2\eta$. These results demonstrate the fact that both the low ($E_{\gamma}\ll m_{\pi}/2$) and high-energy ($E_{\gamma}\gg m_{\pi}/2$) $\gamma$-ray spectral slopes reflect that of the parent proton spectrum, though with differing signs. Indeed, being built up of symmetric functions as discussed in section~\ref{sec:GamFit}, the $\gamma$-ray spectrum $\Phi_{\gamma}$ must itself be a symmetric function about $E_{\gamma}=m_{\pi}/2$, which discloses the origin of the $\sim 0.1$~GeV bump feature.
% TABLE OF RESULTS?

Similar to the above description of the threshold region, we adopt a parametrisation of the $\gamma$-ray spectrum produced from energies above 10~GeV into the cutoff region, of the form,
\begin{equation}\label{gamma_spec}
\Phi_{\gamma}(E_{\gamma})=\frac{A'}{E_{\gamma}^{\alpha '}}\exp{\left [-\left(\frac{E_{\gamma}}{E_{\gamma}^{\rm max}}\right)^{\beta '} \right ]}.
\end{equation}
With this prescription, the $\gamma$-ray and proton spectrum cutoff parameters, $\beta '$ and $\beta$ respectively, were found to follow the relation $\beta '=a\beta/(\beta+b)$. This description was found to provide an accurate description of the cutoff region, defined by the energy flux in the range $(1-10^{-3})\times E_{\gamma}^{2}\Phi_{\gamma}(10~{\rm GeV})$, for all hadronic models, with an accuracy of better than 20\% in all cases. A table of $a$ and $b$ best-fit values for specific proton spectral indices is provided in table~\ref{ab_values}. The spectrum of secondary $\gamma$-rays, therefore, undergoes a slower cutoff than that of the parent proton spectrum. Furthermore, a cutoff with $\beta '>a$ is never expected in the secondary $\gamma$-ray spectrum, regardless of the severity of the proton spectrum cutoff.

\begin{figure}[h]
\includegraphics[scale=0.45]{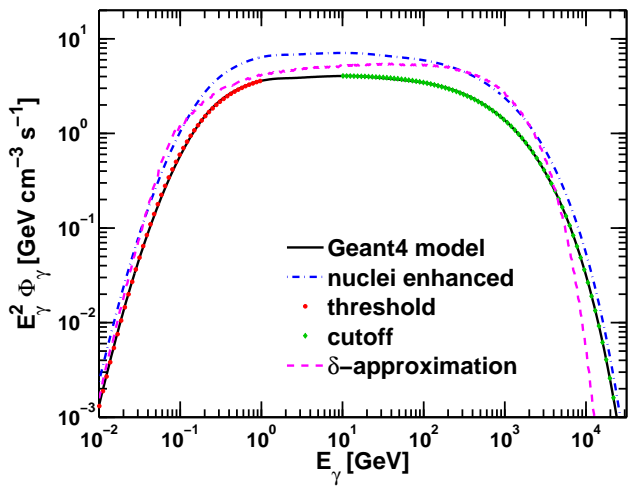}
\caption{An archetypal spectral plot (black line) demonstrating the quality of fits to both the ``threshold'' (red dotted-line region on left of the plot) and ``cutoff'' (green dotted-line region on right of the plot) regions. The model parameters adopted are $\alpha=2.0$, $\beta=1.0$ and $p_{p}^{\rm max}=10^4$~GeV. The blue dash-dot-line is the same result including the nuclear enhancement factor calculated in section~\ref{sec:NucEnhanc}. The magenta dashed-line shows the equivalent result adopting the $\delta$-approximation (no enhancement factor). \label{example_plot}}
\end{figure}

Additional insight into the nature of the threshold and cutoff regions is provided through a consideration of the $\delta$-approximation description. Since different authors adopt different definitions for approximation, we state explicitly the procedure adopted here. For this prescription, only single pion production is considered, with the pion receiving energy $E_{\pi}=KT_{p}$, in the lab frame, through a pion production interaction. We here adopt $K=0.17$, as was motivated in \cite{Kelner2006}. The differential cross section approximation for this description assumes,
\begin{equation}
\frac{d\sigma_{\pi}}{dE_{\pi}}=\sigma_{\rm inel}(T_{p})\delta(E_{\pi}-KT_{p}).
\end{equation}
Subsequent to a pion's production, its isotropic decay gives rise to a $\gamma$-ray spectrum with a top-hat functional form in $\Phi_{\gamma}$, with the end-points of this function being dictated by the kinematics described in eq.~(\ref{eq:EpiEgMAX}). With such top-hats being symmetric about $E_{\gamma}=m_{\pi}/2$, the subsequent spectrum produced from a power-law distribution of protons will, as for the full description, itself be symmetric. The $\delta$-approximation method, therefore, is expected to accurately describe the $\gamma$-ray spectrum deep in threshold region. However, at the high energy end of this region ($E_{\gamma}\approx m_{\pi}/2$), this agreement starts to fail, achieving agreement only within a factor of $\sim$3 at this break energy. Furthermore, with the $\delta$-approximation description neglecting large $K$ interactions, it predicts a cutoff shape which closely follows that of the parent proton spectrum. Thus, the agreement this description achieves in the cutoff region is very poor. A comparison of the $\delta$-approximation result with that produced through the full description is shown in figure~\ref{example_plot} for a particular example case.

\begin{table}[h]
\caption{Table of $a$ and $b$ fit parameters for different proton spectral index values, $\alpha$. \label{ab_values}}
\begin{tabular}{c|cc|cc|cc|cc}
%\toprule 
\toprule
%$\alpha$ & $a$ & $b$ \\
 & \multicolumn{2}{c|}{Geant} & \multicolumn{2}{c|}{Pythia} & \multicolumn{2}{c|}{SIBYLL} & \multicolumn{2}{|c}{QGSJET}\\
$\alpha$ & a& b& a & b & a &b &a &b \\
%\toprule
\hline 
\multicolumn{1}{c|}{1.5} & \multicolumn{1}{c}{1.0} & \multicolumn{1}{c|}{1.0} & \multicolumn{1}{c}{1.1} & \multicolumn{1}{c|}{1.2} & \multicolumn{1}{c}{1.2} &\multicolumn{1}{c|}{1.2} & \multicolumn{1}{c}{1.1} & \multicolumn{1}{c}{1.1}\\
1.75 & 1.1 & 1.1 & 1.2 & 1.3 & 1.3 & 1.3 & 1.2 & 1.2 \\
2.0 & 1.3 & 1.1 & 1.4 & 1.4 & 1.5 & 1.4 & 1.3 & 1.3 \\
2.25 & 1.4 & 1.2 & 1.5 & 1.5 & 1.6 & 1.4 & 1.4 & 1.3 \\
2.5 & 1.5 & 1.1 & 1.7 & 1.7 & 1.7 & 1.5 & 1.5 & 1.4 \\
\toprule
\end{tabular}
\end{table}

\section{ Discussion and Conclusions \label{sec:conclusion}}
The aim of this work is to provide a framework of accurate and simple fitting formulas for the $\gamma$-ray spectra produced in p--p inelastic collisions that cover the energy interval from threshold to very high energies ($T_p\sim1$~PeV), with the added flexibility to switch between different high energy models. As space-based $\gamma$-ray detectors are beginning to probe the $E_\gamma<100$~MeV, improved accuracy of the $pp\to\gamma$ production cross sections at low energies are needed. As a result we have revised here the low energy $\pi^0$ production cross section and differential cross section data below 2 GeV. We use Geant~4.10.0 for energies above 2 GeV and at high energies, where different hadronic interaction model descriptions diverge, we have used Geant~4.10.0, Pythia~8.18, SIBYLL~2.1 and QGSJET-I. Different validations of the MC codes against the $\pi^0$ and $\gamma$-ray experimental data at the LHC energies (see e.g. \cite{Enterria2011,LHCf7TeVPi0, LHCf09TeVGam, LHCf7TeVGam}) show that none of the MC models is able to explain fully all the available experimental data. It is for this reason that we parametrize different models which are representative of different descriptions of hadronic interactions. Note that the latest version of QGSJET in the CORSIKA code is QGSJET-II. In the present work, however, we do not run the CORSIKA code. We use here the fits of $\pi^0$ and $\eta$ spectra for both SIBYLL~2.1 and QGSJET-I provided in \cite{Kelner2006}. Despite this, we have good reason to believe that the parametrization provided here can fit well the spectra of other MC codes that were not considered.

Recently the TOTEM collaboration at the LHC has measured the p--p inelastic cross section at $\sqrt{s}=7$ and 8 TeV. The widely used parametrizations for this cross section in astrophysics e.g. \cite{Kamae2006,Kelner2006}, do not fit well these two data points. Therefore, here we give a parametrization of the p--p inelastic cross section that fits well both the low energy as well as new high energy LHC data.

The inclusive $\pi^0$ production cross section that we have suggested here, is composed of fits of the experimental one-pion and two-pion production data and fits from Geant~4.10.0 $\pi^0$ production multiplicity. At very high energies we have fitted separately the $\pi^0$ production multiplicity from Geant~4.10.0, Pythia~8.18, SIBYLL~2.1 and QGSJET-I. All these fits have an accuracy better than 3~\% for proton energy $T_p^{\rm th}\le T_p\le 1$~PeV. It is very likely that this parametrization continues to accurately describe the inclusive $\pi^0$ production cross section above 1~PeV.

By comparing another widely used low energy parametrization \cite{Dermer1986b} with the experimental $\pi^0$ production cross section data near the kinematic threshold (around $T_p=0.3$ GeV), we show that this parametrization is 70--80 \% higher than the data. Furthermore, we highlight that it fails to describe some of the features around 1--2~GeV, being at least 20 \% higher than the inclusive cross section data around 1.25~GeV provided by \cite{Agakishiev2012inc}.
%The experimental $\pi^0$ production cross section data near the kinematic threshold (around $T_p=300$ MeV) shows that the parametrization provided by \cite{Dermer1986b} is about 70--80 \% higher. 

The $\gamma$-ray production differential cross section is divided into two parts. One part is the value at the peak of the differential cross section (i.e. the value at $E_\gamma=m_\pi/2$), which is only a function of proton energy, namely $A_{\rm max}(T_p)$. The second part is the shape of the spectrum and is a function of $\gamma$-ray and proton energies $F(T_p,E_\gamma)$. This division turns out to be useful since it simplifies the fitting of each part. The $A_{\rm max}(T_p)$ for instance, can be composed into the inclusive $\pi^0$ production cross section and a relatively simple function of proton energy, as it is shown in eq.~(\ref{eq:Amax}). The shape of the spectrum $F(T_p,E_\gamma)$, on the other hand, expressed in terms of the variable $X_\gamma$ defined in eq.~(\ref{eq:YandX}), simplifies the fitting function. As is shown in eq.~(\ref{eq:Fgamma2}), $F(T_p,X_\gamma)$ is a function defined in the range $[0,1]\times[0,1]$ for a given proton energy. Its denominator $\left(1+ \frac{X_\gamma}{C} \right)^{\gamma(T_p)}$ is responsible for the shape near $E_\gamma=m_\pi/2$, whereas the numerator $\left(1- X_\gamma \right)^{\beta(T_p)}$ or $\left(1- \sqrt{X_\gamma} \right)^{\beta={\rm const}}$ is responsible for the shape near the kinematic limit $E_\gamma=E_\gamma^{\rm max}$. 

The accuracy of the $A_{\rm max}(T_p)$ is better than 15~\% at energies $T_p<5$~GeV and better than 3 \% at higher energies. It is remarkable to notice that  $A_{\rm max}(T_p)/\sigma_{\rm inel}$ for $T_p >10$~GeV has a very weak energy dependence, thus, $A_{\rm max}(T_p)$ mimics the $\sigma_{\rm inel}$. 

The accuracy of the $\gamma$-ray production differential cross section $d\sigma/dE_\gamma= A_{\rm max}\times F(T_p,E_\gamma)$ is better than 20~\%, see figures~\ref{fig:G4spec} to \ref{fig:Qspec}. %figures~\ref{fig:G4spec,fig:P8spec,fig:Sspec,fig:Qspec}. 
As the proton energy increases, the $\gamma$-ray spectrum becomes wider and the ``distance'' between $E_\gamma=m_\pi/2$ and $E_\gamma=E_\gamma^{\rm max}$  becomes larger. As a result, at very high energies, the accuracy of the fit for $m_\pi/2 \ll E_\gamma \ll E_\gamma^{\rm max}$ drops to 20~\% or below as can be seen for the case of SIBYLL~2.1 and QGSJET-I.

The $\gamma$-ray spectra for $T_p^{\rm th} \leq T_p <1$~GeV is calculated using experimental $\pi^0$ spectral data. Although, the shape of the $\pi^0$ spectra between experimental data and \cite{Stecker1971,Dermer1986b} isobaric model calculations are very different in the center-of-mass frame, these differences in the $\gamma$-ray spectra are less noticeable. The $\delta$-approximation introduced at higher energies by \cite{aharonian_atoyan2000} (see section~\ref{sec:App}), is another example where the $\pi^0$ spectral shape itself does not play a large role. This quality of the $\gamma$-ray spectrum is evidence that the $\pi^0$ multiplicity and the kinematics of the $\pi^0\to2\gamma$ decay play the greatest contribution.

We observe that the $\gamma$-ray spectrum parametrization provided in \cite{Kamae2006}, does not satisfy the expected kinematic limit at certain proton energies, with the spectrum going substantially beyond the $E_\gamma=E_\gamma^{\rm max}$ limit. A second  unphysical feature of this parametrization is that the $\gamma$-ray spectrum is not symmetric with respect to $E_\gamma=m_\pi/2$ in the $\log(E_\gamma)$ axis, which is expected from the kinematics of $\pi^0\to2\gamma$ decay.

In real astrophysical environments Hydrogen is not the only element present; thus, $\pi^0$ production due to nucleus-nucleus interaction must be included. Traditionally, the nuclear effect has been embedded in the nuclear enhancement factor. This factor however, is meaningful only if the energy of the nuclei are much higher than 0.28~GeV/nucleon. For completeness we have shown explicitly how to calculate this factor using the Glauber theory and the wounded nucleon model as well as using inelastic nucleus-nucleus cross sections from \cite{Sihver1993} with a modification that counts the increasing of this cross section at very high energies. At low energies $T_p<1$~GeV/nucleon the p--p contribution to the $\gamma$-ray spectrum decreases. Nucleus--nucleus interaction on the other hand can still produce $\pi^0$s at this energies very efficiently even when $T_p<0.28$~GeV/nucleon through the subthreshold meson production effect. At such energies the nuclear enhancement factor increases and goes to infinity as we approach $T_p=0.28$~GeV/nucleon beyond which this factor becomes completely meaningless and the contribution of nucleus--nucleus in the $\gamma$-ray spectrum should be calculated separately. Furthermore, the Glauber multi--scattering theory is also expected to break down at these energies due to the internal motion of the nucleons inside the nucleus and the coherent interaction between groups of nucleons from projectile and target nuclei becoming important.

The parametrization that we described here is applied to different proton spectra, see eq.~(\ref{proton_spec}). Simple fits and relations are found between proton and $\gamma$-ray spectral parameters. Generally, by neglecting the cutoff effect, the power-law part of the $\gamma$-ray spectrum is hardened by 4~\% each decadal increase in $\gamma$-ray energy. For the energy region from threshold up to 1~GeV the spectrum can be fitted accurately with a very simple function shown in eq.~(\ref{eq:thresh_g_spec}). The parameters $\eta$ and $\delta$ of the eq.~(\ref{eq:thresh_g_spec}), are closely related to the power-law index $\alpha$ of the proton spectrum as follows: $\eta \approx 0.1+0.9\,\alpha$ and $\delta\approx 2\,\eta$. Near the cutoff region the $\gamma$-ray spectrum can be fitted with the function shown in eq.~(\ref{gamma_spec}). The $\beta '$ parameter of the $\gamma$-ray spectrum cutoff is related to the $\beta$ parameter of the proton spectrum cutoff by $\beta '=a\,\beta/(\beta + b)$; where, $a$ and $b$ are fitting parameters and are functions of the proton power-law index $\alpha$ and their relation is provided in a tabular form in table~\ref{ab_values}.

Recent study of the $\gamma$-ray spectrum due to $\pi^0$ production from the galactic cosmic rays \cite{Profumo2014}, have used a model that combine the model from \cite{Dermer1986a} near threshold and interpolates it to \cite{Kachelries2012} model at very high energies. By using the parametrization provided here, we have recreated \cite{Profumo2014} power-law and broken power-law proton spectra calculations. We find that in case of the power-law proton spectrum, our $\gamma$-ray spectrum peaks around $E_\gamma \approx 0.4$~GeV; whereas, the $\gamma$-ray spectrum from \cite{Profumo2014} has a plateau-like between $0.3<E_\gamma<1$~GeV which is about 10~\% lower than our $\gamma$-ray spectrum peak. For energies between $1\leq E_\gamma \leq 10$~GeV, \cite{Profumo2014} $\gamma$-ray spectrum is higher than our spectrum with the maximum difference about 30~\% at $E_\gamma \approx 1$~GeV. At higher energies the two spectra are comparable. For the broken power-law case the two spectra are comparable except for energies between $0.2<E_\gamma<1$~GeV, where the \cite{Profumo2014} $\gamma$-ray spectrum is lower than our spectrum. The largest difference is about 40--45~\% around $E_\gamma \approx 0.4$~GeV. 

Lastly, it is possible that one can use the $\gamma$-ray spectra to gain information on other secondary particles such are muons, electrons, neutrinos, and their antiparticles. This is possible due to the relations that exist between charged and neutral pion production cross sections far from the p--p kinematic threshold. Therefore, by using the production cross section ratios between $\gamma$-rays and other secondary particles that are described e.g. in \cite{Kelner2006}, one can potentially apply this information to our results to obtain the secondary spectra of other particles also.

\section{Summary}
In this work we combined experimental data on the $pp\to\pi^0$ production below 2~GeV and results from publicly available Monte Carlo codes at higher energies, to parametrize the $\gamma$-ray spectrum due to inelastic p--p collisions. This parametrization is accurate ($\leq 20$~\%), and spans from the p--p kinematic threshold to 1~PeV proton energy, and provides flexibility to switch between different high energy models.

We have introduced a practical way to calculate the nuclear enhancement factor at high energies and we show that this factor increases for $T_p<2$~GeV/nucleon and eventually becomes meaningless as we approach p--p kinematic threshold $T_p\sim 0.28$~GeV/nucleon. At this point nucleus--nucleus continues to efficiently produce $\pi^0$ for $T_p< 0.28$~GeV/nucleon through the subthreshold meson production effect.

We have applied our parametrization to different proton spectra and related $\gamma$-ray and proton spectral parameters. Fitting formulas were found which provided a simple relation between the two spectra parameters.

\begin{acknowledgments}
\label{Acknowledgments}
The authors are grateful to Torbj\"orn Sj\"ostrand and Alberto Ribon for their help and discussions regarding Pythia~8 and Geant~4. Authors are also grateful to Maxim Barkov and Stanislav Kelner for their helpful discussions and Pol Bordas and Ruizhi Yang for carefully checking the results presented here. A.T. acknowledges a Schroedinger fellowship at DIAS. G.S.V. thanks the MPIK and its High Energy Astrophysics Theory Group for their support and hospitality during the first stages of this work, and Asociaci\'on Argentina de Astronom\'ia for the award of the ``Carlos M. Varsavsky Prize'', that greatly contributed to funding her stay in the MPIK. G.S.V. was also supported by Argentinian agencies CONICET and ANPCyT through grants PIP 0078 and PICT 2012-00278 Pr\'estamo BID.  
\end{acknowledgments}

\bibliographystyle{apsrev}
\bibliography{spect,data,highenergy,NucleiXS,SubthreshPi0}

\begin{thebibliography}{106}
\expandafter\ifx\csname natexlab\endcsname\relax\def\natexlab#1{#1}\fi
\expandafter\ifx\csname bibnamefont\endcsname\relax
  \def\bibnamefont#1{#1}\fi
\expandafter\ifx\csname bibfnamefont\endcsname\relax
  \def\bibfnamefont#1{#1}\fi
\expandafter\ifx\csname citenamefont\endcsname\relax
  \def\citenamefont#1{#1}\fi
\expandafter\ifx\csname url\endcsname\relax
  \def\url#1{\texttt{#1}}\fi
\expandafter\ifx\csname urlprefix\endcsname\relax\def\urlprefix{URL }\fi
\providecommand{\bibinfo}[2]{#2}
\providecommand{\eprint}[2][]{\url{#2}}

\bibitem[{\citenamefont{{Yuan} and {Lindenbaum}}(1956)}]{Yuan1956isobar}
\bibinfo{author}{\bibfnamefont{L.~C.} \bibnamefont{{Yuan}}} \bibnamefont{and}
  \bibinfo{author}{\bibfnamefont{S.~J.} \bibnamefont{{Lindenbaum}}},
  \bibinfo{journal}{Physical Review} \textbf{\bibinfo{volume}{103}},
  \bibinfo{pages}{404} (\bibinfo{year}{1956}).

\bibitem[{\citenamefont{{Lindenbaum} and
  {Sternheimer}}(1957)}]{Lindenbaum1957isobar}
\bibinfo{author}{\bibfnamefont{S.~J.} \bibnamefont{{Lindenbaum}}}
  \bibnamefont{and} \bibinfo{author}{\bibfnamefont{R.~M.}
  \bibnamefont{{Sternheimer}}}, \bibinfo{journal}{Physical Review}
  \textbf{\bibinfo{volume}{105}}, \bibinfo{pages}{1874} (\bibinfo{year}{1957}).

\bibitem[{\citenamefont{Watson}(1952)}]{Watson1952}
\bibinfo{author}{\bibfnamefont{K.~M.} \bibnamefont{Watson}},
  \bibinfo{journal}{Phys. Rev.} \textbf{\bibinfo{volume}{88}},
  \bibinfo{pages}{1163} (\bibinfo{year}{1952}),
  \urlprefix\url{http://link.aps.org/doi/10.1103/PhysRev.88.1163}.

\bibitem[{\citenamefont{{Gell-Mann} and {Watson}}(1954)}]{Gell-Mann1954}
\bibinfo{author}{\bibfnamefont{M.}~\bibnamefont{{Gell-Mann}}} \bibnamefont{and}
  \bibinfo{author}{\bibfnamefont{K.~M.} \bibnamefont{{Watson}}},
  \bibinfo{journal}{Annual Review of Nuclear and Particle Science}
  \textbf{\bibinfo{volume}{4}}, \bibinfo{pages}{219} (\bibinfo{year}{1954}).

\bibitem[{\citenamefont{{Lewis} et~al.}(1948)\citenamefont{{Lewis},
  {Oppenheimer}, and {Wouthuysen}}}]{Lewis1948fireball}
\bibinfo{author}{\bibfnamefont{H.~W.} \bibnamefont{{Lewis}}},
  \bibinfo{author}{\bibfnamefont{J.~R.} \bibnamefont{{Oppenheimer}}},
  \bibnamefont{and} \bibinfo{author}{\bibfnamefont{S.~A.}
  \bibnamefont{{Wouthuysen}}}, \bibinfo{journal}{Physical Review}
  \textbf{\bibinfo{volume}{73}}, \bibinfo{pages}{127} (\bibinfo{year}{1948}).

\bibitem[{\citenamefont{{Fermi}}(1950)}]{Fermi1950fireball}
\bibinfo{author}{\bibfnamefont{E.}~\bibnamefont{{Fermi}}},
  \bibinfo{journal}{Progress of Theoretical Physics}
  \textbf{\bibinfo{volume}{5}}, \bibinfo{pages}{570} (\bibinfo{year}{1950}).

\bibitem[{\citenamefont{{Umezawa} et~al.}(1951)\citenamefont{{Umezawa},
  {Takahashi}, and {Kamefuchi}}}]{Umezawa1951fireball}
\bibinfo{author}{\bibfnamefont{H.}~\bibnamefont{{Umezawa}}},
  \bibinfo{author}{\bibfnamefont{Y.}~\bibnamefont{{Takahashi}}},
  \bibnamefont{and}
  \bibinfo{author}{\bibfnamefont{S.}~\bibnamefont{{Kamefuchi}}},
  \bibinfo{journal}{Progress of Theoretical Physics}
  \textbf{\bibinfo{volume}{6}}, \bibinfo{pages}{428} (\bibinfo{year}{1951}).

\bibitem[{\citenamefont{{Heisenberg}}(1952)}]{Heisenberg1952fireball}
\bibinfo{author}{\bibfnamefont{W.}~\bibnamefont{{Heisenberg}}},
  \bibinfo{journal}{Naturwissenschaften} \textbf{\bibinfo{volume}{39}},
  \bibinfo{pages}{69} (\bibinfo{year}{1952}).

\bibitem[{\citenamefont{Belen’kji and Landau}(1956)}]{Landau1956fireball}
\bibinfo{author}{\bibfnamefont{S.}~\bibnamefont{Belen’kji}} \bibnamefont{and}
  \bibinfo{author}{\bibfnamefont{L.}~\bibnamefont{Landau}},
  \bibinfo{journal}{Il Nuovo Cimento} \textbf{\bibinfo{volume}{3}},
  \bibinfo{pages}{15} (\bibinfo{year}{1956}), ISSN \bibinfo{issn}{0029-6341},
  \urlprefix\url{http://dx.doi.org/10.1007/BF02745507}.

\bibitem[{\citenamefont{{Cocconi}}(1958)}]{Cocconi1958_2fireball}
\bibinfo{author}{\bibfnamefont{G.}~\bibnamefont{{Cocconi}}},
  \bibinfo{journal}{Physical Review} \textbf{\bibinfo{volume}{111}},
  \bibinfo{pages}{1699} (\bibinfo{year}{1958}).

\bibitem[{\citenamefont{{Stecker} et~al.}(1968)\citenamefont{{Stecker},
  {Tsuruta}, and {Fazio}}}]{Stecker1968}
\bibinfo{author}{\bibfnamefont{F.~W.} \bibnamefont{{Stecker}}},
  \bibinfo{author}{\bibfnamefont{S.}~\bibnamefont{{Tsuruta}}},
  \bibnamefont{and} \bibinfo{author}{\bibfnamefont{G.~G.}
  \bibnamefont{{Fazio}}}, \bibinfo{journal}{\apj}
  \textbf{\bibinfo{volume}{151}}, \bibinfo{pages}{881} (\bibinfo{year}{1968}).

\bibitem[{\citenamefont{{Stecker}}(1971)}]{Stecker1971}
\bibinfo{author}{\bibfnamefont{F.~W.} \bibnamefont{{Stecker}}},
  \bibinfo{journal}{NASA Special Publication} \textbf{\bibinfo{volume}{249}}
  (\bibinfo{year}{1971}).

\bibitem[{\citenamefont{{Badhwar} et~al.}(1977)\citenamefont{{Badhwar},
  {Golden}, and {Stephens}}}]{Badhwar1977}
\bibinfo{author}{\bibfnamefont{G.~D.} \bibnamefont{{Badhwar}}},
  \bibinfo{author}{\bibfnamefont{R.~L.} \bibnamefont{{Golden}}},
  \bibnamefont{and} \bibinfo{author}{\bibfnamefont{S.~A.}
  \bibnamefont{{Stephens}}}, \bibinfo{journal}{\prd}
  \textbf{\bibinfo{volume}{15}}, \bibinfo{pages}{820} (\bibinfo{year}{1977}).

\bibitem[{\citenamefont{{Stephens} and {Badhwar}}(1981)}]{Stephens1981}
\bibinfo{author}{\bibfnamefont{S.~A.} \bibnamefont{{Stephens}}}
  \bibnamefont{and} \bibinfo{author}{\bibfnamefont{G.~D.}
  \bibnamefont{{Badhwar}}}, \bibinfo{journal}{\apss}
  \textbf{\bibinfo{volume}{76}}, \bibinfo{pages}{213} (\bibinfo{year}{1981}).

\bibitem[{\citenamefont{{Tan} and {Ng}}(1983)}]{TanNg1983}
\bibinfo{author}{\bibfnamefont{L.~C.} \bibnamefont{{Tan}}} \bibnamefont{and}
  \bibinfo{author}{\bibfnamefont{L.~K.} \bibnamefont{{Ng}}},
  \bibinfo{journal}{Journal of Physics G Nuclear Physics}
  \textbf{\bibinfo{volume}{9}}, \bibinfo{pages}{227} (\bibinfo{year}{1983}).

\bibitem[{\citenamefont{{Feynman}}(1969)}]{Feynman1969scaling}
\bibinfo{author}{\bibfnamefont{R.~P.} \bibnamefont{{Feynman}}},
  \bibinfo{journal}{Physical Review Letters} \textbf{\bibinfo{volume}{23}},
  \bibinfo{pages}{1415} (\bibinfo{year}{1969}).

\bibitem[{\citenamefont{{Koba} et~al.}(1972)\citenamefont{{Koba}, {Nielsen},
  and {Olesen}}}]{KNOscaling}
\bibinfo{author}{\bibfnamefont{Z.}~\bibnamefont{{Koba}}},
  \bibinfo{author}{\bibfnamefont{H.~B.} \bibnamefont{{Nielsen}}},
  \bibnamefont{and} \bibinfo{author}{\bibfnamefont{P.}~\bibnamefont{{Olesen}}},
  \bibinfo{journal}{Nuclear Physics B} \textbf{\bibinfo{volume}{40}},
  \bibinfo{pages}{317} (\bibinfo{year}{1972}).

\bibitem[{\citenamefont{{Blattnig} et~al.}(2000)\citenamefont{{Blattnig},
  {Swaminathan}, {Kruger}, {Ngom}, and {Norbury}}}]{Blattnig2000}
\bibinfo{author}{\bibfnamefont{S.~R.} \bibnamefont{{Blattnig}}},
  \bibinfo{author}{\bibfnamefont{S.~R.} \bibnamefont{{Swaminathan}}},
  \bibinfo{author}{\bibfnamefont{A.~T.} \bibnamefont{{Kruger}}},
  \bibinfo{author}{\bibfnamefont{M.}~\bibnamefont{{Ngom}}}, \bibnamefont{and}
  \bibinfo{author}{\bibfnamefont{J.~W.} \bibnamefont{{Norbury}}},
  \bibinfo{journal}{\prd} \textbf{\bibinfo{volume}{62}}, \bibinfo{eid}{094030}
  (\bibinfo{year}{2000}), \eprint{arXiv:hep-ph/0010170}.

\bibitem[{\citenamefont{{Sato} et~al.}(2012)\citenamefont{{Sato}, {Shibata},
  and {Yamazaki}}}]{Sato2012}
\bibinfo{author}{\bibfnamefont{H.}~\bibnamefont{{Sato}}},
  \bibinfo{author}{\bibfnamefont{T.}~\bibnamefont{{Shibata}}},
  \bibnamefont{and}
  \bibinfo{author}{\bibfnamefont{R.}~\bibnamefont{{Yamazaki}}},
  \bibinfo{journal}{Astroparticle Physics} \textbf{\bibinfo{volume}{36}},
  \bibinfo{pages}{83} (\bibinfo{year}{2012}), \eprint{1205.2145}.

\bibitem[{\citenamefont{{Dermer}}(1986{\natexlab{a}})}]{Dermer1986a}
\bibinfo{author}{\bibfnamefont{C.~D.} \bibnamefont{{Dermer}}},
  \bibinfo{journal}{\aap} \textbf{\bibinfo{volume}{157}}, \bibinfo{pages}{223}
  (\bibinfo{year}{1986}{\natexlab{a}}).

\bibitem[{\citenamefont{{B{\"a}hr} et~al.}(2008)\citenamefont{{B{\"a}hr},
  {Gieseke}, {Gigg}, {Grellscheid}, {Hamilton}, {Latunde-Dada}, {Pl{\"a}tzer},
  {Richardson}, {Seymour}, {Sherstnev} et~al.}}]{Herwig}
\bibinfo{author}{\bibfnamefont{M.}~\bibnamefont{{B{\"a}hr}}},
  \bibinfo{author}{\bibfnamefont{S.}~\bibnamefont{{Gieseke}}},
  \bibinfo{author}{\bibfnamefont{M.~A.} \bibnamefont{{Gigg}}},
  \bibinfo{author}{\bibfnamefont{D.}~\bibnamefont{{Grellscheid}}},
  \bibinfo{author}{\bibfnamefont{K.}~\bibnamefont{{Hamilton}}},
  \bibinfo{author}{\bibfnamefont{O.}~\bibnamefont{{Latunde-Dada}}},
  \bibinfo{author}{\bibfnamefont{S.}~\bibnamefont{{Pl{\"a}tzer}}},
  \bibinfo{author}{\bibfnamefont{P.}~\bibnamefont{{Richardson}}},
  \bibinfo{author}{\bibfnamefont{M.~H.} \bibnamefont{{Seymour}}},
  \bibinfo{author}{\bibfnamefont{A.}~\bibnamefont{{Sherstnev}}},
  \bibnamefont{et~al.}, \bibinfo{journal}{European Physical Journal C}
  \textbf{\bibinfo{volume}{58}}, \bibinfo{pages}{639} (\bibinfo{year}{2008}),
  \eprint{0803.0883}.

\bibitem[{\citenamefont{{Gleisberg} et~al.}(2009)\citenamefont{{Gleisberg},
  {H{\"o}che}, {Krauss}, {Sch{\"o}nherr}, {Schumann}, {Siegert}, and
  {Winter}}}]{Sherpa1}
\bibinfo{author}{\bibfnamefont{T.}~\bibnamefont{{Gleisberg}}},
  \bibinfo{author}{\bibfnamefont{S.}~\bibnamefont{{H{\"o}che}}},
  \bibinfo{author}{\bibfnamefont{F.}~\bibnamefont{{Krauss}}},
  \bibinfo{author}{\bibfnamefont{M.}~\bibnamefont{{Sch{\"o}nherr}}},
  \bibinfo{author}{\bibfnamefont{S.}~\bibnamefont{{Schumann}}},
  \bibinfo{author}{\bibfnamefont{F.}~\bibnamefont{{Siegert}}},
  \bibnamefont{and} \bibinfo{author}{\bibfnamefont{J.}~\bibnamefont{{Winter}}},
  \bibinfo{journal}{Journal of High Energy Physics}
  \textbf{\bibinfo{volume}{2}}, \bibinfo{eid}{007} (\bibinfo{year}{2009}),
  \eprint{0811.4622}.

\bibitem[{\citenamefont{{Sj{\"o}strand}
  et~al.}(2006)\citenamefont{{Sj{\"o}strand}, {Mrenna}, and
  {Skands}}}]{Pythia1}
\bibinfo{author}{\bibfnamefont{T.}~\bibnamefont{{Sj{\"o}strand}}},
  \bibinfo{author}{\bibfnamefont{S.}~\bibnamefont{{Mrenna}}}, \bibnamefont{and}
  \bibinfo{author}{\bibfnamefont{P.}~\bibnamefont{{Skands}}},
  \bibinfo{journal}{Journal of High Energy Physics}
  \textbf{\bibinfo{volume}{5}}, \bibinfo{eid}{026} (\bibinfo{year}{2006}),
  \eprint{hep-ph/0603175}.

\bibitem[{\citenamefont{{Sj{\"o}strand}
  et~al.}(2008)\citenamefont{{Sj{\"o}strand}, {Mrenna}, and
  {Skands}}}]{Pythia2}
\bibinfo{author}{\bibfnamefont{T.}~\bibnamefont{{Sj{\"o}strand}}},
  \bibinfo{author}{\bibfnamefont{S.}~\bibnamefont{{Mrenna}}}, \bibnamefont{and}
  \bibinfo{author}{\bibfnamefont{P.}~\bibnamefont{{Skands}}},
  \bibinfo{journal}{Computer Physics Communications}
  \textbf{\bibinfo{volume}{178}}, \bibinfo{pages}{852} (\bibinfo{year}{2008}),
  \eprint{0710.3820}.

\bibitem[{\citenamefont{{Engel}}(1995)}]{phojet1}
\bibinfo{author}{\bibfnamefont{R.}~\bibnamefont{{Engel}}},
  \bibinfo{journal}{Zeitschrift fur Physik C Particles and Fields}
  \textbf{\bibinfo{volume}{66}}, \bibinfo{pages}{203} (\bibinfo{year}{1995}).

\bibitem[{\citenamefont{{Engel} et~al.}(1995)\citenamefont{{Engel}, {Ranft},
  and {Roesler}}}]{phojet2}
\bibinfo{author}{\bibfnamefont{R.}~\bibnamefont{{Engel}}},
  \bibinfo{author}{\bibfnamefont{J.}~\bibnamefont{{Ranft}}}, \bibnamefont{and}
  \bibinfo{author}{\bibfnamefont{S.}~\bibnamefont{{Roesler}}},
  \bibinfo{journal}{\prd} \textbf{\bibinfo{volume}{52}}, \bibinfo{pages}{1459}
  (\bibinfo{year}{1995}), \eprint{hep-ph/9502319}.

\bibitem[{\citenamefont{{Engel} and {Ranft}}(1996)}]{phojet3}
\bibinfo{author}{\bibfnamefont{R.}~\bibnamefont{{Engel}}} \bibnamefont{and}
  \bibinfo{author}{\bibfnamefont{J.}~\bibnamefont{{Ranft}}},
  \bibinfo{journal}{\prd} \textbf{\bibinfo{volume}{54}}, \bibinfo{pages}{4244}
  (\bibinfo{year}{1996}), \eprint{hep-ph/9509373}.

\bibitem[{\citenamefont{{Bopp} et~al.}(2008)\citenamefont{{Bopp}, {Ranft},
  {Engel}, and {Roesler}}}]{dpmjet}
\bibinfo{author}{\bibfnamefont{F.~W.} \bibnamefont{{Bopp}}},
  \bibinfo{author}{\bibfnamefont{J.}~\bibnamefont{{Ranft}}},
  \bibinfo{author}{\bibfnamefont{R.}~\bibnamefont{{Engel}}}, \bibnamefont{and}
  \bibinfo{author}{\bibfnamefont{S.}~\bibnamefont{{Roesler}}},
  \bibinfo{journal}{\prc} \textbf{\bibinfo{volume}{77}}, \bibinfo{eid}{014904}
  (\bibinfo{year}{2008}), \eprint{hep-ph/0505035}.

\bibitem[{\citenamefont{{Werner} et~al.}(2006)\citenamefont{{Werner}, {Liu},
  and {Pierog}}}]{epos}
\bibinfo{author}{\bibfnamefont{K.}~\bibnamefont{{Werner}}},
  \bibinfo{author}{\bibfnamefont{F.-M.} \bibnamefont{{Liu}}}, \bibnamefont{and}
  \bibinfo{author}{\bibfnamefont{T.}~\bibnamefont{{Pierog}}},
  \bibinfo{journal}{\prc} \textbf{\bibinfo{volume}{74}}, \bibinfo{eid}{044902}
  (\bibinfo{year}{2006}), \eprint{hep-ph/0506232}.

\bibitem[{\citenamefont{{Engel} et~al.}(1992)\citenamefont{{Engel}, {Gaisser},
  {Lipari}, and {Stanev}}}]{SIBYLL1992}
\bibinfo{author}{\bibfnamefont{J.}~\bibnamefont{{Engel}}},
  \bibinfo{author}{\bibfnamefont{T.~K.} \bibnamefont{{Gaisser}}},
  \bibinfo{author}{\bibfnamefont{P.}~\bibnamefont{{Lipari}}}, \bibnamefont{and}
  \bibinfo{author}{\bibfnamefont{T.}~\bibnamefont{{Stanev}}},
  \bibinfo{journal}{\prd} \textbf{\bibinfo{volume}{46}}, \bibinfo{pages}{5013}
  (\bibinfo{year}{1992}).

\bibitem[{\citenamefont{{Fletcher} et~al.}(1994)\citenamefont{{Fletcher},
  {Gaisser}, {Lipari}, and {Stanev}}}]{SIBYLL1994}
\bibinfo{author}{\bibfnamefont{R.~S.} \bibnamefont{{Fletcher}}},
  \bibinfo{author}{\bibfnamefont{T.~K.} \bibnamefont{{Gaisser}}},
  \bibinfo{author}{\bibfnamefont{P.}~\bibnamefont{{Lipari}}}, \bibnamefont{and}
  \bibinfo{author}{\bibfnamefont{T.}~\bibnamefont{{Stanev}}},
  \bibinfo{journal}{\prd} \textbf{\bibinfo{volume}{50}}, \bibinfo{pages}{5710}
  (\bibinfo{year}{1994}).

\bibitem[{\citenamefont{{Ahn} et~al.}(2009)\citenamefont{{Ahn}, {Engel},
  {Gaisser}, {Lipari}, and {Stanev}}}]{SIBYLL2009}
\bibinfo{author}{\bibfnamefont{E.-J.} \bibnamefont{{Ahn}}},
  \bibinfo{author}{\bibfnamefont{R.}~\bibnamefont{{Engel}}},
  \bibinfo{author}{\bibfnamefont{T.~K.} \bibnamefont{{Gaisser}}},
  \bibinfo{author}{\bibfnamefont{P.}~\bibnamefont{{Lipari}}}, \bibnamefont{and}
  \bibinfo{author}{\bibfnamefont{T.}~\bibnamefont{{Stanev}}},
  \bibinfo{journal}{\prd} \textbf{\bibinfo{volume}{80}}, \bibinfo{eid}{094003}
  (\bibinfo{year}{2009}), \eprint{0906.4113}.

\bibitem[{\citenamefont{{Kalmykov} and {Ostapchenko}}(1993)}]{QSJET1993}
\bibinfo{author}{\bibfnamefont{N.~N.} \bibnamefont{{Kalmykov}}}
  \bibnamefont{and} \bibinfo{author}{\bibfnamefont{S.~S.}
  \bibnamefont{{Ostapchenko}}}, \bibinfo{journal}{Physics of Atomic Nuclei}
  \textbf{\bibinfo{volume}{56}}, \bibinfo{pages}{346} (\bibinfo{year}{1993}).

\bibitem[{\citenamefont{{Kalmykov} et~al.}(1997)\citenamefont{{Kalmykov},
  {Ostapchenko}, and {Pavlov}}}]{GGSJET11997}
\bibinfo{author}{\bibfnamefont{N.~N.} \bibnamefont{{Kalmykov}}},
  \bibinfo{author}{\bibfnamefont{S.~S.} \bibnamefont{{Ostapchenko}}},
  \bibnamefont{and} \bibinfo{author}{\bibfnamefont{A.~I.}
  \bibnamefont{{Pavlov}}}, \bibinfo{journal}{Nuclear Physics B Proceedings
  Supplements} \textbf{\bibinfo{volume}{52}}, \bibinfo{pages}{17}
  (\bibinfo{year}{1997}).

\bibitem[{\citenamefont{{Ostapchenko}}(2006{\natexlab{a}})}]{QGSJET-IIc}
\bibinfo{author}{\bibfnamefont{S.}~\bibnamefont{{Ostapchenko}}},
  \bibinfo{journal}{\prd} \textbf{\bibinfo{volume}{74}}, \bibinfo{eid}{014026}
  (\bibinfo{year}{2006}{\natexlab{a}}), \eprint{hep-ph/0505259}.

\bibitem[{\citenamefont{{Ostapchenko}}(2006{\natexlab{b}})}]{QGSJET-IIa}
\bibinfo{author}{\bibfnamefont{S.}~\bibnamefont{{Ostapchenko}}},
  \bibinfo{journal}{Nuclear Physics B Proceedings Supplements}
  \textbf{\bibinfo{volume}{151}}, \bibinfo{pages}{143}
  (\bibinfo{year}{2006}{\natexlab{b}}), \eprint{hep-ph/0412332}.

\bibitem[{\citenamefont{{Ostapchenko}}(2007)}]{QGSJET-IId}
\bibinfo{author}{\bibfnamefont{S.}~\bibnamefont{{Ostapchenko}}}, in
  \emph{\bibinfo{booktitle}{Collicers to Cosmic Rays}}, edited by
  \bibinfo{editor}{\bibfnamefont{M.}~\bibnamefont{{Tripathi}}}
  \bibnamefont{and} \bibinfo{editor}{\bibfnamefont{R.~E.}
  \bibnamefont{{Breedon}}} (\bibinfo{year}{2007}), vol. \bibinfo{volume}{928}
  of \emph{\bibinfo{series}{American Institute of Physics Conference Series}},
  pp. \bibinfo{pages}{118--125}, \eprint{0706.3784}.

\bibitem[{\citenamefont{{Thakuria} and {Boruah}}(2012)}]{Thakuria2012}
\bibinfo{author}{\bibfnamefont{C.~C.} \bibnamefont{{Thakuria}}}
  \bibnamefont{and} \bibinfo{author}{\bibfnamefont{K.}~\bibnamefont{{Boruah}}},
  \bibinfo{journal}{ArXiv e-prints}  (\bibinfo{year}{2012}),
  \bibinfo{note}{comparison between models}, \eprint{1202.3661}.

\bibitem[{\citenamefont{{H{\"a}nssgen} and {Ranft}}(1986)}]{hadrin}
\bibinfo{author}{\bibfnamefont{K.}~\bibnamefont{{H{\"a}nssgen}}}
  \bibnamefont{and} \bibinfo{author}{\bibfnamefont{J.}~\bibnamefont{{Ranft}}},
  \bibinfo{journal}{Computer Physics Communications}
  \textbf{\bibinfo{volume}{39}}, \bibinfo{pages}{53} (\bibinfo{year}{1986}).

\bibitem[{\citenamefont{{Pi}}(1992)}]{fritiof}
\bibinfo{author}{\bibfnamefont{H.}~\bibnamefont{{Pi}}},
  \bibinfo{journal}{Computer Physics Communications}
  \textbf{\bibinfo{volume}{71}}, \bibinfo{pages}{173} (\bibinfo{year}{1992}).

\bibitem[{\citenamefont{{Mori}}(1997)}]{Mori1997}
\bibinfo{author}{\bibfnamefont{M.}~\bibnamefont{{Mori}}},
  \bibinfo{journal}{\apj} \textbf{\bibinfo{volume}{478}}, \bibinfo{pages}{225}
  (\bibinfo{year}{1997}), \eprint{arXiv:astro-ph/9611233}.

\bibitem[{\citenamefont{{Kamae} et~al.}(2005)\citenamefont{{Kamae}, {Abe}, and
  {Koi}}}]{Kamae2005}
\bibinfo{author}{\bibfnamefont{T.}~\bibnamefont{{Kamae}}},
  \bibinfo{author}{\bibfnamefont{T.}~\bibnamefont{{Abe}}}, \bibnamefont{and}
  \bibinfo{author}{\bibfnamefont{T.}~\bibnamefont{{Koi}}},
  \bibinfo{journal}{\apj} \textbf{\bibinfo{volume}{620}}, \bibinfo{pages}{244}
  (\bibinfo{year}{2005}), \eprint{arXiv:astro-ph/0410617}.

\bibitem[{\citenamefont{{Kamae} et~al.}(2006)\citenamefont{{Kamae}, {Karlsson},
  {Mizuno}, {Abe}, and {Koi}}}]{Kamae2006}
\bibinfo{author}{\bibfnamefont{T.}~\bibnamefont{{Kamae}}},
  \bibinfo{author}{\bibfnamefont{N.}~\bibnamefont{{Karlsson}}},
  \bibinfo{author}{\bibfnamefont{T.}~\bibnamefont{{Mizuno}}},
  \bibinfo{author}{\bibfnamefont{T.}~\bibnamefont{{Abe}}}, \bibnamefont{and}
  \bibinfo{author}{\bibfnamefont{T.}~\bibnamefont{{Koi}}},
  \bibinfo{journal}{\apj} \textbf{\bibinfo{volume}{647}}, \bibinfo{pages}{692}
  (\bibinfo{year}{2006}), \eprint{arXiv:astro-ph/0605581}.

\bibitem[{\citenamefont{{Kelner} et~al.}(2006)\citenamefont{{Kelner},
  {Aharonian}, and {Bugayov}}}]{Kelner2006}
\bibinfo{author}{\bibfnamefont{S.~R.} \bibnamefont{{Kelner}}},
  \bibinfo{author}{\bibfnamefont{F.~A.} \bibnamefont{{Aharonian}}},
  \bibnamefont{and} \bibinfo{author}{\bibfnamefont{V.~V.}
  \bibnamefont{{Bugayov}}}, \bibinfo{journal}{\prd}
  \textbf{\bibinfo{volume}{74}}, \bibinfo{eid}{034018} (\bibinfo{year}{2006}),
  \eprint{arXiv:astro-ph/0606058}.

\bibitem[{\citenamefont{{Agostinelli} et~al.}(2003)\citenamefont{{Agostinelli},
  {Allison}, {Amako}, {Apostolakis}, {Araujo}, {Arce}, {Asai}, {Axen},
  {Banerjee}, {Barrand} et~al.}}]{Geant42003}
\bibinfo{author}{\bibfnamefont{S.}~\bibnamefont{{Agostinelli}}},
  \bibinfo{author}{\bibfnamefont{J.}~\bibnamefont{{Allison}}},
  \bibinfo{author}{\bibfnamefont{K.}~\bibnamefont{{Amako}}},
  \bibinfo{author}{\bibfnamefont{J.}~\bibnamefont{{Apostolakis}}},
  \bibinfo{author}{\bibfnamefont{H.}~\bibnamefont{{Araujo}}},
  \bibinfo{author}{\bibfnamefont{P.}~\bibnamefont{{Arce}}},
  \bibinfo{author}{\bibfnamefont{M.}~\bibnamefont{{Asai}}},
  \bibinfo{author}{\bibfnamefont{D.}~\bibnamefont{{Axen}}},
  \bibinfo{author}{\bibfnamefont{S.}~\bibnamefont{{Banerjee}}},
  \bibinfo{author}{\bibfnamefont{G.}~\bibnamefont{{Barrand}}},
  \bibnamefont{et~al.}, \bibinfo{journal}{Nuclear Instruments and Methods in
  Physics Research A} \textbf{\bibinfo{volume}{506}}, \bibinfo{pages}{250}
  (\bibinfo{year}{2003}).

\bibitem[{\citenamefont{{Allison} et~al.}(2006)\citenamefont{{Allison},
  {Amako}, {Apostolakis}, {Araujo}, {Arce Dubois}, {Asai}, {Barrand}, {Capra},
  {Chauvie}, {Chytracek} et~al.}}]{Geant42006}
\bibinfo{author}{\bibfnamefont{J.}~\bibnamefont{{Allison}}},
  \bibinfo{author}{\bibfnamefont{K.}~\bibnamefont{{Amako}}},
  \bibinfo{author}{\bibfnamefont{J.}~\bibnamefont{{Apostolakis}}},
  \bibinfo{author}{\bibfnamefont{H.}~\bibnamefont{{Araujo}}},
  \bibinfo{author}{\bibfnamefont{P.}~\bibnamefont{{Arce Dubois}}},
  \bibinfo{author}{\bibfnamefont{M.}~\bibnamefont{{Asai}}},
  \bibinfo{author}{\bibfnamefont{G.}~\bibnamefont{{Barrand}}},
  \bibinfo{author}{\bibfnamefont{R.}~\bibnamefont{{Capra}}},
  \bibinfo{author}{\bibfnamefont{S.}~\bibnamefont{{Chauvie}}},
  \bibinfo{author}{\bibfnamefont{R.}~\bibnamefont{{Chytracek}}},
  \bibnamefont{et~al.}, \bibinfo{journal}{IEEE Transactions on Nuclear Science}
  \textbf{\bibinfo{volume}{53}}, \bibinfo{pages}{270} (\bibinfo{year}{2006}).

\bibitem[{\citenamefont{{Beringer} et~al.}(2012)\citenamefont{{Beringer},
  {Arguin}, {Barnett}, {Copic}, {Dahl}, {Groom}, {Lin}, {Lys}, {Murayama},
  {Wohl} et~al.}}]{PDG}
\bibinfo{author}{\bibfnamefont{J.}~\bibnamefont{{Beringer}}},
  \bibinfo{author}{\bibfnamefont{J.-F.} \bibnamefont{{Arguin}}},
  \bibinfo{author}{\bibfnamefont{R.~M.} \bibnamefont{{Barnett}}},
  \bibinfo{author}{\bibfnamefont{K.}~\bibnamefont{{Copic}}},
  \bibinfo{author}{\bibfnamefont{O.}~\bibnamefont{{Dahl}}},
  \bibinfo{author}{\bibfnamefont{D.~E.} \bibnamefont{{Groom}}},
  \bibinfo{author}{\bibfnamefont{C.-J.} \bibnamefont{{Lin}}},
  \bibinfo{author}{\bibfnamefont{J.}~\bibnamefont{{Lys}}},
  \bibinfo{author}{\bibfnamefont{H.}~\bibnamefont{{Murayama}}},
  \bibinfo{author}{\bibfnamefont{C.~G.} \bibnamefont{{Wohl}}},
  \bibnamefont{et~al.}, \bibinfo{journal}{\prd} \textbf{\bibinfo{volume}{86}},
  \bibinfo{eid}{010001} (\bibinfo{year}{2012}).

\bibitem[{\citenamefont{Antchev et~al.}(2013{\natexlab{a}})}]{TOTEMxs7TeV}
\bibinfo{author}{\bibfnamefont{G.}~\bibnamefont{Antchev}} \bibnamefont{et~al.}
  (\bibinfo{collaboration}{TOTEM}), \bibinfo{journal}{Europhys.Lett.}
  \textbf{\bibinfo{volume}{101}}, \bibinfo{pages}{21004}
  (\bibinfo{year}{2013}{\natexlab{a}}).

\bibitem[{\citenamefont{Antchev et~al.}(2013{\natexlab{b}})}]{TOTEMxs8TeV}
\bibinfo{author}{\bibfnamefont{G.}~\bibnamefont{Antchev}} \bibnamefont{et~al.}
  (\bibinfo{collaboration}{TOTEM Collaboration}),
  \bibinfo{journal}{Phys.Rev.Lett.} \textbf{\bibinfo{volume}{111}},
  \bibinfo{pages}{012001} (\bibinfo{year}{2013}{\natexlab{b}}).

\bibitem[{\citenamefont{{Dermer}}(1986{\natexlab{b}})}]{Dermer1986b}
\bibinfo{author}{\bibfnamefont{C.~D.} \bibnamefont{{Dermer}}},
  \bibinfo{journal}{\apj} \textbf{\bibinfo{volume}{307}}, \bibinfo{pages}{47}
  (\bibinfo{year}{1986}{\natexlab{b}}).

\bibitem[{\citenamefont{{Bondar} et~al.}(1995)\citenamefont{{Bondar},
  {Cal{\'e}n}, {Carius}, {Ekstr{\"o}m}, {Fransson}, {Gustafsson},
  {H{\"a}ggstr{\"o}m}, {H{\"o}istad}, {Johansson}, {Johansson}
  et~al.}}]{Bondar1995}
\bibinfo{author}{\bibfnamefont{A.}~\bibnamefont{{Bondar}}},
  \bibinfo{author}{\bibfnamefont{H.}~\bibnamefont{{Cal{\'e}n}}},
  \bibinfo{author}{\bibfnamefont{S.}~\bibnamefont{{Carius}}},
  \bibinfo{author}{\bibfnamefont{C.}~\bibnamefont{{Ekstr{\"o}m}}},
  \bibinfo{author}{\bibfnamefont{K.}~\bibnamefont{{Fransson}}},
  \bibinfo{author}{\bibfnamefont{L.}~\bibnamefont{{Gustafsson}}},
  \bibinfo{author}{\bibfnamefont{S.}~\bibnamefont{{H{\"a}ggstr{\"o}m}}},
  \bibinfo{author}{\bibfnamefont{B.}~\bibnamefont{{H{\"o}istad}}},
  \bibinfo{author}{\bibfnamefont{A.}~\bibnamefont{{Johansson}}},
  \bibinfo{author}{\bibfnamefont{T.}~\bibnamefont{{Johansson}}},
  \bibnamefont{et~al.}, \bibinfo{journal}{Physics Letters B}
  \textbf{\bibinfo{volume}{356}}, \bibinfo{pages}{8} (\bibinfo{year}{1995}).

\bibitem[{\citenamefont{{Meyer} et~al.}(1990)\citenamefont{{Meyer}, {Ross},
  {Pollock}, {Berdoz}, {Dohrmann}, {Goodwin}, {Minty}, {Nann}, {Pancella},
  {Pate} et~al.}}]{Meyer1990}
\bibinfo{author}{\bibfnamefont{H.~O.} \bibnamefont{{Meyer}}},
  \bibinfo{author}{\bibfnamefont{M.~A.} \bibnamefont{{Ross}}},
  \bibinfo{author}{\bibfnamefont{R.~E.} \bibnamefont{{Pollock}}},
  \bibinfo{author}{\bibfnamefont{A.}~\bibnamefont{{Berdoz}}},
  \bibinfo{author}{\bibfnamefont{F.}~\bibnamefont{{Dohrmann}}},
  \bibinfo{author}{\bibfnamefont{J.~E.} \bibnamefont{{Goodwin}}},
  \bibinfo{author}{\bibfnamefont{M.~G.} \bibnamefont{{Minty}}},
  \bibinfo{author}{\bibfnamefont{H.}~\bibnamefont{{Nann}}},
  \bibinfo{author}{\bibfnamefont{P.~V.} \bibnamefont{{Pancella}}},
  \bibinfo{author}{\bibfnamefont{S.~F.} \bibnamefont{{Pate}}},
  \bibnamefont{et~al.}, \bibinfo{journal}{Physical Review Letters}
  \textbf{\bibinfo{volume}{65}}, \bibinfo{pages}{2846} (\bibinfo{year}{1990}).

\bibitem[{\citenamefont{Meyer et~al.}(1992)\citenamefont{Meyer, Horowitz, Nann,
  Pancella, Pate, Pollock, Przewoski, Rinckel, Ross, and Sperisen}}]{Meyer1992}
\bibinfo{author}{\bibfnamefont{H.}~\bibnamefont{Meyer}},
  \bibinfo{author}{\bibfnamefont{C.}~\bibnamefont{Horowitz}},
  \bibinfo{author}{\bibfnamefont{H.}~\bibnamefont{Nann}},
  \bibinfo{author}{\bibfnamefont{P.}~\bibnamefont{Pancella}},
  \bibinfo{author}{\bibfnamefont{S.}~\bibnamefont{Pate}},
  \bibinfo{author}{\bibfnamefont{R.}~\bibnamefont{Pollock}},
  \bibinfo{author}{\bibfnamefont{B.~V.} \bibnamefont{Przewoski}},
  \bibinfo{author}{\bibfnamefont{T.}~\bibnamefont{Rinckel}},
  \bibinfo{author}{\bibfnamefont{M.}~\bibnamefont{Ross}}, \bibnamefont{and}
  \bibinfo{author}{\bibfnamefont{F.}~\bibnamefont{Sperisen}},
  \bibinfo{journal}{Nuclear Physics A} \textbf{\bibinfo{volume}{539}},
  \bibinfo{pages}{633 } (\bibinfo{year}{1992}), ISSN \bibinfo{issn}{0375-9474},
  \urlprefix\url{http://www.sciencedirect.com/science/article/pii/037594749290%
130C}.

\bibitem[{\citenamefont{{Bilger} et~al.}(2001)\citenamefont{{Bilger},
  {Brodowski}, {Cal{\'e}n}, {Clement}, {Dyring}, {Ekstr{\"o}m}, {F{\"a}ldt},
  {Fransson}, {Greiff}, {Gustafsson} et~al.}}]{Bilger2001}
\bibinfo{author}{\bibfnamefont{R.}~\bibnamefont{{Bilger}}},
  \bibinfo{author}{\bibfnamefont{W.}~\bibnamefont{{Brodowski}}},
  \bibinfo{author}{\bibfnamefont{H.}~\bibnamefont{{Cal{\'e}n}}},
  \bibinfo{author}{\bibfnamefont{H.}~\bibnamefont{{Clement}}},
  \bibinfo{author}{\bibfnamefont{J.}~\bibnamefont{{Dyring}}},
  \bibinfo{author}{\bibfnamefont{C.}~\bibnamefont{{Ekstr{\"o}m}}},
  \bibinfo{author}{\bibfnamefont{G.}~\bibnamefont{{F{\"a}ldt}}},
  \bibinfo{author}{\bibfnamefont{K.}~\bibnamefont{{Fransson}}},
  \bibinfo{author}{\bibfnamefont{J.}~\bibnamefont{{Greiff}}},
  \bibinfo{author}{\bibfnamefont{L.}~\bibnamefont{{Gustafsson}}},
  \bibnamefont{et~al.}, \bibinfo{journal}{Nuclear Physics A}
  \textbf{\bibinfo{volume}{693}}, \bibinfo{pages}{633} (\bibinfo{year}{2001}).

\bibitem[{\citenamefont{Shimizu et~al.}(1982)\citenamefont{Shimizu, Kubota,
  Koiso, Sai, Sakamoto et~al.}}]{Shimizu1982}
\bibinfo{author}{\bibfnamefont{F.}~\bibnamefont{Shimizu}},
  \bibinfo{author}{\bibfnamefont{Y.}~\bibnamefont{Kubota}},
  \bibinfo{author}{\bibfnamefont{H.}~\bibnamefont{Koiso}},
  \bibinfo{author}{\bibfnamefont{F.}~\bibnamefont{Sai}},
  \bibinfo{author}{\bibfnamefont{S.}~\bibnamefont{Sakamoto}},
  \bibnamefont{et~al.}, \bibinfo{journal}{Nucl.Phys.}
  \textbf{\bibinfo{volume}{A386}}, \bibinfo{pages}{571} (\bibinfo{year}{1982}).

\bibitem[{\citenamefont{Rappenecker et~al.}(1995)\citenamefont{Rappenecker,
  Rigney, Didelez, Hourani, Rosier et~al.}}]{Rappenecker1995}
\bibinfo{author}{\bibfnamefont{G.}~\bibnamefont{Rappenecker}},
  \bibinfo{author}{\bibfnamefont{M.}~\bibnamefont{Rigney}},
  \bibinfo{author}{\bibfnamefont{J.}~\bibnamefont{Didelez}},
  \bibinfo{author}{\bibfnamefont{E.}~\bibnamefont{Hourani}},
  \bibinfo{author}{\bibfnamefont{L.}~\bibnamefont{Rosier}},
  \bibnamefont{et~al.}, \bibinfo{journal}{Nucl.Phys.}
  \textbf{\bibinfo{volume}{A590}}, \bibinfo{pages}{763} (\bibinfo{year}{1995}).

\bibitem[{\citenamefont{El-Samad et~al.}(2006)}]{ElSamad2006}
\bibinfo{author}{\bibfnamefont{S.~A.} \bibnamefont{El-Samad}}
  \bibnamefont{et~al.} (\bibinfo{collaboration}{COSY-TOF Collaboration}),
  \bibinfo{journal}{Eur.Phys.J.} \textbf{\bibinfo{volume}{A30}},
  \bibinfo{pages}{443} (\bibinfo{year}{2006}), \eprint{nucl-ex/0609011}.

\bibitem[{\citenamefont{Andreev et~al.}(1994)\citenamefont{Andreev, Kravtsov,
  Makarov, Medvedev, Poromov et~al.}}]{Andree1988}
\bibinfo{author}{\bibfnamefont{V.~P.} \bibnamefont{Andreev}},
  \bibinfo{author}{\bibfnamefont{A.}~\bibnamefont{Kravtsov}},
  \bibinfo{author}{\bibfnamefont{M.}~\bibnamefont{Makarov}},
  \bibinfo{author}{\bibfnamefont{V.}~\bibnamefont{Medvedev}},
  \bibinfo{author}{\bibfnamefont{V.}~\bibnamefont{Poromov}},
  \bibnamefont{et~al.}, \bibinfo{journal}{Phys.Rev.}
  \textbf{\bibinfo{volume}{C50}}, \bibinfo{pages}{15} (\bibinfo{year}{1994}).

\bibitem[{\citenamefont{Ermakov et~al.}(2011)\citenamefont{Ermakov, Medvedev,
  Nikonov, Rogachevsky, Sarantsev et~al.}}]{Ermakov2011}
\bibinfo{author}{\bibfnamefont{K.}~\bibnamefont{Ermakov}},
  \bibinfo{author}{\bibfnamefont{V.}~\bibnamefont{Medvedev}},
  \bibinfo{author}{\bibfnamefont{V.}~\bibnamefont{Nikonov}},
  \bibinfo{author}{\bibfnamefont{O.}~\bibnamefont{Rogachevsky}},
  \bibinfo{author}{\bibfnamefont{A.}~\bibnamefont{Sarantsev}},
  \bibnamefont{et~al.}, \bibinfo{journal}{Eur.Phys.J.}
  \textbf{\bibinfo{volume}{A47}}, \bibinfo{pages}{159} (\bibinfo{year}{2011}),
  \eprint{1109.1111}.

\bibitem[{\citenamefont{{Bugg} et~al.}(1964{\natexlab{a}})\citenamefont{{Bugg},
  {Oxley}, {Zoll}, {Rushbrooke}, {Barnes}, {Kinson}, {Dodd}, {Doran}, and
  {Riddiford}}}]{Bugg1964}
\bibinfo{author}{\bibfnamefont{D.~V.} \bibnamefont{{Bugg}}},
  \bibinfo{author}{\bibfnamefont{A.~J.} \bibnamefont{{Oxley}}},
  \bibinfo{author}{\bibfnamefont{J.~A.} \bibnamefont{{Zoll}}},
  \bibinfo{author}{\bibfnamefont{J.~G.} \bibnamefont{{Rushbrooke}}},
  \bibinfo{author}{\bibfnamefont{V.~E.} \bibnamefont{{Barnes}}},
  \bibinfo{author}{\bibfnamefont{J.~B.} \bibnamefont{{Kinson}}},
  \bibinfo{author}{\bibfnamefont{W.~P.} \bibnamefont{{Dodd}}},
  \bibinfo{author}{\bibfnamefont{G.~A.} \bibnamefont{{Doran}}},
  \bibnamefont{and}
  \bibinfo{author}{\bibfnamefont{L.}~\bibnamefont{{Riddiford}}},
  \bibinfo{journal}{Physical Review} \textbf{\bibinfo{volume}{133}},
  \bibinfo{pages}{1017} (\bibinfo{year}{1964}{\natexlab{a}}).

\bibitem[{\citenamefont{{Agakishiev}
  et~al.}(2012{\natexlab{a}})\citenamefont{{Agakishiev}, {Alvarez-Pol},
  {Balanda}, and {\it et. al.}}}]{Agakishiev2012}
\bibinfo{author}{\bibfnamefont{G.}~\bibnamefont{{Agakishiev}}},
  \bibinfo{author}{\bibfnamefont{H.}~\bibnamefont{{Alvarez-Pol}}},
  \bibinfo{author}{\bibfnamefont{A.}~\bibnamefont{{Balanda}}},
  \bibnamefont{and} \bibinfo{author}{\bibnamefont{{\it et. al.}}},
  \bibinfo{journal}{European Physical Journal A} \textbf{\bibinfo{volume}{48}},
  \bibinfo{pages}{74} (\bibinfo{year}{2012}{\natexlab{a}}), \eprint{1203.1333}.

\bibitem[{\citenamefont{{Johanson} et~al.}(2002)\citenamefont{{Johanson},
  {Bilger}, {Brodowski}, {Cal{\'e}n}, {Clement}, {Ekstr{\"o}m}, {Fransson},
  {Greiff}, {Gustafsson}, {H{\"a}ggstr{\"o}m} et~al.}}]{Johanson2002}
\bibinfo{author}{\bibfnamefont{J.}~\bibnamefont{{Johanson}}},
  \bibinfo{author}{\bibfnamefont{R.}~\bibnamefont{{Bilger}}},
  \bibinfo{author}{\bibfnamefont{W.}~\bibnamefont{{Brodowski}}},
  \bibinfo{author}{\bibfnamefont{H.}~\bibnamefont{{Cal{\'e}n}}},
  \bibinfo{author}{\bibfnamefont{H.}~\bibnamefont{{Clement}}},
  \bibinfo{author}{\bibfnamefont{C.}~\bibnamefont{{Ekstr{\"o}m}}},
  \bibinfo{author}{\bibfnamefont{K.}~\bibnamefont{{Fransson}}},
  \bibinfo{author}{\bibfnamefont{J.}~\bibnamefont{{Greiff}}},
  \bibinfo{author}{\bibfnamefont{L.}~\bibnamefont{{Gustafsson}}},
  \bibinfo{author}{\bibfnamefont{S.}~\bibnamefont{{H{\"a}ggstr{\"o}m}}},
  \bibnamefont{et~al.}, \bibinfo{journal}{Nuclear Physics A}
  \textbf{\bibinfo{volume}{712}}, \bibinfo{pages}{75} (\bibinfo{year}{2002}).

\bibitem[{\citenamefont{{Skorodko} et~al.}(2009)\citenamefont{{Skorodko},
  {Bashkanov}, {Bogoslawsky}, {Cal{\'e}n}, {Clement}, {Doroshkevich},
  {Demiroers}, {Ekstr{\"o}m}, {Fransson}, {Gustafsson} et~al.}}]{Skorodko2009}
\bibinfo{author}{\bibfnamefont{T.}~\bibnamefont{{Skorodko}}},
  \bibinfo{author}{\bibfnamefont{M.}~\bibnamefont{{Bashkanov}}},
  \bibinfo{author}{\bibfnamefont{D.}~\bibnamefont{{Bogoslawsky}}},
  \bibinfo{author}{\bibfnamefont{H.}~\bibnamefont{{Cal{\'e}n}}},
  \bibinfo{author}{\bibfnamefont{H.}~\bibnamefont{{Clement}}},
  \bibinfo{author}{\bibfnamefont{E.}~\bibnamefont{{Doroshkevich}}},
  \bibinfo{author}{\bibfnamefont{L.}~\bibnamefont{{Demiroers}}},
  \bibinfo{author}{\bibfnamefont{C.}~\bibnamefont{{Ekstr{\"o}m}}},
  \bibinfo{author}{\bibfnamefont{K.}~\bibnamefont{{Fransson}}},
  \bibinfo{author}{\bibfnamefont{L.}~\bibnamefont{{Gustafsson}}},
  \bibnamefont{et~al.}, \bibinfo{journal}{Physics Letters B}
  \textbf{\bibinfo{volume}{679}}, \bibinfo{pages}{30} (\bibinfo{year}{2009}),
  \eprint{0906.3087}.

\bibitem[{\citenamefont{{Pauly} et~al.}(2007)\citenamefont{{Pauly},
  {Celsius-Wasa Collaboration}, {Jacewicz}, {Koch}, {Bashkanov}, {Bogoslawsky},
  {Cal{\'e}n}, {Capellaro}, {Clement}, {Demir{\"o}rs} et~al.}}]{Pauly2007}
\bibinfo{author}{\bibfnamefont{C.}~\bibnamefont{{Pauly}}},
  \bibinfo{author}{\bibnamefont{{Celsius-Wasa Collaboration}}},
  \bibinfo{author}{\bibfnamefont{M.}~\bibnamefont{{Jacewicz}}},
  \bibinfo{author}{\bibfnamefont{I.}~\bibnamefont{{Koch}}},
  \bibinfo{author}{\bibfnamefont{M.}~\bibnamefont{{Bashkanov}}},
  \bibinfo{author}{\bibfnamefont{D.}~\bibnamefont{{Bogoslawsky}}},
  \bibinfo{author}{\bibfnamefont{H.}~\bibnamefont{{Cal{\'e}n}}},
  \bibinfo{author}{\bibfnamefont{F.}~\bibnamefont{{Capellaro}}},
  \bibinfo{author}{\bibfnamefont{H.}~\bibnamefont{{Clement}}},
  \bibinfo{author}{\bibfnamefont{L.}~\bibnamefont{{Demir{\"o}rs}}},
  \bibnamefont{et~al.}, \bibinfo{journal}{Physics Letters B}
  \textbf{\bibinfo{volume}{649}}, \bibinfo{pages}{122} (\bibinfo{year}{2007}),
  \eprint{arXiv:nucl-ex/0602006}.

\bibitem[{\citenamefont{{Adlarson} et~al.}(2012)\citenamefont{{Adlarson},
  {Adolph}, {Augustyniak}, {Bashkanov}, {Bednarski}, {Bergmann},
  {Ber{\l}owski}, {Bhatt}, {Brinkmann}, {B{\"u}scher} et~al.}}]{Adlarson2012}
\bibinfo{author}{\bibfnamefont{P.}~\bibnamefont{{Adlarson}}},
  \bibinfo{author}{\bibfnamefont{C.}~\bibnamefont{{Adolph}}},
  \bibinfo{author}{\bibfnamefont{W.}~\bibnamefont{{Augustyniak}}},
  \bibinfo{author}{\bibfnamefont{M.}~\bibnamefont{{Bashkanov}}},
  \bibinfo{author}{\bibfnamefont{T.}~\bibnamefont{{Bednarski}}},
  \bibinfo{author}{\bibfnamefont{F.~S.} \bibnamefont{{Bergmann}}},
  \bibinfo{author}{\bibfnamefont{M.}~\bibnamefont{{Ber{\l}owski}}},
  \bibinfo{author}{\bibfnamefont{H.}~\bibnamefont{{Bhatt}}},
  \bibinfo{author}{\bibfnamefont{K.-T.} \bibnamefont{{Brinkmann}}},
  \bibinfo{author}{\bibfnamefont{M.}~\bibnamefont{{B{\"u}scher}}},
  \bibnamefont{et~al.}, \bibinfo{journal}{Physics Letters B}
  \textbf{\bibinfo{volume}{706}}, \bibinfo{pages}{256} (\bibinfo{year}{2012}),
  \eprint{1107.0879}.

\bibitem[{\citenamefont{Eisner et~al.}(1965)\citenamefont{Eisner, Hart,
  Louttit, and Morris}}]{Eisner1965}
\bibinfo{author}{\bibfnamefont{A.~M.} \bibnamefont{Eisner}},
  \bibinfo{author}{\bibfnamefont{E.~L.} \bibnamefont{Hart}},
  \bibinfo{author}{\bibfnamefont{R.~I.} \bibnamefont{Louttit}},
  \bibnamefont{and} \bibinfo{author}{\bibfnamefont{T.~W.}
  \bibnamefont{Morris}}, \bibinfo{journal}{Phys. Rev.}
  \textbf{\bibinfo{volume}{138}}, \bibinfo{pages}{B670} (\bibinfo{year}{1965}),
  \urlprefix\url{http://link.aps.org/doi/10.1103/PhysRev.138.B670}.

\bibitem[{\citenamefont{Pickup et~al.}(1962)\citenamefont{Pickup, Robinson, and
  Salant}}]{Pickup1962}
\bibinfo{author}{\bibfnamefont{E.}~\bibnamefont{Pickup}},
  \bibinfo{author}{\bibfnamefont{D.~K.} \bibnamefont{Robinson}},
  \bibnamefont{and} \bibinfo{author}{\bibfnamefont{E.~O.}
  \bibnamefont{Salant}}, \bibinfo{journal}{Phys. Rev.}
  \textbf{\bibinfo{volume}{125}}, \bibinfo{pages}{2091} (\bibinfo{year}{1962}),
  \urlprefix\url{http://link.aps.org/doi/10.1103/PhysRev.125.2091}.

\bibitem[{\citenamefont{{Agakichiev} et~al.}(2010)\citenamefont{{Agakichiev},
  {Balanda}, {Belver}, {Belyaev}, {Blanco}, {B{\"o}hmer}, {Boyard},
  {Braun-Munzinger}, {Cabanelas}, {Castro} et~al.}}]{Agakichiev2010}
\bibinfo{author}{\bibfnamefont{G.}~\bibnamefont{{Agakichiev}}},
  \bibinfo{author}{\bibfnamefont{A.}~\bibnamefont{{Balanda}}},
  \bibinfo{author}{\bibfnamefont{D.}~\bibnamefont{{Belver}}},
  \bibinfo{author}{\bibfnamefont{A.~V.} \bibnamefont{{Belyaev}}},
  \bibinfo{author}{\bibfnamefont{A.}~\bibnamefont{{Blanco}}},
  \bibinfo{author}{\bibfnamefont{M.}~\bibnamefont{{B{\"o}hmer}}},
  \bibinfo{author}{\bibfnamefont{J.~L.} \bibnamefont{{Boyard}}},
  \bibinfo{author}{\bibfnamefont{P.}~\bibnamefont{{Braun-Munzinger}}},
  \bibinfo{author}{\bibfnamefont{P.}~\bibnamefont{{Cabanelas}}},
  \bibinfo{author}{\bibfnamefont{E.}~\bibnamefont{{Castro}}},
  \bibnamefont{et~al.}, \bibinfo{journal}{Physics Letters B}
  \textbf{\bibinfo{volume}{690}}, \bibinfo{pages}{118} (\bibinfo{year}{2010}),
  \eprint{0910.5875}.

\bibitem[{\citenamefont{{Agakishiev}
  et~al.}(2012{\natexlab{b}})\citenamefont{{Agakishiev}, {Balanda}, {Belver},
  {Belyaev}, {Blanco}, {B{\"o}hmer}, {Boyard}, {Cabanelas}, {Castro}, {Chen}
  et~al.}}]{Agakishiev2012inc}
\bibinfo{author}{\bibfnamefont{G.}~\bibnamefont{{Agakishiev}}},
  \bibinfo{author}{\bibfnamefont{A.}~\bibnamefont{{Balanda}}},
  \bibinfo{author}{\bibfnamefont{D.}~\bibnamefont{{Belver}}},
  \bibinfo{author}{\bibfnamefont{A.}~\bibnamefont{{Belyaev}}},
  \bibinfo{author}{\bibfnamefont{A.}~\bibnamefont{{Blanco}}},
  \bibinfo{author}{\bibfnamefont{M.}~\bibnamefont{{B{\"o}hmer}}},
  \bibinfo{author}{\bibfnamefont{J.~L.} \bibnamefont{{Boyard}}},
  \bibinfo{author}{\bibfnamefont{P.}~\bibnamefont{{Cabanelas}}},
  \bibinfo{author}{\bibfnamefont{E.}~\bibnamefont{{Castro}}},
  \bibinfo{author}{\bibfnamefont{J.~C.} \bibnamefont{{Chen}}},
  \bibnamefont{et~al.}, \bibinfo{journal}{European Physical Journal A}
  \textbf{\bibinfo{volume}{48}}, \bibinfo{pages}{64}
  (\bibinfo{year}{2012}{\natexlab{b}}), \eprint{1112.3607}.

\bibitem[{\citenamefont{{AbdEl Samad, S. and
  other}}(2003)}]{El-Samad2003pi0Spec}
\bibinfo{author}{\bibnamefont{{AbdEl Samad, S. and other}}}
  (\bibinfo{collaboration}{{COSY-TOF Collaboration}}),
  \bibinfo{journal}{European Physical Journal A} \textbf{\bibinfo{volume}{17}},
  \bibinfo{pages}{595} (\bibinfo{year}{2003}), \eprint{nucl-ex/0212024}.

\bibitem[{\citenamefont{{Zloma{\'n}czuk}
  et~al.}(2000)\citenamefont{{Zloma{\'n}czuk}, {Bilger}, {Brodowski},
  {Cal{\'e}n}, {Clement}, {Dyring}, {Ekstr{\"o}m}, {F{\"a}ldt}, {Fransson},
  {Gustafsson} et~al.}}]{Zloma2000pi0Spec}
\bibinfo{author}{\bibfnamefont{J.}~\bibnamefont{{Zloma{\'n}czuk}}},
  \bibinfo{author}{\bibfnamefont{R.}~\bibnamefont{{Bilger}}},
  \bibinfo{author}{\bibfnamefont{W.}~\bibnamefont{{Brodowski}}},
  \bibinfo{author}{\bibfnamefont{H.}~\bibnamefont{{Cal{\'e}n}}},
  \bibinfo{author}{\bibfnamefont{H.}~\bibnamefont{{Clement}}},
  \bibinfo{author}{\bibfnamefont{J.}~\bibnamefont{{Dyring}}},
  \bibinfo{author}{\bibfnamefont{C.}~\bibnamefont{{Ekstr{\"o}m}}},
  \bibinfo{author}{\bibfnamefont{G.}~\bibnamefont{{F{\"a}ldt}}},
  \bibinfo{author}{\bibfnamefont{K.}~\bibnamefont{{Fransson}}},
  \bibinfo{author}{\bibfnamefont{L.}~\bibnamefont{{Gustafsson}}},
  \bibnamefont{et~al.}, \bibinfo{journal}{Nuclear Physics A}
  \textbf{\bibinfo{volume}{663}}, \bibinfo{pages}{452} (\bibinfo{year}{2000}).

\bibitem[{\citenamefont{{Andreev} et~al.}(1994)\citenamefont{{Andreev},
  {Kravtsov}, {Makarov}, {Medvedev}, {Poromov}, {Sarantsev}, {Sherman},
  {Sokolov}, and {Sokornov}}}]{Andreev1994pi0Spec}
\bibinfo{author}{\bibfnamefont{V.~P.} \bibnamefont{{Andreev}}},
  \bibinfo{author}{\bibfnamefont{A.~V.} \bibnamefont{{Kravtsov}}},
  \bibinfo{author}{\bibfnamefont{M.~M.} \bibnamefont{{Makarov}}},
  \bibinfo{author}{\bibfnamefont{V.~I.} \bibnamefont{{Medvedev}}},
  \bibinfo{author}{\bibfnamefont{V.~I.} \bibnamefont{{Poromov}}},
  \bibinfo{author}{\bibfnamefont{V.~V.} \bibnamefont{{Sarantsev}}},
  \bibinfo{author}{\bibfnamefont{S.~G.} \bibnamefont{{Sherman}}},
  \bibinfo{author}{\bibfnamefont{G.~L.} \bibnamefont{{Sokolov}}},
  \bibnamefont{and} \bibinfo{author}{\bibfnamefont{A.~B.}
  \bibnamefont{{Sokornov}}}, \bibinfo{journal}{\prc}
  \textbf{\bibinfo{volume}{50}}, \bibinfo{pages}{15} (\bibinfo{year}{1994}).

\bibitem[{\citenamefont{{Comptour} et~al.}(1994)\citenamefont{{Comptour},
  {Augerat}, {Berthot}, {Bertin}, {Bouyakhlef}, {Fonvieille}, {Audit},
  {Babinet}, {Fournier}, {Gosset} et~al.}}]{Comptour1994pi0Spec}
\bibinfo{author}{\bibfnamefont{C.}~\bibnamefont{{Comptour}}},
  \bibinfo{author}{\bibfnamefont{J.}~\bibnamefont{{Augerat}}},
  \bibinfo{author}{\bibfnamefont{J.}~\bibnamefont{{Berthot}}},
  \bibinfo{author}{\bibfnamefont{P.}~\bibnamefont{{Bertin}}},
  \bibinfo{author}{\bibfnamefont{K.}~\bibnamefont{{Bouyakhlef}}},
  \bibinfo{author}{\bibfnamefont{H.}~\bibnamefont{{Fonvieille}}},
  \bibinfo{author}{\bibfnamefont{G.}~\bibnamefont{{Audit}}},
  \bibinfo{author}{\bibfnamefont{R.}~\bibnamefont{{Babinet}}},
  \bibinfo{author}{\bibfnamefont{G.}~\bibnamefont{{Fournier}}},
  \bibinfo{author}{\bibfnamefont{J.}~\bibnamefont{{Gosset}}},
  \bibnamefont{et~al.}, \bibinfo{journal}{Nuclear Physics A}
  \textbf{\bibinfo{volume}{579}}, \bibinfo{pages}{369} (\bibinfo{year}{1994}).

\bibitem[{\citenamefont{{Bugg} et~al.}(1964{\natexlab{b}})\citenamefont{{Bugg},
  {Oxley}, {Zoll}, {Rushbrooke}, {Barnes}, {Kinson}, {Dodd}, {Doran}, and
  {Riddiford}}}]{Bugg1964pi0Spec}
\bibinfo{author}{\bibfnamefont{D.~V.} \bibnamefont{{Bugg}}},
  \bibinfo{author}{\bibfnamefont{A.~J.} \bibnamefont{{Oxley}}},
  \bibinfo{author}{\bibfnamefont{J.~A.} \bibnamefont{{Zoll}}},
  \bibinfo{author}{\bibfnamefont{J.~G.} \bibnamefont{{Rushbrooke}}},
  \bibinfo{author}{\bibfnamefont{V.~E.} \bibnamefont{{Barnes}}},
  \bibinfo{author}{\bibfnamefont{J.~B.} \bibnamefont{{Kinson}}},
  \bibinfo{author}{\bibfnamefont{W.~P.} \bibnamefont{{Dodd}}},
  \bibinfo{author}{\bibfnamefont{G.~A.} \bibnamefont{{Doran}}},
  \bibnamefont{and}
  \bibinfo{author}{\bibfnamefont{L.}~\bibnamefont{{Riddiford}}},
  \bibinfo{journal}{Physical Review} \textbf{\bibinfo{volume}{133}},
  \bibinfo{pages}{1017} (\bibinfo{year}{1964}{\natexlab{b}}).

\bibitem[{\citenamefont{{Sarantsev} et~al.}(2004)\citenamefont{{Sarantsev},
  {Ermakov}, {Medvedev}, {Oposhnyan}, {Rogachevsky}, and
  {Sherman}}}]{Sarantsev2004pi0Spec}
\bibinfo{author}{\bibfnamefont{V.~V.} \bibnamefont{{Sarantsev}}},
  \bibinfo{author}{\bibfnamefont{K.~N.} \bibnamefont{{Ermakov}}},
  \bibinfo{author}{\bibfnamefont{V.~I.} \bibnamefont{{Medvedev}}},
  \bibinfo{author}{\bibfnamefont{T.~S.} \bibnamefont{{Oposhnyan}}},
  \bibinfo{author}{\bibfnamefont{O.~V.} \bibnamefont{{Rogachevsky}}},
  \bibnamefont{and} \bibinfo{author}{\bibfnamefont{S.~G.}
  \bibnamefont{{Sherman}}}, \bibinfo{journal}{European Physical Journal A}
  \textbf{\bibinfo{volume}{21}}, \bibinfo{pages}{303} (\bibinfo{year}{2004}).

\bibitem[{\citenamefont{{Kachelrie{\ss}} and
  {Ostapchenko}}(2012)}]{Kachelries2012}
\bibinfo{author}{\bibfnamefont{M.}~\bibnamefont{{Kachelrie{\ss}}}}
  \bibnamefont{and}
  \bibinfo{author}{\bibfnamefont{S.}~\bibnamefont{{Ostapchenko}}},
  \bibinfo{journal}{\prd} \textbf{\bibinfo{volume}{86}}, \bibinfo{eid}{043004}
  (\bibinfo{year}{2012}), \eprint{1206.4705}.

\bibitem[{\citenamefont{{Glauber}}(1955)}]{Glauber1955}
\bibinfo{author}{\bibfnamefont{R.~J.} \bibnamefont{{Glauber}}},
  \bibinfo{journal}{Physical Review} \textbf{\bibinfo{volume}{100}},
  \bibinfo{pages}{242} (\bibinfo{year}{1955}).

\bibitem[{\citenamefont{{Franco} and {Glauber}}(1966)}]{Franco1966}
\bibinfo{author}{\bibfnamefont{V.}~\bibnamefont{{Franco}}} \bibnamefont{and}
  \bibinfo{author}{\bibfnamefont{R.~J.} \bibnamefont{{Glauber}}},
  \bibinfo{journal}{Physical Review} \textbf{\bibinfo{volume}{142}},
  \bibinfo{pages}{1195} (\bibinfo{year}{1966}).

\bibitem[{\citenamefont{{Glauber} and {Matthiae}}(1970)}]{Glauber1970}
\bibinfo{author}{\bibfnamefont{R.~J.} \bibnamefont{{Glauber}}}
  \bibnamefont{and}
  \bibinfo{author}{\bibfnamefont{G.}~\bibnamefont{{Matthiae}}},
  \bibinfo{journal}{Nuclear Physics B} \textbf{\bibinfo{volume}{21}},
  \bibinfo{pages}{135} (\bibinfo{year}{1970}).

\bibitem[{\citenamefont{{Bia{\l}{\l}as}
  et~al.}(1976)\citenamefont{{Bia{\l}{\l}as}, {Bleszy{\'n}ski}, and
  {Czy{\.z}}}}]{Bialas1976}
\bibinfo{author}{\bibfnamefont{A.}~\bibnamefont{{Bia{\l}{\l}as}}},
  \bibinfo{author}{\bibfnamefont{M.}~\bibnamefont{{Bleszy{\'n}ski}}},
  \bibnamefont{and}
  \bibinfo{author}{\bibfnamefont{W.}~\bibnamefont{{Czy{\.z}}}},
  \bibinfo{journal}{Nuclear Physics B} \textbf{\bibinfo{volume}{111}},
  \bibinfo{pages}{461} (\bibinfo{year}{1976}).

\bibitem[{\citenamefont{{Bradt} and {Peters}}(1950)}]{Bradt1950}
\bibinfo{author}{\bibfnamefont{H.~L.} \bibnamefont{{Bradt}}} \bibnamefont{and}
  \bibinfo{author}{\bibfnamefont{B.}~\bibnamefont{{Peters}}},
  \bibinfo{journal}{Physical Review} \textbf{\bibinfo{volume}{77}},
  \bibinfo{pages}{54} (\bibinfo{year}{1950}).

\bibitem[{\citenamefont{{Karol}}(1975)}]{Karol1975}
\bibinfo{author}{\bibfnamefont{P.~J.} \bibnamefont{{Karol}}},
  \bibinfo{journal}{\prc} \textbf{\bibinfo{volume}{11}}, \bibinfo{pages}{1203}
  (\bibinfo{year}{1975}).

\bibitem[{\citenamefont{{Letaw} et~al.}(1983)\citenamefont{{Letaw},
  {Silberberg}, and {Tsao}}}]{Letaw1983}
\bibinfo{author}{\bibfnamefont{J.~R.} \bibnamefont{{Letaw}}},
  \bibinfo{author}{\bibfnamefont{R.}~\bibnamefont{{Silberberg}}},
  \bibnamefont{and} \bibinfo{author}{\bibfnamefont{C.~H.}
  \bibnamefont{{Tsao}}}, \bibinfo{journal}{\apjs}
  \textbf{\bibinfo{volume}{51}}, \bibinfo{pages}{271} (\bibinfo{year}{1983}).

\bibitem[{\citenamefont{{Kox} et~al.}(1987)\citenamefont{{Kox}, {Gamp},
  {Perrin}, {Arvieux}, {Bertholet}, {Bruandet}, {Buenerd}, {Cherkaoui}, {Cole},
  {El-Masri} et~al.}}]{Kox1987}
\bibinfo{author}{\bibfnamefont{S.}~\bibnamefont{{Kox}}},
  \bibinfo{author}{\bibfnamefont{A.}~\bibnamefont{{Gamp}}},
  \bibinfo{author}{\bibfnamefont{C.}~\bibnamefont{{Perrin}}},
  \bibinfo{author}{\bibfnamefont{J.}~\bibnamefont{{Arvieux}}},
  \bibinfo{author}{\bibfnamefont{R.}~\bibnamefont{{Bertholet}}},
  \bibinfo{author}{\bibfnamefont{J.~F.} \bibnamefont{{Bruandet}}},
  \bibinfo{author}{\bibfnamefont{M.}~\bibnamefont{{Buenerd}}},
  \bibinfo{author}{\bibfnamefont{R.}~\bibnamefont{{Cherkaoui}}},
  \bibinfo{author}{\bibfnamefont{A.~J.} \bibnamefont{{Cole}}},
  \bibinfo{author}{\bibfnamefont{Y.}~\bibnamefont{{El-Masri}}},
  \bibnamefont{et~al.}, \bibinfo{journal}{\prc} \textbf{\bibinfo{volume}{35}},
  \bibinfo{pages}{1678} (\bibinfo{year}{1987}).

\bibitem[{\citenamefont{qing Shen et~al.}(1989)\citenamefont{qing Shen, Wang,
  Feng, long Zhan, tai Zhu, and pu~Feng}}]{Shen1989}
\bibinfo{author}{\bibfnamefont{W.}~\bibnamefont{qing Shen}},
  \bibinfo{author}{\bibfnamefont{B.}~\bibnamefont{Wang}},
  \bibinfo{author}{\bibfnamefont{J.}~\bibnamefont{Feng}},
  \bibinfo{author}{\bibfnamefont{W.}~\bibnamefont{long Zhan}},
  \bibinfo{author}{\bibfnamefont{Y.}~\bibnamefont{tai Zhu}}, \bibnamefont{and}
  \bibinfo{author}{\bibfnamefont{E.}~\bibnamefont{pu~Feng}},
  \bibinfo{journal}{Nuclear Physics A} \textbf{\bibinfo{volume}{491}},
  \bibinfo{pages}{130 } (\bibinfo{year}{1989}), ISSN \bibinfo{issn}{0375-9474},
  \urlprefix\url{http://www.sciencedirect.com/science/article/pii/037594748990%
2091}.

\bibitem[{\citenamefont{{Webber} et~al.}(1990)\citenamefont{{Webber}, {Kish},
  and {Schrier}}}]{Webber1990}
\bibinfo{author}{\bibfnamefont{W.~R.} \bibnamefont{{Webber}}},
  \bibinfo{author}{\bibfnamefont{J.~C.} \bibnamefont{{Kish}}},
  \bibnamefont{and} \bibinfo{author}{\bibfnamefont{D.~A.}
  \bibnamefont{{Schrier}}}, \bibinfo{journal}{\prc}
  \textbf{\bibinfo{volume}{41}}, \bibinfo{pages}{520} (\bibinfo{year}{1990}).

\bibitem[{\citenamefont{{Sihver} et~al.}(1993)\citenamefont{{Sihver}, {Tsao},
  {Silberberg}, {Kanai}, and {Barghouty}}}]{Sihver1993}
\bibinfo{author}{\bibfnamefont{L.}~\bibnamefont{{Sihver}}},
  \bibinfo{author}{\bibfnamefont{C.~H.} \bibnamefont{{Tsao}}},
  \bibinfo{author}{\bibfnamefont{R.}~\bibnamefont{{Silberberg}}},
  \bibinfo{author}{\bibfnamefont{T.}~\bibnamefont{{Kanai}}}, \bibnamefont{and}
  \bibinfo{author}{\bibfnamefont{A.~F.} \bibnamefont{{Barghouty}}},
  \bibinfo{journal}{\prc} \textbf{\bibinfo{volume}{47}}, \bibinfo{pages}{1225}
  (\bibinfo{year}{1993}).

\bibitem[{\citenamefont{{Tripathi} et~al.}(1996)\citenamefont{{Tripathi},
  {Cucinotta}, and {Wilson}}}]{Tripathi1996}
\bibinfo{author}{\bibfnamefont{R.~K.} \bibnamefont{{Tripathi}}},
  \bibinfo{author}{\bibfnamefont{F.~A.} \bibnamefont{{Cucinotta}}},
  \bibnamefont{and} \bibinfo{author}{\bibfnamefont{J.~W.}
  \bibnamefont{{Wilson}}}, \bibinfo{journal}{Nuclear Instruments and Methods in
  Physics Research B} \textbf{\bibinfo{volume}{117}}, \bibinfo{pages}{347}
  (\bibinfo{year}{1996}).

\bibitem[{\citenamefont{{Tripathi} et~al.}(1997)\citenamefont{{Tripathi},
  {Wilson}, and {Cucinotta}}}]{Tripathi1997}
\bibinfo{author}{\bibfnamefont{R.~K.} \bibnamefont{{Tripathi}}},
  \bibinfo{author}{\bibfnamefont{J.~W.} \bibnamefont{{Wilson}}},
  \bibnamefont{and} \bibinfo{author}{\bibfnamefont{F.~A.}
  \bibnamefont{{Cucinotta}}}, \bibinfo{journal}{Nuclear Instruments and Methods
  in Physics Research B} \textbf{\bibinfo{volume}{129}}, \bibinfo{pages}{11}
  (\bibinfo{year}{1997}).

\bibitem[{\citenamefont{{Tripathi} et~al.}(1999)\citenamefont{{Tripathi},
  {Cucinotta}, and {Wilson}}}]{Tripathi1999}
\bibinfo{author}{\bibfnamefont{R.~K.} \bibnamefont{{Tripathi}}},
  \bibinfo{author}{\bibfnamefont{F.~A.} \bibnamefont{{Cucinotta}}},
  \bibnamefont{and} \bibinfo{author}{\bibfnamefont{J.~W.}
  \bibnamefont{{Wilson}}}, \bibinfo{journal}{Nuclear Instruments and Methods in
  Physics Research B} \textbf{\bibinfo{volume}{155}}, \bibinfo{pages}{349}
  (\bibinfo{year}{1999}).

\bibitem[{\citenamefont{{Gaisser} and {Honda}}(2002)}]{Gaisser2002}
\bibinfo{author}{\bibfnamefont{T.~K.} \bibnamefont{{Gaisser}}}
  \bibnamefont{and} \bibinfo{author}{\bibfnamefont{M.}~\bibnamefont{{Honda}}},
  \bibinfo{journal}{Annual Review of Nuclear and Particle Science}
  \textbf{\bibinfo{volume}{52}}, \bibinfo{pages}{153} (\bibinfo{year}{2002}),
  \eprint{hep-ph/0203272}.

\bibitem[{\citenamefont{{Meyer}}(1985)}]{Meyer1985}
\bibinfo{author}{\bibfnamefont{J.-P.} \bibnamefont{{Meyer}}},
  \bibinfo{journal}{\apjs} \textbf{\bibinfo{volume}{57}}, \bibinfo{pages}{173}
  (\bibinfo{year}{1985}).

\bibitem[{\citenamefont{{Cavallo} and {Gould}}(1971)}]{Cavallo1971}
\bibinfo{author}{\bibfnamefont{G.}~\bibnamefont{{Cavallo}}} \bibnamefont{and}
  \bibinfo{author}{\bibfnamefont{R.~J.} \bibnamefont{{Gould}}},
  \bibinfo{journal}{Nuovo Cimento B Serie} \textbf{\bibinfo{volume}{2}},
  \bibinfo{pages}{77} (\bibinfo{year}{1971}).

\bibitem[{\citenamefont{{Mori}}(2009)}]{Mori2009}
\bibinfo{author}{\bibfnamefont{M.}~\bibnamefont{{Mori}}},
  \bibinfo{journal}{Astroparticle Physics} \textbf{\bibinfo{volume}{31}},
  \bibinfo{pages}{341} (\bibinfo{year}{2009}), \eprint{0903.3260}.

\bibitem[{\citenamefont{{Kachelriess} et~al.}(2014)\citenamefont{{Kachelriess},
  {Moskalenko}, and {Ostapchenko}}}]{Kachelriess2014}
\bibinfo{author}{\bibfnamefont{M.}~\bibnamefont{{Kachelriess}}},
  \bibinfo{author}{\bibfnamefont{I.~V.} \bibnamefont{{Moskalenko}}},
  \bibnamefont{and} \bibinfo{author}{\bibfnamefont{S.~S.}
  \bibnamefont{{Ostapchenko}}}, \bibinfo{journal}{ArXiv e-prints}
  (\bibinfo{year}{2014}), \eprint{1406.0035}.

\bibitem[{\citenamefont{{Braun-Munzinger}
  et~al.}(1984)\citenamefont{{Braun-Munzinger}, {Paul}, {Ricken}, {Stachel},
  {Zhang}, {Young}, {Obenshain}, and {Grosse}}}]{BraunMunzinger1984}
\bibinfo{author}{\bibfnamefont{P.}~\bibnamefont{{Braun-Munzinger}}},
  \bibinfo{author}{\bibfnamefont{P.}~\bibnamefont{{Paul}}},
  \bibinfo{author}{\bibfnamefont{L.}~\bibnamefont{{Ricken}}},
  \bibinfo{author}{\bibfnamefont{J.}~\bibnamefont{{Stachel}}},
  \bibinfo{author}{\bibfnamefont{P.~H.} \bibnamefont{{Zhang}}},
  \bibinfo{author}{\bibfnamefont{G.~R.} \bibnamefont{{Young}}},
  \bibinfo{author}{\bibfnamefont{F.~E.} \bibnamefont{{Obenshain}}},
  \bibnamefont{and} \bibinfo{author}{\bibfnamefont{E.}~\bibnamefont{{Grosse}}},
  \bibinfo{journal}{Physical Review Letters} \textbf{\bibinfo{volume}{52}},
  \bibinfo{pages}{255} (\bibinfo{year}{1984}).

\bibitem[{\citenamefont{{Noll} et~al.}(1984)\citenamefont{{Noll}, {Grosse},
  {Braun-Munzinger}, {Dabrowski}, {Heckwolf}, {Klepper}, {Michel},
  {M{\"u}ller}, {Stelzer}, {Brendel} et~al.}}]{Noll1984}
\bibinfo{author}{\bibfnamefont{H.}~\bibnamefont{{Noll}}},
  \bibinfo{author}{\bibfnamefont{E.}~\bibnamefont{{Grosse}}},
  \bibinfo{author}{\bibfnamefont{P.}~\bibnamefont{{Braun-Munzinger}}},
  \bibinfo{author}{\bibfnamefont{H.}~\bibnamefont{{Dabrowski}}},
  \bibinfo{author}{\bibfnamefont{H.}~\bibnamefont{{Heckwolf}}},
  \bibinfo{author}{\bibfnamefont{O.}~\bibnamefont{{Klepper}}},
  \bibinfo{author}{\bibfnamefont{C.}~\bibnamefont{{Michel}}},
  \bibinfo{author}{\bibfnamefont{W.~F.~J.} \bibnamefont{{M{\"u}ller}}},
  \bibinfo{author}{\bibfnamefont{H.}~\bibnamefont{{Stelzer}}},
  \bibinfo{author}{\bibfnamefont{C.}~\bibnamefont{{Brendel}}},
  \bibnamefont{et~al.}, \bibinfo{journal}{Physical Review Letters}
  \textbf{\bibinfo{volume}{52}}, \bibinfo{pages}{1284} (\bibinfo{year}{1984}).

\bibitem[{\citenamefont{{Badal{\'a}} et~al.}(1993)\citenamefont{{Badal{\'a}},
  {Barbera}, {Palmeri}, {Pappalardo}, {Riggi}, {Russo}, {Agodi}, {Alba},
  {Bellia}, {Coniglione} et~al.}}]{Badala1993}
\bibinfo{author}{\bibfnamefont{A.}~\bibnamefont{{Badal{\'a}}}},
  \bibinfo{author}{\bibfnamefont{R.}~\bibnamefont{{Barbera}}},
  \bibinfo{author}{\bibfnamefont{A.}~\bibnamefont{{Palmeri}}},
  \bibinfo{author}{\bibfnamefont{G.~S.} \bibnamefont{{Pappalardo}}},
  \bibinfo{author}{\bibfnamefont{F.}~\bibnamefont{{Riggi}}},
  \bibinfo{author}{\bibfnamefont{A.~C.} \bibnamefont{{Russo}}},
  \bibinfo{author}{\bibfnamefont{C.}~\bibnamefont{{Agodi}}},
  \bibinfo{author}{\bibfnamefont{R.}~\bibnamefont{{Alba}}},
  \bibinfo{author}{\bibfnamefont{G.}~\bibnamefont{{Bellia}}},
  \bibinfo{author}{\bibfnamefont{R.}~\bibnamefont{{Coniglione}}},
  \bibnamefont{et~al.}, \bibinfo{journal}{\prc} \textbf{\bibinfo{volume}{47}},
  \bibinfo{pages}{231} (\bibinfo{year}{1993}).

\bibitem[{\citenamefont{{Badal{\`a}} et~al.}(1996)\citenamefont{{Badal{\`a}},
  {Barbera}, {Palmeri}, {Pappalardo}, {Riggi}, {Russo}, {Russo}, and
  {Turrisi}}}]{Badala1996}
\bibinfo{author}{\bibfnamefont{A.}~\bibnamefont{{Badal{\`a}}}},
  \bibinfo{author}{\bibfnamefont{R.}~\bibnamefont{{Barbera}}},
  \bibinfo{author}{\bibfnamefont{A.}~\bibnamefont{{Palmeri}}},
  \bibinfo{author}{\bibfnamefont{G.~S.} \bibnamefont{{Pappalardo}}},
  \bibinfo{author}{\bibfnamefont{F.}~\bibnamefont{{Riggi}}},
  \bibinfo{author}{\bibfnamefont{A.~C.} \bibnamefont{{Russo}}},
  \bibinfo{author}{\bibfnamefont{G.}~\bibnamefont{{Russo}}}, \bibnamefont{and}
  \bibinfo{author}{\bibfnamefont{R.}~\bibnamefont{{Turrisi}}},
  \bibinfo{journal}{\prc} \textbf{\bibinfo{volume}{53}}, \bibinfo{pages}{1782}
  (\bibinfo{year}{1996}).

\bibitem[{\citenamefont{Behnke}(2014)}]{Behnke2014HADES}
\bibinfo{author}{\bibfnamefont{C.}~\bibnamefont{Behnke}}
  (\bibinfo{collaboration}{HADES}), \bibinfo{journal}{J.Phys.Conf.Ser.}
  \textbf{\bibinfo{volume}{503}}, \bibinfo{pages}{012015}
  (\bibinfo{year}{2014}).

\bibitem[{\citenamefont{{d'Enterria} et~al.}(2011)\citenamefont{{d'Enterria},
  {Engel}, {Pierog}, {Ostapchenko}, and {Werner}}}]{Enterria2011}
\bibinfo{author}{\bibfnamefont{D.}~\bibnamefont{{d'Enterria}}},
  \bibinfo{author}{\bibfnamefont{R.}~\bibnamefont{{Engel}}},
  \bibinfo{author}{\bibfnamefont{T.}~\bibnamefont{{Pierog}}},
  \bibinfo{author}{\bibfnamefont{S.}~\bibnamefont{{Ostapchenko}}},
  \bibnamefont{and} \bibinfo{author}{\bibfnamefont{K.}~\bibnamefont{{Werner}}},
  \bibinfo{journal}{Astroparticle Physics} \textbf{\bibinfo{volume}{35}},
  \bibinfo{pages}{98} (\bibinfo{year}{2011}), \eprint{1101.5596}.

\bibitem[{\citenamefont{{Adriani}
  et~al.}(2012{\natexlab{a}})\citenamefont{{Adriani}, {Bonechi}, {Bongi},
  {Castellini}, {D'Alessandro}, {Fukatsu}, {Haguenauer}, {Iso}, {Itow},
  {Kasahara} et~al.}}]{LHCf7TeVPi0}
\bibinfo{author}{\bibfnamefont{O.}~\bibnamefont{{Adriani}}},
  \bibinfo{author}{\bibfnamefont{L.}~\bibnamefont{{Bonechi}}},
  \bibinfo{author}{\bibfnamefont{M.}~\bibnamefont{{Bongi}}},
  \bibinfo{author}{\bibfnamefont{G.}~\bibnamefont{{Castellini}}},
  \bibinfo{author}{\bibfnamefont{R.}~\bibnamefont{{D'Alessandro}}},
  \bibinfo{author}{\bibfnamefont{K.}~\bibnamefont{{Fukatsu}}},
  \bibinfo{author}{\bibfnamefont{M.}~\bibnamefont{{Haguenauer}}},
  \bibinfo{author}{\bibfnamefont{T.}~\bibnamefont{{Iso}}},
  \bibinfo{author}{\bibfnamefont{Y.}~\bibnamefont{{Itow}}},
  \bibinfo{author}{\bibfnamefont{K.}~\bibnamefont{{Kasahara}}},
  \bibnamefont{et~al.}, \bibinfo{journal}{\prd} \textbf{\bibinfo{volume}{86}},
  \bibinfo{eid}{092001} (\bibinfo{year}{2012}{\natexlab{a}}),
  \eprint{1205.4578}.

\bibitem[{\citenamefont{{Adriani}
  et~al.}(2012{\natexlab{b}})\citenamefont{{Adriani}, {Bonechi}, {Bongi},
  {Castellini}, {D'Alessandro}, {Fukatsu}, {Haguenauer}, {Iso}, {Itow},
  {Kasahara} et~al.}}]{LHCf09TeVGam}
\bibinfo{author}{\bibfnamefont{O.}~\bibnamefont{{Adriani}}},
  \bibinfo{author}{\bibfnamefont{L.}~\bibnamefont{{Bonechi}}},
  \bibinfo{author}{\bibfnamefont{M.}~\bibnamefont{{Bongi}}},
  \bibinfo{author}{\bibfnamefont{G.}~\bibnamefont{{Castellini}}},
  \bibinfo{author}{\bibfnamefont{R.}~\bibnamefont{{D'Alessandro}}},
  \bibinfo{author}{\bibfnamefont{K.}~\bibnamefont{{Fukatsu}}},
  \bibinfo{author}{\bibfnamefont{M.}~\bibnamefont{{Haguenauer}}},
  \bibinfo{author}{\bibfnamefont{T.}~\bibnamefont{{Iso}}},
  \bibinfo{author}{\bibfnamefont{Y.}~\bibnamefont{{Itow}}},
  \bibinfo{author}{\bibfnamefont{K.}~\bibnamefont{{Kasahara}}},
  \bibnamefont{et~al.}, \bibinfo{journal}{Physics Letters B}
  \textbf{\bibinfo{volume}{715}}, \bibinfo{pages}{298}
  (\bibinfo{year}{2012}{\natexlab{b}}), \eprint{1207.7183}.

\bibitem[{\citenamefont{{Menjo} et~al.}(2012)\citenamefont{{Menjo}, {Adriani},
  {Bongi}, {Castellini}, {D'Alessandro}, {Fukatsu}, {Haguenauer}, {Itow},
  {Kasahara}, {Kawade} et~al.}}]{LHCf7TeVGam}
\bibinfo{author}{\bibfnamefont{H.}~\bibnamefont{{Menjo}}},
  \bibinfo{author}{\bibfnamefont{O.}~\bibnamefont{{Adriani}}},
  \bibinfo{author}{\bibfnamefont{M.}~\bibnamefont{{Bongi}}},
  \bibinfo{author}{\bibfnamefont{G.}~\bibnamefont{{Castellini}}},
  \bibinfo{author}{\bibfnamefont{R.}~\bibnamefont{{D'Alessandro}}},
  \bibinfo{author}{\bibfnamefont{K.}~\bibnamefont{{Fukatsu}}},
  \bibinfo{author}{\bibfnamefont{M.}~\bibnamefont{{Haguenauer}}},
  \bibinfo{author}{\bibfnamefont{Y.}~\bibnamefont{{Itow}}},
  \bibinfo{author}{\bibfnamefont{K.}~\bibnamefont{{Kasahara}}},
  \bibinfo{author}{\bibfnamefont{K.}~\bibnamefont{{Kawade}}},
  \bibnamefont{et~al.}, \bibinfo{journal}{Nuclear Instruments and Methods in
  Physics Research A} \textbf{\bibinfo{volume}{692}}, \bibinfo{pages}{224}
  (\bibinfo{year}{2012}).

\bibitem[{\citenamefont{{Aharonian} and {Atoyan}}(2000)}]{aharonian_atoyan2000}
\bibinfo{author}{\bibfnamefont{F.~A.} \bibnamefont{{Aharonian}}}
  \bibnamefont{and} \bibinfo{author}{\bibfnamefont{A.~M.}
  \bibnamefont{{Atoyan}}}, \bibinfo{journal}{\aap}
  \textbf{\bibinfo{volume}{362}}, \bibinfo{pages}{937} (\bibinfo{year}{2000}),
  \eprint{arXiv:astro-ph/0009009}.

\bibitem[{\citenamefont{{Carlson} and {Profumo}}(2014)}]{Profumo2014}
\bibinfo{author}{\bibfnamefont{E.}~\bibnamefont{{Carlson}}} \bibnamefont{and}
  \bibinfo{author}{\bibfnamefont{S.}~\bibnamefont{{Profumo}}},
  \bibinfo{journal}{ArXiv e-prints}  (\bibinfo{year}{2014}),
  \eprint{1405.7685}.

\end{thebibliography}

\end{document}